\documentclass[floatfix,superscriptaddress,twocolumn, aps, prb]{revtex4-2}
\usepackage{dcolumn}
\usepackage{graphicx}
\usepackage{bm}
\usepackage{physics}
\usepackage[mathlines]{lineno}
\usepackage{setspace}
\usepackage[utf8]{inputenc}
\usepackage[english]{babel}
\usepackage[T1]{fontenc}
\usepackage[dvipsnames]{xcolor}
\usepackage{amssymb}
\usepackage{amsmath}
\usepackage{xcolor}
\usepackage{chemist}
\usepackage{graphicx}
\graphicspath{ {./images/} }
\usepackage{chemformula}
\usepackage{siunitx}

\begin{document}

\title{Structural evolution of iron oxides melts at Earth's outer-core pressures}

\author{C\'eline Cr\'episson}
\email{celine.crepisson@physics.ox.ac.uk}
\affiliation{Department of Physics, Clarendon Laboratory, University of Oxford, Parks Road, Oxford OX1 3PU, UK}

\author{Mila Fitzgerald}
\affiliation{Department of Physics, Clarendon Laboratory, University of Oxford, Parks Road, Oxford OX1 3PU, UK}

\author{Domenic Peake}
\affiliation{Department of Physics, Clarendon Laboratory, University of Oxford, Parks Road, Oxford OX1 3PU, UK}

\author{Patrick Heighway}
\affiliation{Department of Physics, Clarendon Laboratory, University of Oxford, Parks Road, Oxford OX1 3PU, UK}

\author{Thomas Stevens}
\affiliation{Department of Physics, Clarendon Laboratory, University of Oxford, Parks Road, Oxford OX1 3PU, UK}

\author{Adrien Descamps}
\affiliation{School of Mathematics and Physics, Queen’s University Belfast, University Road, Belfast BT7 1NN, United Kingdom}

\author{David McGonegle}
\affiliation{AWE, Aldermaston, Reading, RG7 4PR, United Kingdom}

\author{Alexis Amouretti}
\affiliation{Graduate School of Engineering, University of Osaka, Suita, Osaka 565-0871, Japan}

\author{Karim K. Alaa El-Din}
\affiliation{Department of Physics, Clarendon Laboratory, University of Oxford, Parks Road, Oxford OX1 3PU, UK}

\author{Michal Andrzejewski}
\affiliation{European XFEL, Holzkoppel 4, 22869 Schenefeld, Germany}

\author{Sam Azadi}
\affiliation{Department of Physics, Clarendon Laboratory, University of Oxford, Parks Road, Oxford OX1 3PU, UK}
\affiliation{Department of Physics and Astronomy, University of Manchester, Oxford Road, Manchester M13 9PL, United Kingdom}

\author{Erik Brambrink}
\affiliation{European XFEL, Holzkoppel 4, 22869 Schenefeld, Germany}

\author{Carolina Camarda}
\affiliation{European XFEL, Holzkoppel 4, 22869 Schenefeld, Germany}

\author{David A. Chin}
\affiliation{University of Rochester Laboratory for Laser Energetics, Rochester, NY, USA}

\author{Samuele Di Dio Cafiso}
\affiliation{Helmholtz-Zentrum Dresden-Rossendorf (HZDR), Bautzner Landstraße 400, Dresden, 01328, Germany}

\author{Ana Coutinho Dutra}
\affiliation{Department of Physics, Clarendon Laboratory, University of Oxford, Parks Road, Oxford OX1 3PU, UK}

\author{Hauke Höppner}
\affiliation{Helmholtz-Zentrum Dresden-Rossendorf (HZDR), Bautzner Landstraße 400, Dresden, 01328, Germany}

\author{Kohdai Yamamoto}
\affiliation{Graduate School of Engineering, University of Osaka, Suita, Osaka 565-0871, Japan}

\author{Zuzana Konôpková}
\affiliation{European XFEL, Holzkoppel 4, 22869 Schenefeld, Germany}

\author{Motoaki Nakatsutsumi}
\affiliation{European XFEL, Holzkoppel 4, 22869 Schenefeld, Germany}

\author{Norimasa Ozaki}
\affiliation{Graduate School of Engineering, University of Osaka, Suita, Osaka 565-0871, Japan}
\affiliation{Institute of Laser Engineering, University of Osaka, Suita, Osaka 565-0871, Japan}

\author{Danae N. Polsin}
\affiliation{University of Rochester Laboratory for Laser Energetics, Rochester, NY, USA}

\author{Jan-Patrick Schwinkendorf}
\affiliation{Helmholtz-Zentrum Dresden-Rossendorf (HZDR), Bautzner Landstraße 400, Dresden, 01328, Germany}

\author{Georgiy Shoulga}
\affiliation{Helmholtz-Zentrum Dresden-Rossendorf (HZDR), Bautzner Landstraße 400, Dresden, 01328, Germany}

\author{Cornelius Strohm}
\affiliation{European XFEL, Holzkoppel 4, 22869 Schenefeld, Germany}

\author{Minxue Tang}
\affiliation{European XFEL, Holzkoppel 4, 22869 Schenefeld, Germany}

\author{Harry Taylor}
\affiliation{Department of Physics, Clarendon Laboratory, University of Oxford, Parks Road, Oxford OX1 3PU, UK}

\author{Monika Toncian}
\affiliation{Helmholtz-Zentrum Dresden-Rossendorf (HZDR), Bautzner Landstraße 400, Dresden, 01328, Germany}

\author{Yizhen Wang}
\affiliation{Department of Physics, Clarendon Laboratory, University of Oxford, Parks Road, Oxford OX1 3PU, UK}

\author{Jin Yao}
\affiliation{National Thin Film Cluster Facility for Advanced Functional Materials, Department of Physics, University of Oxford, Parks Road, Oxford OX1 3PU, UK}

\author{Gianluca Gregori}
\affiliation{Department of Physics, Clarendon Laboratory, University of Oxford, Parks Road, Oxford OX1 3PU, UK}

\author{Justin S. Wark}
\affiliation{Department of Physics, Clarendon Laboratory, University of Oxford, Parks Road, Oxford OX1 3PU, UK}

\author{Karen Appel}
\affiliation{European XFEL, Holzkoppel 4, 22869 Schenefeld, Germany}

\author{Marion Harmand}
\affiliation{PIMM, Arts et Metiers Institute of Technology, CNRS, Cnam,
HESAM University, 151 boulevard de l’H\^opital, 75013 Paris, France}

\author{Sam M. Vinko}
\affiliation{Department of Physics, Clarendon Laboratory, University of Oxford, Parks Road, Oxford OX1 3PU, UK}

\date{\today}

\begin{abstract}
Oxygen and other light elements comprise up to 5 wt\% of the Earth’s outer-core, and may significantly influence its physical properties and the operation of the geodynamo. Here we report {\it in situ} x-ray diffraction measurements of Fe, Fe + 4.5 FeO (atomic proportion), and Fe$_2$O$_3$ melts at 177–438~GPa, achieved using laser-driven shock compression at an x-ray free-electron laser. The melts exhibit Fe–O coordination numbers between 4.0(0.4) and 4.5(0.4), indicating predominantly four-fold coordination environments. These coordination states are significantly smaller than those of Fe-bearing lower-mantle phases such as bridgmanite and ferropericlase. Shorter Fe-Fe interatomic distances in compressed iron oxide melts drive the denser packing relative to ambient melts, while the structural differences between Fe + 4.5 FeO and Fe$_2$O$_3$ melts under shock indicate that the oxidation state modulates oxygen solubility in liquid Fe. At $\sim$177~GPa ($\sim$380~km below the core-mantle boundary), Fe$_2$O$_3$ melts exhibit higher Fe–O coordination, suggesting that local variations in oxygen content could contribute to the stratification in the uppermost outer-core inferred from seismological and geomagnetic observations.
\end{abstract}

\maketitle

\section{Introduction}

 \begin{figure*}
     \centering
     \includegraphics[width=1\textwidth]{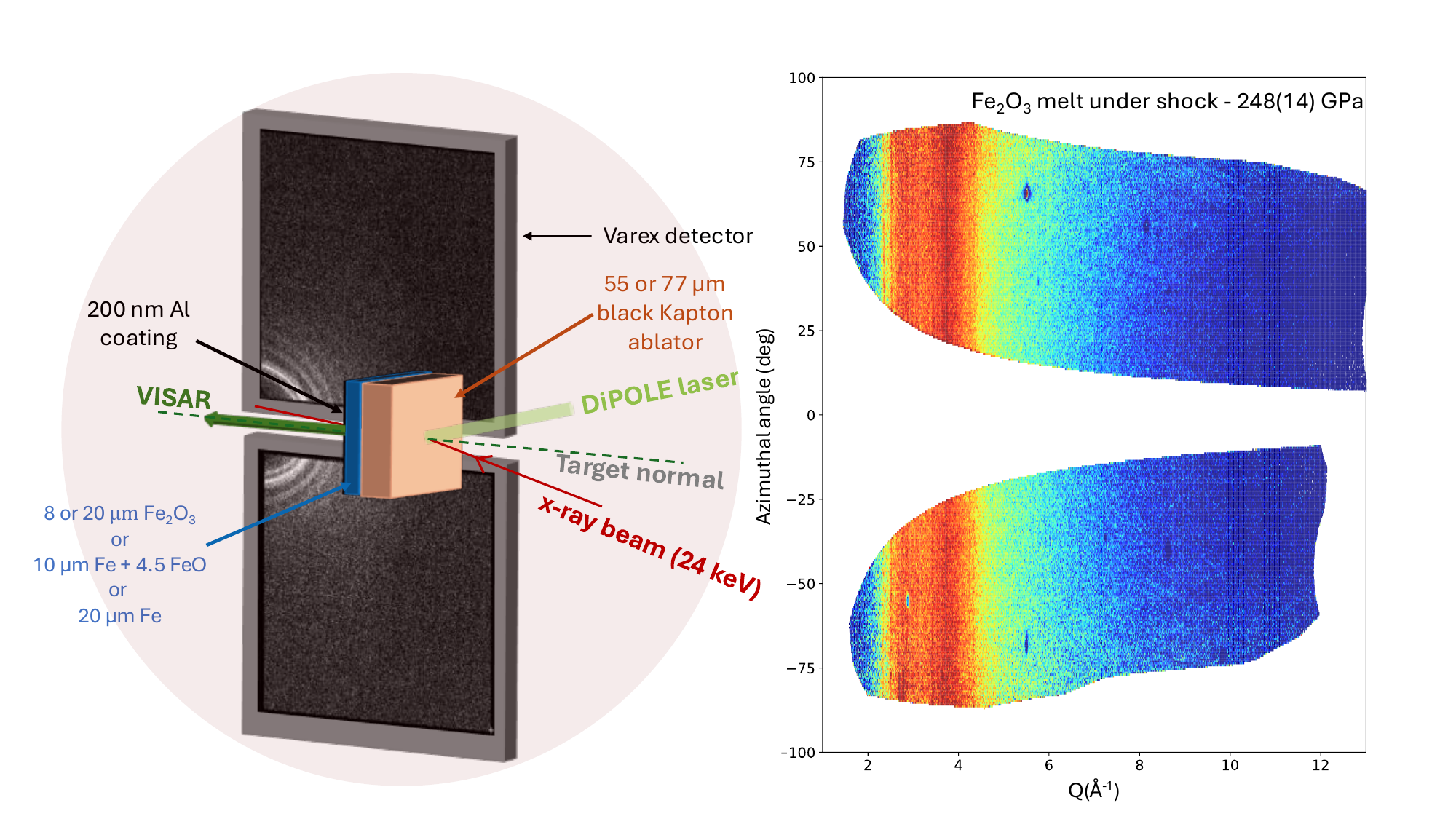}
     \caption{Left: Experimental setup at the HED endstation of the EuXFEL, showing a 2D raw diffraction image of an Fe$_2$O$_3$ target at ambient conditions prior to shock. The DiPOLE 100-X laser is shown in light green, the VISAR probe in dark green, and the x-ray beam in red. Fe$_2$O$_3$ and Fe + 4.5 FeO layers were deposited on a black Kapton ablator via plasma sputtering, while the Fe foil was glued to the ablator. Right: Dewarped 2D XRD image from the Varex detectors for Fe$_2$O$_3$ melt under shock probed around 0.1~ns before the shock exits the target.}
     \label{Fig:set_up}
\end{figure*}

The Earth’s iron-rich outer-core contains up to $\sim$5\% oxygen, together with other light elements such as S, Si, C, and H, incorporated into liquid Fe during core formation through segregation of metal and silicates from a primordial magma ocean~\cite{hirose21,rubie2007}. Similar enrichment has been inferred for Mars, where seismological observations suggest that both the liquid and solid cores contain O, S and C~\cite{bi_2025}. Within the Earth, convection of this light-element-bearing liquid iron sustains the geodynamo, generating the magnetic field that shields the planet’s surface~\cite{buffet2000}. As a magnetic field plays a central role in preserving planetary atmospheres and surface environments, its detection is often considered a primary indicator of potential habitability in exoplanets. Recent studies further suggest that most super-Earths host at least partially molten metallic cores~\cite{white}, comparable to Earth’s. Understanding how such cores evolve and generate magnetic fields is therefore essential for assessing the interior dynamics and habitability of terrestrial exoplanets.

Light elements dissolved in the liquid core strongly influence its transport properties—such as viscosity and thermal and electrical conductivities—which are key parameters governing magnetic field generation~\cite{buffet2000}. During planetary evolution, crystallization of the solid inner core expels these light elements into the overlying liquid, where their upward diffusion and accumulation provides a sustained source of compositional buoyancy that drives outer-core convection~\cite{buffet2000}.

Seismological analyses of body waves and normal modes reveal anomalously low velocities at the top of the Earth’s outer-core, suggesting the presence of a stably stratified, non-convecting layer that could extend up to $\sim$450~km in thickness~\cite{kaneshima}, though its existence remains debated~\cite{irving}. Independent evidence from geomagnetic fluctuations similarly indicates a stratified layer approximately 140~km thick~\cite{buffet2014,buffet2016}. The origin of this stable region remains uncertain, with both thermal~\cite{mound,greenwood} and compositional~\cite{gubbins,helfrich,brodholtbadro} mechanisms proposed. A preferential enrichment in oxygen at the top of the outer-core relative to other light elements~\cite{brodholtbadro}, combined with non-ideal mixing behaviour~\cite{helfrich,yokoo2022}, has been suggested as a possible explanation for the observed reductions in seismic velocity and density. Oxygen partitioning models further imply that an early O-rich layer may have formed just beneath the core–mantle boundary, indicating that the early core was initially undersaturated in oxygen~\cite{pozzo_2019}.
Changes in the Fe–O bonding environment within the outer-core have also been inferred from the B1–B2 phase transition of FeO near 240~GPa~\cite{ozawa}, while {\it ab initio} molecular dynamics simulations of Fe$_{0.96}$O$_{0.04}$ indicate pressure-induced modifications of Fe–O coordination consistent with evolving bonding behaviour under core conditions~\cite{posner}.

Despite extensive theoretical work, our current understanding of melt structure of Fe–O under Earth’s outer-core conditions remains entirely based on {\it ab initio} calculations~\cite{posner,alfe,ohmura,ohmura_2022}. Our previous work on Fe$_2$O$_3$ did not allow to retrieve interatomic distances and coordination numbers due to reduced Q range~\cite{crepissonPRB}, and no other experimental measurement exists for iron oxide melts at pressures corresponding to the outer-core (136–330~GPa)~\cite{dziewonski}. Structural information for pure Fe melt has been obtained up to 275(9)~GPa under laser-driven shock compression~\cite{singh}, whereas Fe$_{0.92}$O melt has been examined only up to $\sim$90~GPa by x-ray diffraction under static compression~\cite{morard2022}. The scarcity of experimental data primarily reflects the extremely high melting temperatures of Fe–O systems, which make {\it in situ} structural measurements challenging under static compression. Laser-driven shock compression, by contrast, enables access to extreme pressure–temperature conditions albeit on nanosecond timescales. Although kinetic effects can lead to differences from static phase equilibria (as seen recently in Fe$_2$O$_3$~\cite{amourettiPRL}), melting occurs on sub-nanosecond timescales above the liquidus~\cite{renganathan}, allowing the melt structure relevant to planetary interior conditions to be probed reliably using dynamic compression techniques.

In this work, we use laser-driven shock compression to experimentally determine the evolution of the Fe–O bonding environment across the Earth’s outer-core pressure range in Fe$^{3+}$ and Fe$^{2+}$-rich iron oxide melts. We report measurements of the melt structure and density of Fe, Fe + 4.5 FeO (55.0(0.4) at\% Fe and 45.0(0.4) at\% O, equivalent to Fe + 4.5 FeO in atomic proportion, assuming stoichiometric FeO composition, with Fe and FeO homogeneously distributed across the sample), and Fe$_2$O$_3$ from 177 (14) up to 400(40) GPa, i.e., pressures directly relevant to the Earth’s outer-core. The primary diagnostic is {\it in situ} x-ray diffraction (XRD) measured in transmission from laser-driven, shock-compressed samples at the High Energy Density endstation of the European X-ray Free-Electron Laser (EuXFEL)~\cite{zastrau}. The XRD intensities are fully corrected for solid angle, polarization, the different apparent thickness due to different angles of the Al filter placed over the detectors, self-attenuation, ablator contribution, and residual ambient material, as detailed in the Methods section. A Velocity Interferometer System for Any Reflector (VISAR) measured the shock breakout time, and hydrodynamic simulations were used to derive the corresponding pressures. The experimental setup is shown in Fig.~\ref{Fig:set_up}.

\section{Results}

\begin{figure}
     \centering
     \includegraphics[width=1\columnwidth]{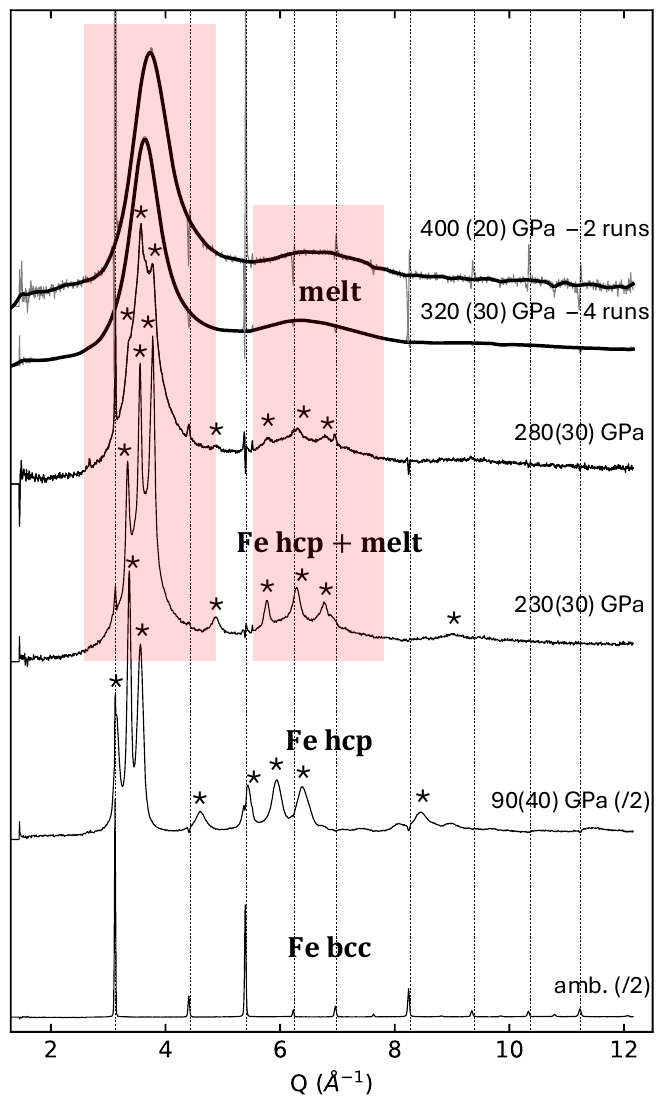}
     \caption{Fully corrected, azimuthally integrated 1D XRD lineout of Fe under laser-driven shock compression, showing the transformation from bcc to hcp Fe, followed by partial and then complete melting above 280(30)~GPa. Vertical dashed lines mark residual ambient bcc peaks, and black stars indicate the main hcp peaks under pressure. Lattice parameters are provided in Section VII and Table 1 of the Supplemental Material. Removal of remaining ambient material produces small dips at the positions of ambient peaks, reflecting sample texture and resulting in differences in relative peak intensity between the shocked and ambient samples. The diffuse scattering region is highlighted in red.}
     \label{Fig:Fe_intensity}
\end{figure}

\begin{figure}
     \centering
     \includegraphics[width=0.9\columnwidth]{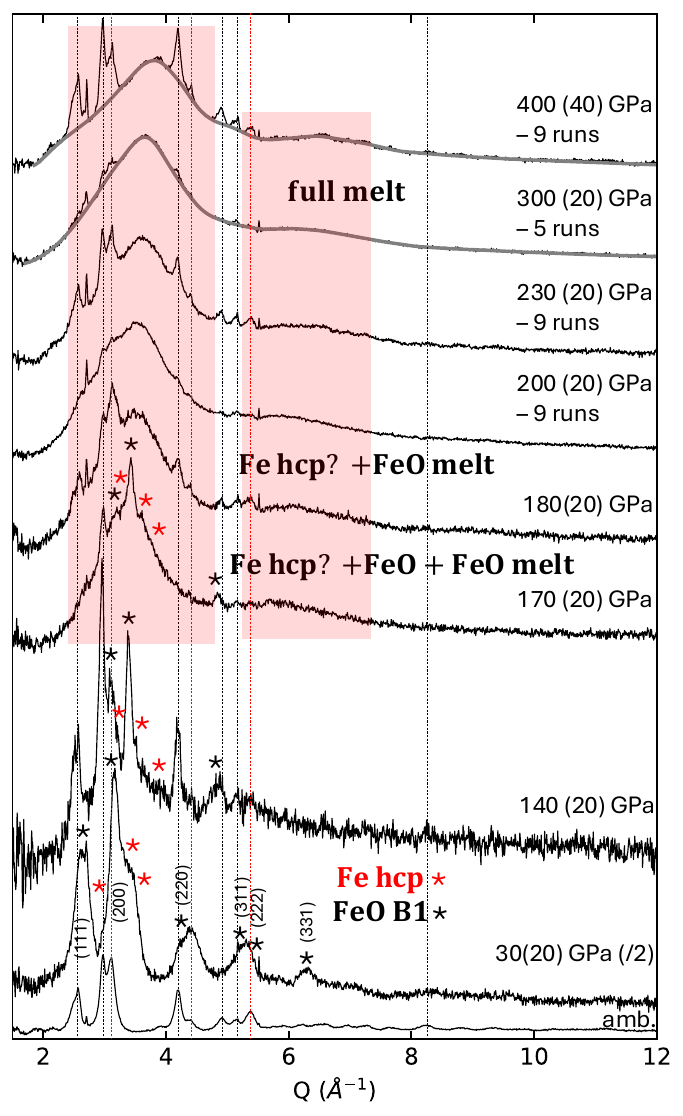}
     \caption{Fully corrected, azimuthally integrated 1D XRD lineout of Fe + 4.5 FeO under laser-driven shock compression. No phase transition is observed in FeO, which remains in the B1 (rocksalt) structure. Above 170(20)~GPa, the disappearance of FeO peaks indicates melting along the Hugoniot. Hcp Fe is seen at 30(20)~GPa, but the hcp reflections become indistinct at higher pressures. Only data above 280 GPa (above Fe melting point under shock as seen in Fig.~\ref{Fig:Fe_intensity}) are taken as fully molten Fe + 4.5 FeO. Vertical dashed lines in black and red mark residual ambient FeO and bcc Fe peaks, respectively; black stars denote the main FeO peaks, and red stars indicate hcp Fe peaks, under pressure. The diffuse scattering region is highlighted in red.}
     \label{Fig:FeO_intensity}
\end{figure}

\begin{figure}
     \centering
     \includegraphics[width=0.9\columnwidth]{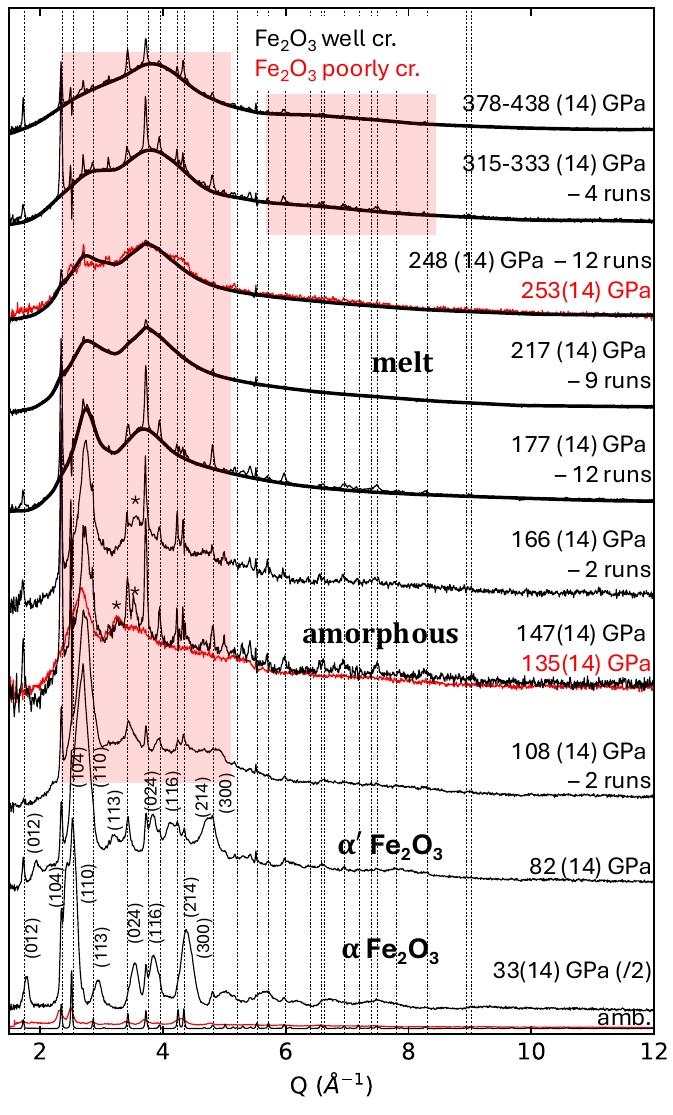}
     \caption{Fully corrected, azimuthally integrated 1D XRD lineout of Fe$_2$O$_3$ under laser-driven shock compression, showing: (1) the $\alpha$ to $\alpha^\prime$-Fe$_2$O$_3$ phase transition between 33(14) and 82(14)~GPa; (2) amorphization from 108(14)~GPa; and (3) melting from 177(14) or 217(14)~GPa. Two samples of differing crystallinity were used: a well-crystalline sample (black) for most shots, and a poorly crystalline sample (red) for two shots.
     Vertical dashed lines mark residual ambient Fe$_2$O$_3$ peaks, while black stars denote peaks visible in the amorphous phase. Removal of remaining ambient material produces small dips at the positions of ambient peaks, reflecting sample texture and resulting in variations in relative peak intensity between shocked and ambient samples. The diffuse scattering region is highlighted in red.}
     \label{Fig:Fe2O3_intensity}
\end{figure}

\subsection{Phase evolution and melting of Fe, Fe + 4.5 FeO and Fe$_2$O$_3$ along the Hugoniot}

Fe transforms from a body-centred cubic (bcc) to a hexagonal close-packed (hcp) structure under shock compression at 13~GPa~\cite{bancroft}. In Fig.~\ref{Fig:Fe_intensity}, coexistence of hcp Fe and partial melt is observed from 230(30)~GPa, while complete melting occurs above 280(30)~GPa. Fully molten Fe has previously been reported under shock at 258(8)~GPa~\cite{singh} and 273(2)~GPa~\cite{turneaure}, consistent with our pressure within the uncertainties.

As seen in Fig.~\ref{Fig:FeO_intensity}, FeO shows no phase transition along the Hugoniot and remains in the B1 (rocksalt) structure up to 170(20)~GPa. This behaviour agrees with static compression phase diagrams~\cite{fisher,greenberg_FeO} and the pressure–temperature relation along the FeO Hugoniot~\cite{jeanloz}. A detailed analysis of volume evolution will be presented elsewhere. At 170(20)~GPa, diffuse scattering appears around 3.4 and 6.2~\AA$^{-1}$ (highlighted in red in Fig.~\ref{Fig:FeO_intensity}), indicating partial melting with Bragg peaks of the FeO B1 structure still visible. At 180(20)~GPa, all FeO Bragg peaks disappear, suggesting that the Hugoniot crosses the FeO liquidus between 170(20) and 180(20)~GPa. The bcc Fe present in the initial sample is observed to have transformed into hcp Fe at 30(20)~GPa (bcc to hcp Fe transition is known to occur at 13~GPa~\cite{bancroft}). At higher pressures, Fe reflections are no longer detected, possibly due to very small domains size for hcp Fe (2-15~nm~\cite{hawreliak2008}) and compatibility strains between the Fe grains and the FeO matrix which could reduce the crystallinity of the Fe near its boundaries. Consequently, only data above 280~GPa where Fe is fully molten (see Fig.~\ref{Fig:Fe_intensity}) are taken to represent complete Fe + 4.5 FeO melting.

In Fig.~\ref{Fig:Fe2O3_intensity}, we see that Fe$_2$O$_3$ transforms from the $\alpha$ to the $\alpha^\prime$-Fe$_2$O$_3$ phase below 97(14)~GPa. This is consistent with previous shock-compression results where this isostructural transition occurred at 50–62~GPa, possibly alongside a spin transition~\cite{amourettiPRL}. None of the four phase transitions reported under static compression~\cite{bykova} are observed. Diffuse scattering (highlighted in red in Fig.~\ref{Fig:Fe2O3_intensity}) and the disappearance of $\alpha^\prime$ Bragg peaks is seen from 108(14)~GPa, indicating amorphization. This is consistent with an earlier study, where amorphization occurred at 122(3)~GPa~\cite{crepissonPRB}, at temperatures too low to induce equilibrium melting (1200~K at 108~GPa, according to SESAME 7440~\cite{barnes_sesame_1987}). At 147(14) and 166(14)~GPa, two peaks at 3.2 and 3.5~\AA$^{-1}$ (marked by black stars in Fig.~\ref{Fig:Fe2O3_intensity}) appear alongside the amorphous phase. These peaks intensify upon release and may correspond to residual $\alpha^\prime$; further discussion is provided in Sections IX and X of the Supplemental Material.

The comparison of two Fe$_2$O$_3$ samples of identical composition but differing initial crystallinity (Fig.~\ref{Fig:Fe2O3_intensity}) allows for a clear distinction between amorphous and molten phases. At 135–147(14)~GPa, the samples display distinct diffraction signatures, consistent with an amorphous or partially amorphous phase that retains features of the original crystallinity. In contrast, at 248–253(14)~GPa, both samples exhibit nearly identical signals, indicative of complete melting. We thus conclude that Fe$_2$O$_3$ is amorphous at 147(14)~GPa, and likely also at 166(14)~GPa, as the corresponding patterns are similar (see Fig.~\ref{Fig:Fe2O3_intensity} and section X of the Supplemental Material for more details). From 217(14)~GPa onwards Fe$_2$O$_3$ is fully molten [XRD signatures at 217(14) and 248(14)~GPa being nearly identical].

The position of the main amorphous peak at $\sim$2.74~\AA$^{-1}$ remains constant from 108(14) to 166(14)~GPa, consistent with previous observations for amorphous Fe$_2$O$_3$~\cite{crepissonPRB}. Between 166(14) and 177(14)~GPa, this peak broadens and decreases markedly in intensity, accompanied by the disappearance of all Bragg reflections. This behaviour indicates that the Fe$_2$O$_3$ Hugoniot crosses the solidus or liquidus near this pressure. A similar decrease and broadening of the first diffraction peak was previously observed between 145(12) and 151(12)~GPa~\cite{crepissonPRB}. The higher transition pressure observed here may be due to the use of a black Kapton ablator here instead of Parylene-N as used previously~\cite{crepissonPRB}. Parylene-N could have allowed for slight laser preheating, leading to melting at lower pressures along the Hugoniot.

Whether the structural change between 166(14) and 177(14)~GPa corresponds to partial or complete melting remains uncertain. The Fe$_2$O$_3$ data point at 177(14)~GPa and $\sim$3,800~K (estimated from SESAME 7440~\cite{barnes_sesame_1987}) lies close to the FeO melting curve of Fu and Hirose~\cite{fu_hirose}, but is significantly below the temperatures reported by Dobrosavljević {\it et al.}~\cite{dobrosavljevic}. The state of the sample at this pressure therefore remains ambiguous, and further comparison of the two Fe$_2$O$_3$ samples at 177(14)~GPa will be required to resolve whether full melting occurs.

 \begin{figure*}
     \centering
     \includegraphics[width=1\textwidth]{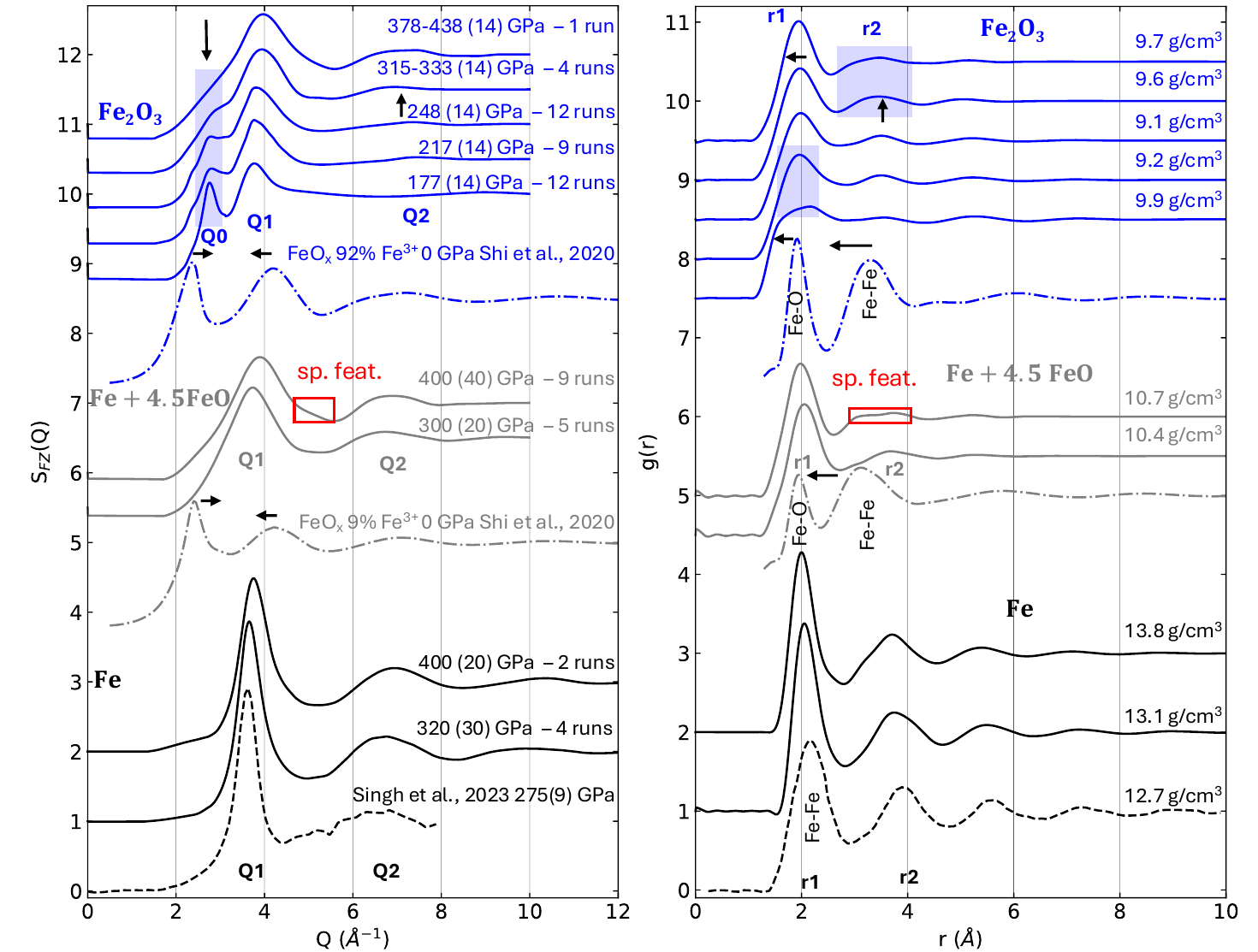}
     \caption{Faber–Ziman structure factor $S_{FZ}(Q)$ and corresponding pair distribution function $g(r)$ for Fe, Fe + 4.5 FeO, and Fe$_2$O$_3$ melts. Data are shifted vertically by 0.5 or 1 for clarity. The density was refined during the calculation of $g(r)$ using an iterative procedure~\cite{eggert}, as detailed in Methods. Denser packing of the melts is primarily explained by a decrease in Fe–Fe interatomic distances in Fe + 4.5 FeO and Fe$_2$O$_3$ melts under shock, relative to the 0~GPa (ambient) melt from Shi {\it et al.}~\cite{shi}. For Fe and Fe + 4.5 FeO, the $g(r)$ shifts toward shorter interatomic distances with increasing pressure. In Fe$_2$O$_3$, two distinct changes (highlighted in blue) are observed: first, between 177(14) and 217(14)~GPa, where the Q0 peak decreases and the first oscillation (r1) changes shape and width; and second, between 315–333(14)~GPa, where the Q2 feature becomes pronounced, r2 increases, and r1 becomes sharper and more intense. The spurious features (sp. feat.) in high-pressure Fe + 4.5 FeO melt data (marked in red) arise from incomplete subtraction of ambient contributions.}
     \label{Fig:SQ_gr}
\end{figure*}

\subsection{Structure and density of Fe, Fe + 4.5 FeO and Fe$_2$O$_3$ melts under shock compression}

To quantify the structural evolution and density of the melts under extreme conditions, we derived Faber–Ziman structure factors $S_{FZ}(Q)$ in Fig.~\ref{Fig:SQ_gr}, from the smoothed XRD lineouts shown in Figs.~\ref{Fig:Fe_intensity}-\ref{Fig:Fe2O3_intensity}. The corresponding pair distribution functions $g(r)$ and melt densities were then obtained using established iterative procedures~\cite{kaplow,eggert,sanloup}. This approach enables direct assessment of interatomic distances and coordination environments in the shocked liquids. Full details of the liquid diffraction analysis are provided in the Supplemental Material, including assessment of uncertainties based on the study of a commercial metallic glass from Goodfellow (Fe$_{78}$B$_{13}$Si$_9$), and comparison of present results on Fe melt with previous experimental and theoretical works.

The structure factors of Fe melt are consistent with those reported by Singh {\it et al.}~\cite{singh}, obtained from XRD measurements under laser-driven shock compression on a synchrotron (see Fig.~\ref{Fig:SQ_gr}). Liquid diffraction peaks Q1 and Q2 (displayed in Fig.~\ref{Fig:SQ_gr}) shift toward higher scattering vectors with increasing pressure, corresponding in the pair distribution function to a decrease in interatomic distances (r1, r2 displayed in Fig.~\ref{Fig:SQ_gr}) and thus shorter Fe–Fe bonds. We obtain Fe–Fe interatomic distances of 2.05(0.03)~\si{\angstrom} at 320(30)~GPa and 2.00(0.03)~\si{\angstrom} at 400(20)~GPa, in agreement with {\it ab initio} molecular dynamics simulations of liquid Fe under pressure~\cite{gonzalez}, which report 2.05~\si{\angstrom} at 323~GPa and 6370~K (Fig.~\ref{Fig:CN}). The Fe–Fe coordination numbers (CN) derived here are 12.3(0.4) at 320(30)~GPa, while the data point at 400(20) GPa has too low shock fraction to extract CN ($\sim$49\% resulting in greater uncertainty in the corrected XRD intensity according to Wark {\it et al.}~\cite{wark}).

The structure factor of Fe + 4.5 FeO melt under shock differs markedly from that of FeO melt at ambient pressure, as shown in Fig.~\ref{Fig:SQ_gr}. The two diffraction peaks characteristic of ambient FeO-type melts~\cite{leydier2010,shi} merge into a single feature, Q1. Similarly, the first two oscillations in the ambient pair distribution function, attributed to Fe–Fe and Fe–O interatomic distances, coalesce into a single peak (r1), reflecting a pronounced decrease in the Fe–Fe distance. Both r1 and r2 shift toward shorter interatomic distances with increasing pressure, consistent with denser atomic packing in the melt. The pair distribution function remains otherwise similar between 300(20) and 400(40)~GPa, apart from this systematic contraction, indicating that the Fe–O bonding environment is largely preserved with pressure. At 300(20)~GPa, we find the Fe-O CN to be 4.2(0.4), a Fe–O distance of 1.95(0.03)~\si{\angstrom}, and an Fe–Fe distance of 2.27(0.03)~\si{\angstrom} (Fig.~\ref{Fig:CN}).

Similarly to what is observed for Fe + 4.5 FeO melt, the structure factor of Fe$_2$O$_3$ melt under shock is very different from Fe$_2$O$_3$-type melt at ambient pressure (Fig.~\ref{Fig:SQ_gr}). The two principal diffraction peaks, Q0 and Q1, evolve in opposite directions: Q0 diminishes with pressure and nearly vanishes at 378–438(14)~GPa, whereas Q1 shifts from 3.77 to 3.97~\si{\angstrom}$^{-1}$ between 177(14) and 378–438(14)~GPa. The absence of a significant shift in Q0, together with the emergence of a clear Q2 feature above 315–333(14)~GPa, results in a structure factor that closely resembles that of the Fe + 4.5 FeO melt at similar pressures.

In the corresponding pair distribution functions (Fig.~\ref{Fig:SQ_gr}), Fe–O and Fe–Fe contributions merge into a single peak (r1), indicating increased structural compaction akin to that seen in Fe + 4.5 FeO melt. Between 177(14) and 217(14)~GPa, the broadening and reshaping of r1 reflect a rise in the Fe–O CN from 4.1(0.4) to 4.4(0.4), and a lengthening of the Fe–O bond from 1.61(0.03) to 1.73(0.03)~\si{\angstrom}. The Fe–Fe distances remain at 2.30(0.03)~\si{\angstrom}, as can be seen from Fig.~\ref{Fig:CN}. From 217(14) to 315–333(14)~GPa, the structure stabilizes, with r1 remaining nearly constant (1.96–1.98~\si{\angstrom}). At higher pressures, r1 shifts to shorter distances (down to 1.94~\si{\angstrom}) and r2 becomes more pronounced, indicating a more densely packed network. These changes correspond to a reduction in the Fe-O CN to 4.0(0.4), a modest expansion of Fe–O bonds to 1.83(0.03)~\si{\angstrom}, and a contraction of Fe–Fe distances to 2.17(0.03)~\si{\angstrom}.

\section{Discussion}

\begin{figure*}
     \includegraphics[width=1\textwidth]{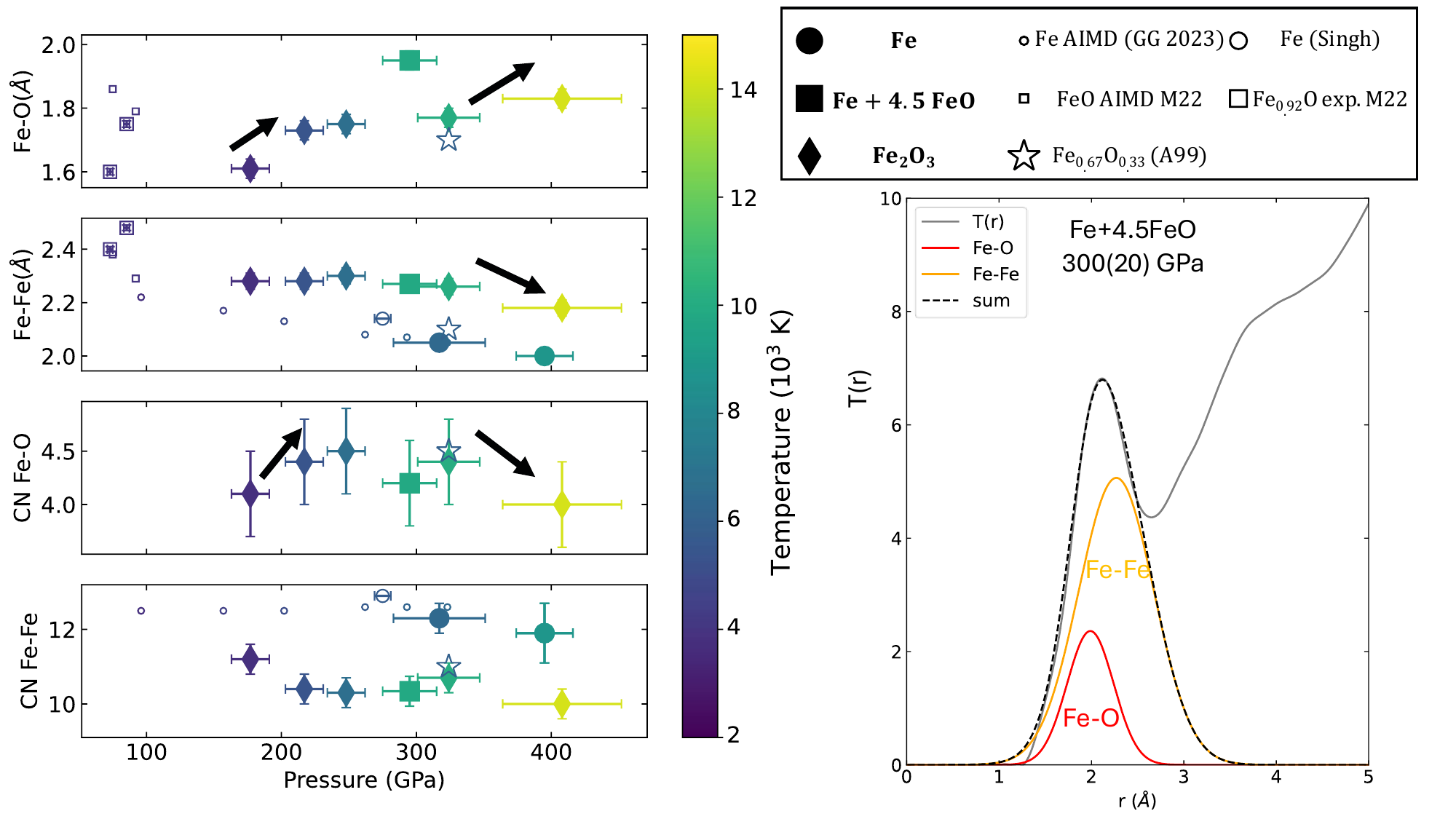}
     \caption{Interatomic distances and coordination numbers (CN) for Fe, Fe + 4.5 FeO, and Fe$_2$O$_3$ melts.
     Fits were performed on $\mathcal{T}(r)$, as detailed in the Methods.
     The O–O contribution is negligible and therefore excluded.
     The Fe-O CN for iron oxide melts under shock ranges between 4.0(0.4) and 4.5(0.4), with Fe–O bond lengths between 1.61 and 1.95~\si{\angstrom}. Two structural changes are evident for Fe$_2$O$_3$ between 177(14) and 217(14)~GPa, and again above 315–333(14)~GPa. These correspond to variations in Fe–O distance and Fe-O CN, and are indicated by black arrows. The results agree well with previous {\it ab initio} molecular dynamics simulations studies (AIMD) of Fe (GG 2023)~\cite{gonzalez}, Fe$_{0.92}$O (M22)~\cite{morard2022}, and Fe$_{0.67}$O$_{0.33}$ (A99)~\cite{alfe} melts, as well as experimental (exp.) shock data for Fe (Singh)~\cite{singh} and Fe$_{0.92}$O (M22)~\cite{morard2022}. Temperature estimates are based on SESAME 7440~\cite{barnes_sesame_1987} for Fe$_2$O$_3$, a linear extrapolation of the FeO Hugoniot~\cite{jeanloz} for Fe + 4.5 FeO, and the DFT-calculated Hugoniot for Fe~\cite{harmand}.}
     \label{Fig:CN}
\end{figure*}

\begin{figure}
     \centering
     \includegraphics[width=1\columnwidth]{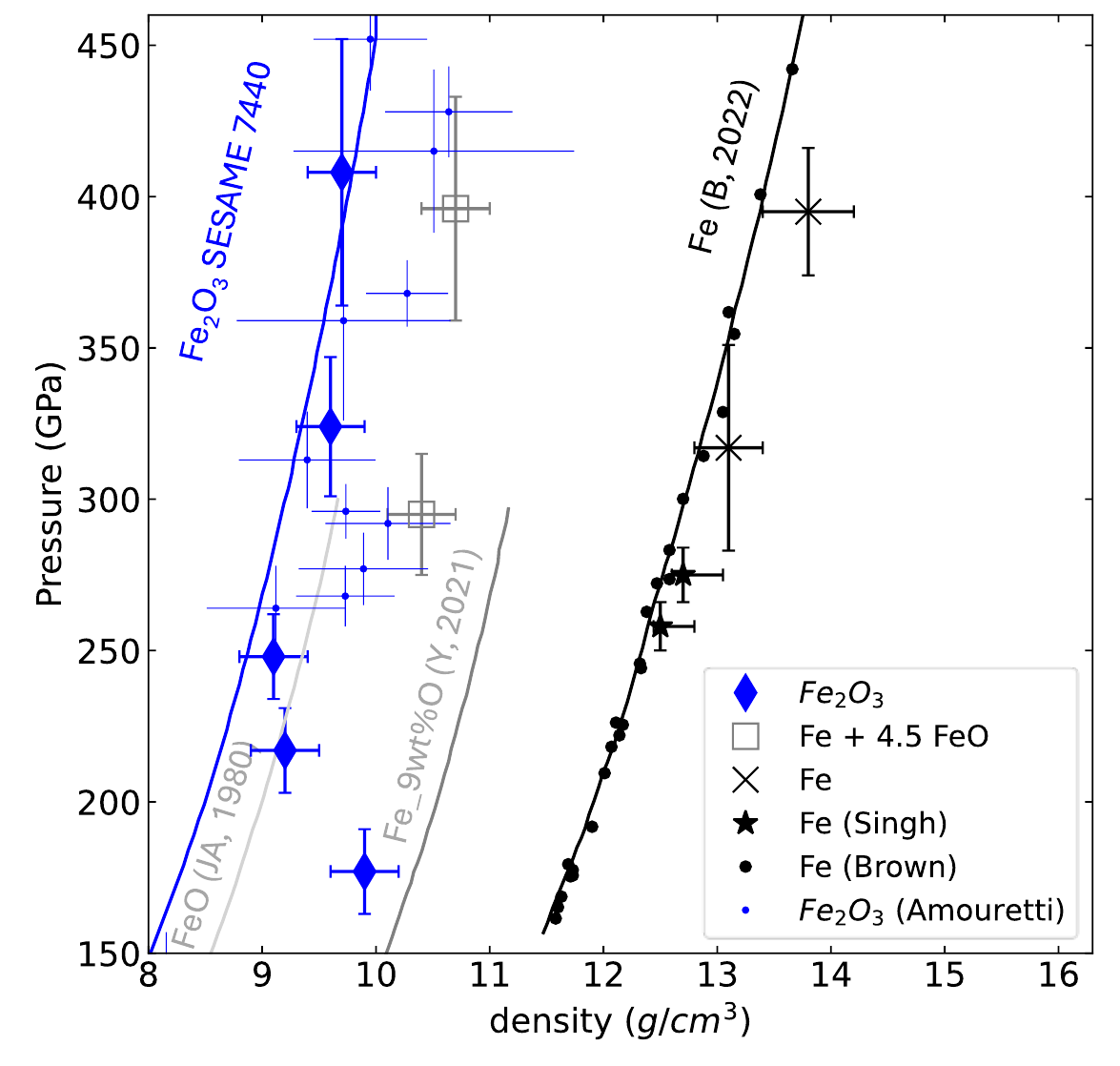}
     \caption{Densities extracted from liquid diffraction analyses. For Fe, results are compared with gas-gun data (Brown)~\cite{brown}, laser-driven shock data (Singh)~\cite{singh}, and theoretical calculations (B,2022)~\cite{benedict}. For Fe + 4.5 FeO, results are compared with gas-gun data for Fe$_0.94$O~\cite{jeanloz} and Fe containing 9 wt\% O (Y,2021)~\cite{young}. For Fe$_2$O$_3$, results are compared with SESAME 7440~\cite{barnes_sesame_1987} and laser-shock data (Amouretti)~\cite{LULI}.}
     \label{Fig:density}
\end{figure}

We resolve the Fe–O bonding environment between 177–438~GPa, and 3,800–14,100~K based on temperature estimates from SESAME 7440~\cite{barnes_sesame_1987}, for Fe$_2$O$_3$ and a linear extrapolation of the FeO Hugoniot~\cite{jeanloz} for Fe + 4.5 FeO. As we show in Fig.~\ref{Fig:CN}, under these extreme conditions the Fe-O CN ranges between 4.0(0.4) and 4.5(0.4), and the Fe-Fe CN ranges between 10.0(0.4) and 11.2(0.4), for both Fe + 4.5 FeO and Fe$_2$O$_3$ melts. These values are consistent with {\it ab initio} molecular dynamics simulations of Fe$_{0.67}$O$_{0.33}$ melt by Alfè {\it et al.}~\cite{alfe}, which reported an Fe-O CN of 4.5 and an Fe-Fe CN of 11 at 324~GPa and 6,000~K.

Our results suggest that the previously proposed increase in Fe–O coordination number from 6 to 8 in liquid FeO near 240~GPa, by analogy with the B1–B2 solid-state transition in FeO, is unlikely to hold~\cite{ozawa}. At ambient pressure, the Fe-O CN ranges from 4.5 in FeO-type melts to 5 in Fe$_2$O$_3$-type compositions~\cite{shi}. Theoretical calculations indicate that these values arise from mixed 4- and 5-fold contributions in FeO-type melts and from mixed 4-, 5-, and 6-fold environments in Fe$_2$O$_3$-type melts~\cite{shi}. Although further modelling is needed to quantify individual contributions under shock conditions, the present Fe-O CN values of 4.0–4.5(0.4) likely reflect similar coordination distributions involving 4-, 5-, and possibly 6-fold Fe–O environments.

This finding implies that the Fe–O bonding environment in iron oxide melts differs markedly from that in major lower-mantle Fe-bearing minerals such as ferropericlase and bridgmanite, where Fe is surrounded by 6 or 8 oxygen atoms. Such structural distinctions are crucial for constraining oxygen solubility between lower-mantle solid phases and for evaluating the stability of Fe$_n$O phases proposed within the inner core~\cite{liu} and liquid outer-core throughout Earth's evolution.

Comparison of the pair distribution functions for the ambient-pressure melt and the shock-compressed data shows a clear structural change. The increase in density with pressure (Fig.~\ref{Fig:density}) mainly results from a strong shortening of the Fe–Fe interatomic distance by around 1~\si{\angstrom}, from 3.1–3.3~\si{\angstrom} in ambient iron oxide melts~\cite{shi} to $\sim$2.30~\si{\angstrom} under shock compression.
In both Fe + 4.5 FeO and Fe$_2$O$_3$ melts, the Fe–Fe distance remains nearly constant between 177(14)-333(14)~GPa at $\sim$2.30~\si{\angstrom}. These values are consistently larger than those in pure Fe melt (Fig.~\ref{Fig:CN}).
{\it Ab initio} molecular dynamics simulations of Fe$_{0.67}$O$_{0.33}$ at 324~GPa and 6,000~K~\cite{alfe} give Fe–Fe distances about 0.2~\si{\angstrom} shorter than our results (Fig.~\ref{Fig:CN}). Distances between 2.3-2.5~\si{\angstrom} are also reported for Fe$_{0.92}$O at $\sim$90~GPa and 4,000~K~\cite{morard2022}. The slightly longer Fe–Fe bonds measured here may reflect the higher temperatures reached along the Hugoniot, where temperature increases significantly with pressure.

At 378–438(14)~GPa and approximately 14,100~K, a distinct shift of the r1 feature in the pair distribution function of Fe$_2$O$_3$ melt is observed in Fig.~\ref{Fig:SQ_gr}. This shift corresponds to a reduction of the Fe–Fe interatomic distance by about 0.1~\si{\angstrom} (Fig.~\ref{Fig:CN}). The shortening of the Fe–Fe distance is accompanied by a decrease in the Fe–O coordination number from 4.4(0.4) to 4.0(0.4) and an increase in the Fe–O bond length from 1.77(0.03) to 1.83(0.03)~\si{\angstrom}.
Between 315–333(14)~GPa, the structure factor and pair distribution function of Fe$_2$O$_3$ melt begin to resemble those of the Fe + 4.5 FeO melt. This convergence reflects a concurrent reduction in Fe–O coordination and elongation of the Fe–O bond, which approach the values observed for Fe + 4.5 FeO melt at 300(20)~GPa (Fig.~\ref{Fig:CN}). These structural changes may signal a modification in oxidation state, potentially associated with partial dissociation of Fe$_2$O$_3$ into FeO and O under extreme conditions.
At ambient pressure, the oxidation enthalpy of Fe$_2$O$_3$ is $-5.83$~eV per mole of O$_2$, indicating that dissociation into 2 FeO + O requires 2.92 eV (equivalent to $\sim$23,200K)~\cite{gautam2018}. This temperature is considerably higher than that reached under shock at 378–438(14)~GPa (around 14,100K). However, recent studies suggest that the energy required for phase dissociation decreases with increasing pressure (e.g., MgSi$_2$O$_3$~\cite{umemoto}). It is therefore possible that, under the present conditions, Fe$_2$O$_3$ partially dissociates into an FeO-type melt and oxygen at lower temperatures around 1~eV.

The Fe–O interatomic distances exhibit a complex evolution, varying between 1.61-1.95(0.03)~\si{\angstrom} across the temperature-pressure range investigated here, as shown in Fig.~\ref{Fig:CN}. Values reported for Fe$_{0.92}$O at 90~GPa and 4,000~K~\cite{morard2022} are within this range. The Fe–O distance obtained previously for Fe$_{0.67}$O$_{0.33}$ at 324~GPa and 6,000~K~\cite{alfe} is also consistent with our measurements.

We first note that the Fe–O interatomic distance in the Fe~+4.5FeO melt is larger than in all Fe$_2$O$_3$ compositions examined here. Moreover, Fig.~\ref{Fig:SQ_gr} shows that up to at least 315–333(14)~GPa, the structure factor and pair distribution function of the Fe + 4.5 FeO and Fe$_2$O$_3$ melts remain distinctly different. As illustrated in Fig.~\ref{Fig:CN}, this difference arises from the larger Fe–O bond length [1.95(0.03)~\si{\angstrom}] and slightly lower Fe–O coordination number [4.2(0.4)] in the Fe + 4.5 FeO melt.
A similar contrast between Fe$_2$O$_3$ and FeO melts is observed at ambient pressure~\cite{shi}. At 0~GPa, Fe$^{3+}$-rich melts exhibit a higher Fe–O coordination number (5) and a shorter Fe–O bond length (1.92~\si{\angstrom}), whereas Fe$^{2+}$-rich melts show lower coordination (4.5) and longer Fe–O bonds (1.95~\si{\angstrom})~\cite{shi}. These structural differences between Fe$^{3+}$- and Fe$^{2+}$-rich melts thus persist under extreme pressures and temperatures.
Such compositional contrasts could contribute to lateral heterogeneities near the top of the Earth’s outer-core, where regions of the lower mantle may be relatively enriched in Fe$^{3+}$ compared with Fe$^{2+}$\cite{frostnature,xu2015}. In addition, the oxidation state of the primitive magma ocean and the lower mantle likely evolved through Earth’s history~\cite{wade2005}, potentially affecting the solubility of oxygen in deep planetary reservoirs.

The Fe–O interatomic distance in Fe$_2$O$_3$ melt increases overall along the Hugoniot, from 1.61(0.03) to 1.83(0.03)~\si{\angstrom}. As shown in Fig.~\ref{Fig:CN}, the most pronounced increases occur between 177(14) and 217(14)~GPa, and again above 315–333(14)~GPa. The increases observed at 177(14) and 217(14)~GPa are accompanied by a rise in the Fe–O coordination number from 4.1(0.4) to 4.4(0.4).
A similar structural change has been reported for B$_2$O$_3$ between 4.3-5.7~GPa~\cite{brazhkin}. In that system, the structure factor exhibits a decrease in the first diffraction peak, similar to the reduction of the Q0 peak observed here between 177-217(14)~GPa (Fig.~\ref{Fig:SQ_gr}). This behaviour is uncommon among oxide melts under compression. Interestingly, in B$_2$O$_3$ the increase in the B–O coordination number from 3 to 4 and the elongation of the B–O bond from 1.40 to 1.45~\si{\angstrom} are associated with a viscosity decrease of nearly four orders of magnitude~\cite{brazhkin}.

A similar trend has been predicted by {\it ab initio} molecular dynamics simulations for Fe$_{0.96}$O$_{0.04}$ at a density of 8~g.cm$^{-3}$~\cite{posner}, where an increase in Fe–O coordination and bond length influences oxygen solubility in liquid Fe. The change in Fe–O bonding environment observed here between 177(14) and 217(14)~GPa may therefore be of particular geophysical interest, as it corresponds to a depth of roughly 380~km below the core–mantle boundary~\cite{dziewonski}. A variation in oxygen incorporation into liquid Fe, possibly accompanied by changes in viscosity or oxygen solubility, could help explain the presence of a relatively thick, stratified layer inferred from seismological and geomagnetic observations~\cite{kaneshima}.
Nevertheless, it remains uncertain whether Fe$_2$O$_3$ is fully molten or partially amorphous at 177(14)~GPa. Partial amorphization could account for the anomalously high density measured at this pressure, as shown in Fig.~\ref{Fig:density}. 

Laser-driven shock compression at XFEL facilities now enables high-accuracy liquid diffraction measurements of complex polyatomic melts. Our measurements of iron oxide at Earth's outer-core pressures reveal a Fe-O coordination number of 4.0–4.5(0.4), distinct from major lower-mantle minerals, with significant structural differences between Fe$^{3+}$- and Fe$^{2+}$-rich melts. These findings constrain oxygen solubility and the local environment in the Earth's outer-core throughout its geological evolution. The anomalous diffuse scattering observed in Fe$_2$O$_3$ at 177(14) GPa (corresponding to a depth of $\sim$380~km) warrants further investigation to elucidate the underlying structural transformations.

\section{Methods}
\subsection{Targets}

Seven distinct target configurations were used in the shock-compression experiments.

Two Fe$_2$O$_3$ samples were prepared:
(1) a well-crystallized 20~\textmu{}m Fe$_2$O$_3$ layer deposited from an Fe sputtering target onto a 55 or 77~\textmu{}m black Kapton ablator using plasma sputtering, and
(2) a poorly crystalline 8~\textmu{}m Fe$_2$O$_3$ layer deposited from an Fe$_2$O$_3$ sputtering target onto a 55~\textmu{}m black Kapton ablator by the same method.
In both cases, a 200–400~nm Al coating was applied to enhance reflectivity for VISAR measurements.
X-ray diffraction (XRD) and Electron MicroProbe Analyses (EMPA) (Supplemental Material Section~I) confirm that both samples consist of pure Fe$_2$O$_3$. Scanning electron microscopy reveals a homogeneous and fully dense starting material.

The Fe + 4.5 FeO targets consist of a 10~\textmu{}m layer containing 80.7(0.3)wt\% Fe and 18.9(0.2)wt\% O (55.0(0.4) at\% Fe and 45.0(0.4) at\% O), deposited via plasma sputtering from an Fe target onto a 77~\textmu{}m black Kapton ablator.
An additional 200–400~nm Al coating was applied for VISAR reflectivity.
Bragg–Brentano XRD and Electron Microprobe analyses show that the initial material contains intergrown Fe (bcc) and FeO (B1) crystals, corresponding to Fe + 4.5 FeO in atomic proportions if we assume FeO stoichiometry.

The Fe target comprises a 20~\textmu{}m Goodfellow Fe foil (99.99\% purity) bonded to a 77~\textmu{}m black Kapton ablator.
The adhesive layer is approximately 5~\textmu{}m thick, with minor variations.

The Cu target used to verify pressure calibration consists of a 25~\textmu{}m Goodfellow Cu foil attached to a 55~\textmu{}m black Kapton ablator using a similar adhesive layer of about 5~\textmu{}m.

Isolated black Kapton targets were employed to calibrate the ablator pressure at specific laser intensities.
These calibration targets consist of 77~\textmu{}m black Kapton coated with a 200~nm Al layer on the VISAR side.

Finally a 25~\textmu{}m commercial metallic glass from Goodfellow (Fe$_{78}$B$_{13}$Si$_9$) previously measured at beamline ID15 (Diamond Light Source) at 76 keV~\cite{sapnik} was used to verify XRD data corrections and help to estimate uncertainties related to liquid diffraction analysis (described in Section XIII of the Supplemental Material).

\subsection{Experimental set-up}

The experiment was performed at the High Energy Density (HED) endstation of the EuXFEL~\cite{zastrau}, in IC2. We used the high-repetition-rate DiPOLE 100-X optical laser~\cite{zastrau,mason} as the shock compression driver. The frequency-doubled 515~nm laser beam was incident at 22.5$^\circ$ to the target normal and focused to a spot of $\sim$250~\textmu{}m in diameter. Flat-top pulses of 4, 6, and 10~ns in duration delivered up to 50~J of energy to the target. Representative pulse profiles are provided in Section~III of the Supplemental Material.

The x-ray probe beam, produced in self-amplified spontaneous emission (SASE) mode, had a photon energy of 24~keV and was focused to a spotsize of 26$\times$15~\textmu{}m. It was directed at 22.5$^\circ$ from the target normal and intersected the target at the centre of the laser focal spot. Two-dimensional X-ray Diffraction (XRD) patterns were recorded using two Varex detectors. Their azimuthal and angular coverage are illustrated in Fig.~\ref{Fig:set_up}. Before each laser shot, a preshot measurement was taken with the x-ray beam attenuated to 5\% transmission and displaced laterally by 20~\textmu{}m from the subsequent shot position. These preshots provided reference data under unshocked conditions.

\subsection{Integration and correction of XRD intensity}

X-ray diffraction (XRD) intensities were renormalized to account for fluctuations in the XFEL beam using the in-chamber IPM diodes~\cite{zastrau}. The data were fully corrected for solid angle, polarization, and self-attenuation. Fluorescence from a 25~\textmu{}m Rh foil (absorption edge at 23.22 keV) was employed to correct for the different apparent thickness due to different angles of the Al filter placed over the Varex detectors through a flat-fielding procedure similar to that previously applied for the same setup~\cite{gormanSn}. Two-dimensional XRD images were azimuthally integrated using the \texttt{pyFAI} library~\cite{kieffer}.

To validate this procedure, we compared the azimuthally integrated XRD signal of a 25~\textmu{}m-thick commercial metallic glass from Goodfellows (Fe$_{78}$B$_{13}$Si$_9$) measured with the present setup with data acquired at beamline ID15 (Diamond Light Source) at 76 keV~\cite{sapnik}. The two datasets show excellent agreement from 1.46 to 12.4~\AA$^{-1}$, with only minor discrepancy below 2.5~\AA$^{-1}$, as shown in fig. S27 of the Supplemental Material.

For shock-compressed data, the residual contribution from the unshocked sample was removed following the procedure described by Wark {\it et al.}~\cite{wark}. A sum of preshot measurements or a shot taken without DiPOLE laser was used as the ambient reference. The removal is complicated by grain-to-grain variations in Fe and Fe$_2$O$_3$, which are textured and show changing Bragg peak intensities between shots. The Fe + 4.5 FeO samples also exhibited broad Bragg peaks, making ambient subtraction less precise. Nevertheless, the shocked fraction exceeds 79\% for all liquid data, except for Fe at 400(20)~GPa, where the shocked fraction is 49\%.

The contribution from the black Kapton ablator was minimized using a structureless analogue for ambient black Kapton (C$_{22}$H$_{10}$N$_2$O$_5$), since no direct ambient measurement was performed. The ambient black Kapton contribution is particularly small in runs with a high shocked fraction (0–20\% unshocked material remaining).

To remove the ablator contribution under shock, we measured fully shocked 77~\textmu{}m black Kapton samples coated with 200~nm Al. The shocked state of black Kapton differs depending on the presence of an adjacent target layer, which induces a reshock and thus higher pressures and lower temperatures compared with isolated black Kapton. However, no significant variation in the compressed black Kapton XRD signal was observed between 21(1) and 122(4)~GPa along the Hugoniot. We therefore consider the signal from isolated black Kapton at 122(4)~GPa a suitable proxy for the ablator response. Diamond formation was detected in black Kapton at 56(2), 64(3), and 72(3)~GPa, but no diamond peaks are expected under the conditions corresponding to the molten data.

All corrections described above were applied to each individual dataset prior to summation. Identical XRD signals from multiple runs were then combined to enhance the signal-to-noise ratio, a step crucial for reliable liquid diffraction analysis.

\subsection{Determination of interatomic distance and coordination number}

For the monoatomic melt (Fe), interatomic distances were determined directly from the maximum of a skewed Gaussian fit to the first oscillation of the pair distribution function. The coordination number (CN) was obtained by integrating the area up to the first minimum of the radial distribution function, $\mathcal{R}(r)$, defined as
$$
\mathcal{R}(r) = 4\pi r^2 g(r) n_0 ,
$$
where $n_0$ is the atomic density.

For the polyatomic melts (Fe + 4.5 FeO and Fe$_2$O$_3$), the Fe–O and Fe–Fe contributions were each represented by a single Gaussian function. The parameters of these functions were optimized to reproduce the first oscillation of
$$
\mathcal{T}(r) = 4\pi r g(r) n_0,
$$
as shown in Fig.~\ref{Fig:CN}. The fitting procedure follows the approach described by Heinen and Drewitt~\cite{liquiddiffract}. The O–O contribution was neglected, as it is small compared with the Fe–O and Fe–Fe components due to the difference in atomic number between O and Fe (8 versus 26).

In this model, the Gaussian area corresponds to the coordination number, and the fitted position [after correction for the offset between $g(r)$ and $\mathcal{T}(r)$] yields the interatomic distance. All Gaussian fits are presented in Section XIII of the Supplemental Material. Reliable fitting could not be achieved for the Fe + 4.5 FeO dataset at 400(40)~GPa, owing to the large fraction of unshocked material, which introduces spurious features in the pair distribution function (Fig.~\ref{Fig:SQ_gr}).

\subsection{Pressure determination}

Where possible, a Velocity Interferometer System for Any Reflector (VISAR)~\cite{barker,descamps_VISAR} was used to determine the breakout time, i.e. the time when the shock exits the target. However, the setup was unable to provide access to the velocity history. A full summary of the VISAR measurements is given in the Supplemental Material. 

A calibration curve relating the ablator pressure to laser intensity was established using isolated 77~\textmu{}m-thick black Kapton samples coated with 200~nm of Al and verify using Cu targets under shock (details are provided in Section II of the Supplemental Material). Breakout times were measured by VISAR, and pressures were derived from the Hugoniot relation for polyimide~\cite{katagiri}. The main sources of uncertainty include variations in black Kapton thickness (measured by touch probe to be between 76.6-77.3~\textmu{}m), the coefficients of the polyimide Hugoniot equation, and the interpolation of data to fit a power-law function. The maximum uncertainty on the pressure is estimated at $\pm$14~GPa. This calibration curve was used to determine the ablator pressure corresponding to a given laser intensity.

Pressures within the Fe, FeO (as a proxy for Fe + 4.5 FeO), and Fe$_2$O$_3$ targets were derived using the impedance-matching method~\cite{zeldovitch}. The equation of state (EOS) for the black Kapton ablator was taken from SESAME~7770, while those for FeO and Fe$_2$O$_3$ were based on prior experimental data~\cite{jeanloz,LULI,Mcqueen}. For Fe, SESAME~2140 and 2145 were used.

Hydrodynamic simulations were performed using the code MULTI~\cite{ramis} to independently estimate the shock pressures in Fe$_2$O$_3$ and Fe + 4.5 FeO. The simulations used SESAME~7440~\cite{barnes_sesame_1987} for Fe$_2$O$_3$ and the FeO EOS of Jeanloz and Ahrens~\cite{jeanloz}, with real pulse profiles as input. A numerical intensity factor ($\sim$0.5–0.7) was applied to the experimental laser intensity to match the measured breakout times. The black Kapton ablator was again modelled using SESAME~7770. When breakout times were unavailable, the intensity multiplier from the closest measured shot was adopted.

Pressures derived from impedance matching and hydrodynamic simulations agree within 20~GPa for Fe + 4.5 FeO, and within 14~GPa for Fe$_2$O$_3$. These values are used as an estimate for pressure uncertainty. For the Fe targets, only the impedance-matching method was applied, as the non-uniform glue thickness between the Fe foil and the black Kapton ablator introduced additional variability in breakout times. The dominant uncertainty arises from the ablator pressure calibration ($\pm$14~GPa). Comparisons between SESAME~2140 and 2145 EOS models produced smaller discrepancies than those due to the uncertainty on the ablator pressure. A complete summary of the pressures determined for all data points is provided in the Supplemental Material.

\section{Acknowledgements}

C.Cr\'episson., A.C.D., J.S.W., and S.M.V. acknowledge support from EPSRC under research grant EP/W010097/1.
M.F., Y.W., G.G., and S.M.V. acknowledge support from EPSRC and First Light Fusion under the AMPLIFI prosperity partnership, grant EP/X025373/1.
P.G.H. and J.S.W. acknowledge support from EPSRC under research grant EP/X031624/1.
D.J.P., H.T., and T.S. acknowledge support from AWE via the Oxford Centre for High Energy Density Science (OxCHEDS).
K.A. and C.Camarda acknowledge the financial support by Deutsche Forschungsgemeinschaft (DFG, German Research Foundation) via project AP262/3-1 and STE1079/10-1 (project number 521549147).
A.A., K.Y., and N.O. acknowledge supports from Japan Society for the Promotion of Science (JSPS) KAKENHI (Grant Nos. 23K20038 and 25H00618), JSPS Core-to-Core Program (JPJSCCA20230003), and MEXT Quantum Leap Flagship Program (JPMXS0118067246).
This material is based upon work supported by the Department of Energy [National Nuclear Security Administration] University of Rochester “National Inertial Confinement Fusion Program” under Award Number(s) DE-NA0004144.
We acknowledge the EPSRC National Thin Film Facility for Advanced Functional Materials (NTCF), hosted by the Department of Physics at the University of Oxford. The NTCF was funded by ESPRC (EP/M022900/1), the Wolfson Foundation and the University of Oxford. Part of Al coating was realized at DESY, Hamburg, Germany by Michael R\"{o}per.
Samples were measured using the Co x-ray diffractometer and Electron Microprobe at the Department of Earth Sciences at the University of Oxford assisted by K. Sokol and A. Matzen.
Sample preparation for analysis was performed at the polishing lab at the Department of Earth Sciences at the University of Oxford assisted by E. Donald. 
We acknowledge the EuXFEL in Schenefeld, Germany, for provision of X-ray Free-Electron laser beam time at the Scientific Instrument HED (High Energy Density Science), proposal p6746.
The authors are indebted to the Helmholtz International Beamline for Extreme
Fields (HIBEF) user consortium for the provision of instrumentation and staff that enabled this experiment.
We are grateful to J. McHardy who provided the Rh foil and to D. Keen and C. Sanloup for fruitful discussions. 

\section{Data availability}

The data presented here are available upon reasonable request at https://doi.org/10.22003/XFEL.EU-DATA-006746-00.

\bibliography{ref.bib}

\begin{thebibliography}{68}%
\makeatletter
\providecommand \@ifxundefined [1]{%
 \@ifx{#1\undefined}
}%
\providecommand \@ifnum [1]{%
 \ifnum #1\expandafter \@firstoftwo
 \else \expandafter \@secondoftwo
 \fi
}%
\providecommand \@ifx [1]{%
 \ifx #1\expandafter \@firstoftwo
 \else \expandafter \@secondoftwo
 \fi
}%
\providecommand \natexlab [1]{#1}%
\providecommand \enquote  [1]{``#1''}%
\providecommand \bibnamefont  [1]{#1}%
\providecommand \bibfnamefont [1]{#1}%
\providecommand \citenamefont [1]{#1}%
\providecommand \href@noop [0]{\@secondoftwo}%
\providecommand \href [0]{\begingroup \@sanitize@url \@href}%
\providecommand \@href[1]{\@@startlink{#1}\@@href}%
\providecommand \@@href[1]{\endgroup#1\@@endlink}%
\providecommand \@sanitize@url [0]{\catcode `\\12\catcode `\$12\catcode `\&12\catcode `\#12\catcode `\^12\catcode `\_12\catcode `\%12\relax}%
\providecommand \@@startlink[1]{}%
\providecommand \@@endlink[0]{}%
\providecommand \url  [0]{\begingroup\@sanitize@url \@url }%
\providecommand \@url [1]{\endgroup\@href {#1}{\urlprefix }}%
\providecommand \urlprefix  [0]{URL }%
\providecommand \Eprint [0]{\href }%
\providecommand \doibase [0]{https://doi.org/}%
\providecommand \selectlanguage [0]{\@gobble}%
\providecommand \bibinfo  [0]{\@secondoftwo}%
\providecommand \bibfield  [0]{\@secondoftwo}%
\providecommand \translation [1]{[#1]}%
\providecommand \BibitemOpen [0]{}%
\providecommand \bibitemStop [0]{}%
\providecommand \bibitemNoStop [0]{.\EOS\space}%
\providecommand \EOS [0]{\spacefactor3000\relax}%
\providecommand \BibitemShut  [1]{\csname bibitem#1\endcsname}%
\let\auto@bib@innerbib\@empty
\bibitem [{\citenamefont {Hirose}\ \emph {et~al.}(2021)\citenamefont {Hirose}, \citenamefont {Wood},\ and\ \citenamefont {Vočadlo}}]{hirose21}%
  \BibitemOpen
  \bibfield  {author} {\bibinfo {author} {\bibfnamefont {K.}~\bibnamefont {Hirose}}, \bibinfo {author} {\bibfnamefont {B.}~\bibnamefont {Wood}},\ and\ \bibinfo {author} {\bibfnamefont {L.}~\bibnamefont {Vočadlo}},\ }\bibfield  {title} {\bibinfo {title} {Light elements in the {E}arth’s core},\ }\href {https://doi.org/https://doi.org/10.1038/s43017-021-00203-6} {\bibfield  {journal} {\bibinfo  {journal} {Nature Reviews {E}arth and Environment}\ }\textbf {\bibinfo {volume} {2}},\ \bibinfo {pages} {645–658} (\bibinfo {year} {2021})}\BibitemShut {NoStop}%
\bibitem [{\citenamefont {Rubie}\ \emph {et~al.}(2007)\citenamefont {Rubie}, \citenamefont {Nimmo},\ and\ \citenamefont {Melosh}}]{rubie2007}%
  \BibitemOpen
  \bibfield  {author} {\bibinfo {author} {\bibfnamefont {D.}~\bibnamefont {Rubie}}, \bibinfo {author} {\bibfnamefont {F.}~\bibnamefont {Nimmo}},\ and\ \bibinfo {author} {\bibfnamefont {H.}~\bibnamefont {Melosh}},\ }\bibfield  {title} {\bibinfo {title} {Formation of {E}arth’s core},\ }\href {https://doi.org/10.1016/B978-044452748-6.00140-1} {\bibfield  {journal} {\bibinfo  {journal} {Treatise on Geophysics}\ }\textbf {\bibinfo {volume} {9}},\ \bibinfo {pages} {51} (\bibinfo {year} {2007})}\BibitemShut {NoStop}%
\bibitem [{\citenamefont {Bi}\ \emph {et~al.}(2025)\citenamefont {Bi}, \citenamefont {Sun}, \citenamefont {Sun}, \citenamefont {Mao}, \citenamefont {Dai},\ and\ \citenamefont {Hemingway}}]{bi_2025}%
  \BibitemOpen
  \bibfield  {author} {\bibinfo {author} {\bibfnamefont {H.}~\bibnamefont {Bi}}, \bibinfo {author} {\bibfnamefont {D.}~\bibnamefont {Sun}}, \bibinfo {author} {\bibfnamefont {N.}~\bibnamefont {Sun}}, \bibinfo {author} {\bibfnamefont {Z.}~\bibnamefont {Mao}}, \bibinfo {author} {\bibfnamefont {M.}~\bibnamefont {Dai}},\ and\ \bibinfo {author} {\bibfnamefont {D.}~\bibnamefont {Hemingway}},\ }\bibfield  {title} {\bibinfo {title} {Seismic detection of a 600-km solid inner core in {M}ars},\ }\href {https://doi.org/10.1038/s41586-025-09361-9} {\bibfield  {journal} {\bibinfo  {journal} {Nature}\ }\textbf {\bibinfo {volume} {645}},\ \bibinfo {pages} {67–72} (\bibinfo {year} {2025})}\BibitemShut {NoStop}%
\bibitem [{\citenamefont {Buffet}(2000)}]{buffet2000}%
  \BibitemOpen
  \bibfield  {author} {\bibinfo {author} {\bibfnamefont {B.}~\bibnamefont {Buffet}},\ }\bibfield  {title} {\bibinfo {title} {{E}arth's core and the geodynamo},\ }\href {https://doi.org/10.1126/science.288.5473.2007} {\bibfield  {journal} {\bibinfo  {journal} {Science}\ }\textbf {\bibinfo {volume} {288}},\ \bibinfo {pages} {2007} (\bibinfo {year} {2000})}\BibitemShut {NoStop}%
\bibitem [{\citenamefont {White}\ and\ \citenamefont {Li}(2025)}]{white}%
  \BibitemOpen
  \bibfield  {author} {\bibinfo {author} {\bibfnamefont {N.~I.}\ \bibnamefont {White}}\ and\ \bibinfo {author} {\bibfnamefont {J.}~\bibnamefont {Li}},\ }\bibfield  {title} {\bibinfo {title} {Initial thermal states of super‐{E}arth exoplanets and implications for early dynamos},\ }\href {https://doi.org/10.1029/2024JE008550} {\bibfield  {journal} {\bibinfo  {journal} {Journal of Geophysical Research}\ }\textbf {\bibinfo {volume} {130}},\ \bibinfo {pages} {e2024JE008550} (\bibinfo {year} {2025})}\BibitemShut {NoStop}%
\bibitem [{\citenamefont {Kaneshima}(2019)}]{kaneshima}%
  \BibitemOpen
  \bibfield  {author} {\bibinfo {author} {\bibfnamefont {S.}~\bibnamefont {Kaneshima}},\ }\bibfield  {title} {\bibinfo {title} {Array analyses of {S}m{KS} waves and the stratification of {E}arth’s outermost core},\ }\href {https://doi.org/https://doi.org/10.1016/j.pepi.2017.03.006} {\bibfield  {journal} {\bibinfo  {journal} {Physics of the {E}arth and Planetary Interiors}\ }\textbf {\bibinfo {volume} {276}},\ \bibinfo {pages} {234} (\bibinfo {year} {2019})}\BibitemShut {NoStop}%
\bibitem [{\citenamefont {Irving}\ \emph {et~al.}(2018)\citenamefont {Irving}, \citenamefont {Cottaar},\ and\ \citenamefont {Lekić}}]{irving}%
  \BibitemOpen
  \bibfield  {author} {\bibinfo {author} {\bibfnamefont {J.~C.~E.}\ \bibnamefont {Irving}}, \bibinfo {author} {\bibfnamefont {S.}~\bibnamefont {Cottaar}},\ and\ \bibinfo {author} {\bibfnamefont {V.}~\bibnamefont {Lekić}},\ }\bibfield  {title} {\bibinfo {title} {Seismically determined elastic parameters for {E}arth’s outer core},\ }\href {https://doi.org/https://doi.org/10.1126/sciadv.aar2538} {\bibfield  {journal} {\bibinfo  {journal} {Science Advances}\ }\textbf {\bibinfo {volume} {4}} (\bibinfo {year} {2018})}\BibitemShut {NoStop}%
\bibitem [{\citenamefont {Buffet}(2014)}]{buffet2014}%
  \BibitemOpen
  \bibfield  {author} {\bibinfo {author} {\bibfnamefont {B.}~\bibnamefont {Buffet}},\ }\bibfield  {title} {\bibinfo {title} {Geomagnetic fluctuations reveal stable stratification at the top of the {E}arth’s core},\ }\href {https://doi.org/https://doi.org/10.1038/nature13122} {\bibfield  {journal} {\bibinfo  {journal} {Nature}\ }\textbf {\bibinfo {volume} {507}},\ \bibinfo {pages} {484} (\bibinfo {year} {2014})}\BibitemShut {NoStop}%
\bibitem [{\citenamefont {Olson}\ \emph {et~al.}(2016)\citenamefont {Olson}, \citenamefont {Landeau},\ and\ \citenamefont {Reynolds}}]{buffet2016}%
  \BibitemOpen
  \bibfield  {author} {\bibinfo {author} {\bibfnamefont {P.}~\bibnamefont {Olson}}, \bibinfo {author} {\bibfnamefont {M.}~\bibnamefont {Landeau}},\ and\ \bibinfo {author} {\bibfnamefont {E.}~\bibnamefont {Reynolds}},\ }\bibfield  {title} {\bibinfo {title} {Evidence for {MAC} waves at the top of {E}arth’s core and implications for variations in length of day},\ }\href {https://doi.org/https://doi.org/10.1093/gji/ggaa368} {\bibfield  {journal} {\bibinfo  {journal} {Geophysical Journal International}\ }\textbf {\bibinfo {volume} {204}},\ \bibinfo {pages} {1789–1800} (\bibinfo {year} {2016})}\BibitemShut {NoStop}%
\bibitem [{\citenamefont {Mound}\ \emph {et~al.}(2019)\citenamefont {Mound}, \citenamefont {Davies}, \citenamefont {Rost},\ and\ \citenamefont {Aurnou}}]{mound}%
  \BibitemOpen
  \bibfield  {author} {\bibinfo {author} {\bibfnamefont {J.}~\bibnamefont {Mound}}, \bibinfo {author} {\bibfnamefont {C.}~\bibnamefont {Davies}}, \bibinfo {author} {\bibfnamefont {S.}~\bibnamefont {Rost}},\ and\ \bibinfo {author} {\bibfnamefont {J.}~\bibnamefont {Aurnou}},\ }\bibfield  {title} {\bibinfo {title} {Regional stratification at the top of {E}arth's core due to core–mantle boundary heat flux variations},\ }\href {https://doi.org/https://doi.org/10.1038/s41561-019-0381-z} {\bibfield  {journal} {\bibinfo  {journal} {Nature Geoscience}\ }\textbf {\bibinfo {volume} {12}},\ \bibinfo {pages} {575} (\bibinfo {year} {2019})}\BibitemShut {NoStop}%
\bibitem [{\citenamefont {Greenwood}\ \emph {et~al.}(2021)\citenamefont {Greenwood}, \citenamefont {Davies},\ and\ \citenamefont {Mound}}]{greenwood}%
  \BibitemOpen
  \bibfield  {author} {\bibinfo {author} {\bibfnamefont {S.}~\bibnamefont {Greenwood}}, \bibinfo {author} {\bibfnamefont {C.~J.}\ \bibnamefont {Davies}},\ and\ \bibinfo {author} {\bibfnamefont {J.~E.}\ \bibnamefont {Mound}},\ }\bibfield  {title} {\bibinfo {title} {On the evolution of thermally stratified layers at the top of {E}arth’s core},\ }\href {https://doi.org/https://doi.org/10.1016/j.pepi.2021.106763} {\bibfield  {journal} {\bibinfo  {journal} {Physics of the {E}arth and Planetary Interiors}\ }\textbf {\bibinfo {volume} {318}} (\bibinfo {year} {2021})}\BibitemShut {NoStop}%
\bibitem [{\citenamefont {Gubbins}\ and\ \citenamefont {Davies}(2013)}]{gubbins}%
  \BibitemOpen
  \bibfield  {author} {\bibinfo {author} {\bibfnamefont {D.}~\bibnamefont {Gubbins}}\ and\ \bibinfo {author} {\bibfnamefont {C.~J.}\ \bibnamefont {Davies}},\ }\bibfield  {title} {\bibinfo {title} {The stratified layer at the core-mantle boundary caused by barodiffusion of {O}xygen, {S}ulphur and {S}ilicon},\ }\href {https://doi.org/https://doi.org/10.1016/j.pepi.2012.11.001} {\bibfield  {journal} {\bibinfo  {journal} {Physics of the {E}arth and Planetary Interiors}\ }\textbf {\bibinfo {volume} {215}},\ \bibinfo {pages} {21} (\bibinfo {year} {2013})}\BibitemShut {NoStop}%
\bibitem [{\citenamefont {Helffrich}(2014)}]{helfrich}%
  \BibitemOpen
  \bibfield  {author} {\bibinfo {author} {\bibfnamefont {G.}~\bibnamefont {Helffrich}},\ }\bibfield  {title} {\bibinfo {title} {Outer core compositional layering and constraints on core liquid transport properties},\ }\href {https://doi.org/https://doi.org/10.1016/j.epsl.2014.01.039} {\bibfield  {journal} {\bibinfo  {journal} {{E}arth and Planetary Science Letters}\ }\textbf {\bibinfo {volume} {391}},\ \bibinfo {pages} {256} (\bibinfo {year} {2014})}\BibitemShut {NoStop}%
\bibitem [{\citenamefont {Brodholt}\ and\ \citenamefont {Badro}(2017)}]{brodholtbadro}%
  \BibitemOpen
  \bibfield  {author} {\bibinfo {author} {\bibfnamefont {J.}~\bibnamefont {Brodholt}}\ and\ \bibinfo {author} {\bibfnamefont {J.}~\bibnamefont {Badro}},\ }\bibfield  {title} {\bibinfo {title} {Composition of the low seismic velocity {E}' layer at the top of {E}arth’s core},\ }\href {https://doi.org/https://doi.org/10.1002/2017GL074261} {\bibfield  {journal} {\bibinfo  {journal} {Geophysical Research Letters}\ }\textbf {\bibinfo {volume} {44}},\ \bibinfo {pages} {8303–8310} (\bibinfo {year} {2017})}\BibitemShut {NoStop}%
\bibitem [{\citenamefont {Yokoo}\ \emph {et~al.}(2022)\citenamefont {Yokoo}, \citenamefont {Hirose}, \citenamefont {Tagawa}, \citenamefont {Morard},\ and\ \citenamefont {Ohishi}}]{yokoo2022}%
  \BibitemOpen
  \bibfield  {author} {\bibinfo {author} {\bibfnamefont {S.}~\bibnamefont {Yokoo}}, \bibinfo {author} {\bibfnamefont {K.}~\bibnamefont {Hirose}}, \bibinfo {author} {\bibfnamefont {S.}~\bibnamefont {Tagawa}}, \bibinfo {author} {\bibfnamefont {G.}~\bibnamefont {Morard}},\ and\ \bibinfo {author} {\bibfnamefont {Y.}~\bibnamefont {Ohishi}},\ }\bibfield  {title} {\bibinfo {title} {Stratification in planetary cores by liquid immiscibility in {F}e-{S}-{H}},\ }\href {https://doi.org/https://doi.org/10.1038/s41467-022-28274-z} {\bibfield  {journal} {\bibinfo  {journal} {Nature Communications}\ }\textbf {\bibinfo {volume} {13}} (\bibinfo {year} {2022})}\BibitemShut {NoStop}%
\bibitem [{\citenamefont {Pozzo}\ \emph {et~al.}(2019)\citenamefont {Pozzo}, \citenamefont {Davies}, \citenamefont {Gubbins},\ and\ \citenamefont {Alfe}}]{pozzo_2019}%
  \BibitemOpen
  \bibfield  {author} {\bibinfo {author} {\bibfnamefont {M.}~\bibnamefont {Pozzo}}, \bibinfo {author} {\bibfnamefont {C.}~\bibnamefont {Davies}}, \bibinfo {author} {\bibfnamefont {D.}~\bibnamefont {Gubbins}},\ and\ \bibinfo {author} {\bibfnamefont {D.}~\bibnamefont {Alfe}},\ }\bibfield  {title} {\bibinfo {title} {\ch{FeO} content of {E}arth’s liquid core},\ }\href {https://doi.org/10.1103/PhysRevX.9.041018} {\bibfield  {journal} {\bibinfo  {journal} {Physical Review X}\ }\textbf {\bibinfo {volume} {9}},\ \bibinfo {pages} {041018} (\bibinfo {year} {2019})}\BibitemShut {NoStop}%
\bibitem [{\citenamefont {Ozawa}\ \emph {et~al.}(2011)\citenamefont {Ozawa}, \citenamefont {Takahashi}, \citenamefont {Hirose}, \citenamefont {Ohishi},\ and\ \citenamefont {Hirao}}]{ozawa}%
  \BibitemOpen
  \bibfield  {author} {\bibinfo {author} {\bibfnamefont {H.}~\bibnamefont {Ozawa}}, \bibinfo {author} {\bibfnamefont {F.}~\bibnamefont {Takahashi}}, \bibinfo {author} {\bibfnamefont {K.}~\bibnamefont {Hirose}}, \bibinfo {author} {\bibfnamefont {Y.}~\bibnamefont {Ohishi}},\ and\ \bibinfo {author} {\bibfnamefont {N.}~\bibnamefont {Hirao}},\ }\bibfield  {title} {\bibinfo {title} {Phase {T}ransition of {F}e{O} and {S}tratification in {E}arth’s {O}uter {C}ore},\ }\href {https://doi.org/https://doi.org/10.1126/science.1208265} {\bibfield  {journal} {\bibinfo  {journal} {Science}\ }\textbf {\bibinfo {volume} {334}},\ \bibinfo {pages} {792} (\bibinfo {year} {2011})}\BibitemShut {NoStop}%
\bibitem [{\citenamefont {Posner}\ \emph {et~al.}(2017)\citenamefont {Posner}, \citenamefont {Steinle-Neumann}, \citenamefont {Vlček},\ and\ \citenamefont {Rubie}}]{posner}%
  \BibitemOpen
  \bibfield  {author} {\bibinfo {author} {\bibfnamefont {E.~S.}\ \bibnamefont {Posner}}, \bibinfo {author} {\bibfnamefont {G.}~\bibnamefont {Steinle-Neumann}}, \bibinfo {author} {\bibfnamefont {V.}~\bibnamefont {Vlček}},\ and\ \bibinfo {author} {\bibfnamefont {D.~C.}\ \bibnamefont {Rubie}},\ }\bibfield  {title} {\bibinfo {title} {Structural changes and anomalous self-diffusion of oxygen in liquid iron at high pressure},\ }\href {https://doi.org/10.1002/2017GL072926} {\bibfield  {journal} {\bibinfo  {journal} {Geophysical Research Letters}\ }\textbf {\bibinfo {volume} {44}},\ \bibinfo {pages} {3526–3534} (\bibinfo {year} {2017})}\BibitemShut {NoStop}%
\bibitem [{\citenamefont {Alfe}\ \emph {et~al.}(1999)\citenamefont {Alfe}, \citenamefont {Price},\ and\ \citenamefont {Gillan}}]{alfe}%
  \BibitemOpen
  \bibfield  {author} {\bibinfo {author} {\bibfnamefont {D.}~\bibnamefont {Alfe}}, \bibinfo {author} {\bibfnamefont {G.~D.}\ \bibnamefont {Price}},\ and\ \bibinfo {author} {\bibfnamefont {M.~J.}\ \bibnamefont {Gillan}},\ }\bibfield  {title} {\bibinfo {title} {Oxygen in the {E}arth’s core: a first-principles study},\ }\href {https://doi.org/10.1016/S0031-9201(98)00134-4} {\bibfield  {journal} {\bibinfo  {journal} {Physics of the {E}arth and Planetary Interiors}\ }\textbf {\bibinfo {volume} {110}},\ \bibinfo {pages} {191} (\bibinfo {year} {1999})}\BibitemShut {NoStop}%
\bibitem [{\citenamefont {Ohmura}\ \emph {et~al.}(2020)\citenamefont {Ohmura}, \citenamefont {Tsuchiya},\ and\ \citenamefont {Shimojo}}]{ohmura}%
  \BibitemOpen
  \bibfield  {author} {\bibinfo {author} {\bibfnamefont {S.}~\bibnamefont {Ohmura}}, \bibinfo {author} {\bibfnamefont {T.}~\bibnamefont {Tsuchiya}},\ and\ \bibinfo {author} {\bibfnamefont {F.}~\bibnamefont {Shimojo}},\ }\bibfield  {title} {\bibinfo {title} {Structures of liquid iron–light-element mixtures under high pressure},\ }\bibfield  {journal} {\bibinfo  {journal} {Basic Solid State Physics}\ }\textbf {\bibinfo {volume} {257}},\ \href {https://doi.org/10.1002/pssb.202000098} {10.1002/pssb.202000098} (\bibinfo {year} {2020})\BibitemShut {NoStop}%
\bibitem [{\citenamefont {Ohmura}\ \emph {et~al.}(2022)\citenamefont {Ohmura}, \citenamefont {Tsuchiya},\ and\ \citenamefont {Shimojo}}]{ohmura_2022}%
  \BibitemOpen
  \bibfield  {author} {\bibinfo {author} {\bibfnamefont {S.}~\bibnamefont {Ohmura}}, \bibinfo {author} {\bibfnamefont {T.}~\bibnamefont {Tsuchiya}},\ and\ \bibinfo {author} {\bibfnamefont {F.}~\bibnamefont {Shimojo}},\ }\bibfield  {title} {\bibinfo {title} {Ab initio molecular–dynamics study of structural and bonding properties of liquid fe–light–element–o systems under high pressure},\ }\bibfield  {journal} {\bibinfo  {journal} {Front. {E}arth Sci.,}\ }\textbf {\bibinfo {volume} {10}},\ \href {https://doi.org/10.3389/feart.2022.873088} {10.3389/feart.2022.873088} (\bibinfo {year} {2022})\BibitemShut {NoStop}%
\bibitem [{\citenamefont {Crepisson}\ \emph {et~al.}(2025)\citenamefont {Crepisson}, \citenamefont {Amouretti}, \citenamefont {Harmand}, \citenamefont {Sanloup}, \citenamefont {Heighway}, \citenamefont {Azadi}, \citenamefont {McGonegle}, \citenamefont {Campbell}, \citenamefont {Pintor}, \citenamefont {Chin}, \citenamefont {Smith}, \citenamefont {Hansen}, \citenamefont {Forte}, \citenamefont {Gawne}, \citenamefont {Lee}, \citenamefont {Nagler}, \citenamefont {Shi}, \citenamefont {Fiquet}, \citenamefont {Guyot}, \citenamefont {Makita}, \citenamefont {Benuzzi-Mounaix}, \citenamefont {Vinci}, \citenamefont {Miyanishi}, \citenamefont {Ozaki}, \citenamefont {Pikuz}, \citenamefont {Nakamura}, \citenamefont {Keiichi~Sueda}, \citenamefont {Yabashi}, \citenamefont {Wark}, \citenamefont {Polsin},\ and\ \citenamefont {Vinko}}]{crepissonPRB}%
  \BibitemOpen
  \bibfield  {author} {\bibinfo {author} {\bibfnamefont {C.}~\bibnamefont {Crepisson}}, \bibinfo {author} {\bibfnamefont {A.}~\bibnamefont {Amouretti}}, \bibinfo {author} {\bibfnamefont {M.}~\bibnamefont {Harmand}}, \bibinfo {author} {\bibfnamefont {C.}~\bibnamefont {Sanloup}}, \bibinfo {author} {\bibfnamefont {P.}~\bibnamefont {Heighway}}, \bibinfo {author} {\bibfnamefont {S.}~\bibnamefont {Azadi}}, \bibinfo {author} {\bibfnamefont {D.}~\bibnamefont {McGonegle}}, \bibinfo {author} {\bibfnamefont {T.}~\bibnamefont {Campbell}}, \bibinfo {author} {\bibfnamefont {J.}~\bibnamefont {Pintor}}, \bibinfo {author} {\bibfnamefont {D.~A.}\ \bibnamefont {Chin}}, \bibinfo {author} {\bibfnamefont {E.}~\bibnamefont {Smith}}, \bibinfo {author} {\bibfnamefont {L.}~\bibnamefont {Hansen}}, \bibinfo {author} {\bibfnamefont {A.}~\bibnamefont {Forte}}, \bibinfo {author} {\bibfnamefont {T.}~\bibnamefont {Gawne}}, \bibinfo {author} {\bibfnamefont {H.~J.}\ \bibnamefont {Lee}}, \bibinfo {author} {\bibfnamefont {B.}~\bibnamefont
  {Nagler}}, \bibinfo {author} {\bibfnamefont {Y.}~\bibnamefont {Shi}}, \bibinfo {author} {\bibfnamefont {G.}~\bibnamefont {Fiquet}}, \bibinfo {author} {\bibfnamefont {F.}~\bibnamefont {Guyot}}, \bibinfo {author} {\bibfnamefont {M.}~\bibnamefont {Makita}}, \bibinfo {author} {\bibfnamefont {A.}~\bibnamefont {Benuzzi-Mounaix}}, \bibinfo {author} {\bibfnamefont {T.}~\bibnamefont {Vinci}}, \bibinfo {author} {\bibfnamefont {K.}~\bibnamefont {Miyanishi}}, \bibinfo {author} {\bibfnamefont {N.}~\bibnamefont {Ozaki}}, \bibinfo {author} {\bibfnamefont {T.}~\bibnamefont {Pikuz}}, \bibinfo {author} {\bibfnamefont {H.}~\bibnamefont {Nakamura}}, \bibinfo {author} {\bibfnamefont {T.~Y.}\ \bibnamefont {Keiichi~Sueda}}, \bibinfo {author} {\bibfnamefont {M.}~\bibnamefont {Yabashi}}, \bibinfo {author} {\bibfnamefont {J.~S.}\ \bibnamefont {Wark}}, \bibinfo {author} {\bibfnamefont {D.~N.}\ \bibnamefont {Polsin}},\ and\ \bibinfo {author} {\bibfnamefont {S.~M.}\ \bibnamefont {Vinko}},\ }\bibfield  {title} {\bibinfo {title}
  {Shock-driven amorphization and melting in \ch{Fe2O3}},\ }\href {https://doi.org/https://doi.org/10.1103/PhysRevB.111.024209} {\bibfield  {journal} {\bibinfo  {journal} {Physical Review B}\ }\textbf {\bibinfo {volume} {111}},\ \bibinfo {pages} {111, 024209} (\bibinfo {year} {2025})}\BibitemShut {NoStop}%
\bibitem [{\citenamefont {Dziewonski}\ and\ \citenamefont {Anderson}(1981)}]{dziewonski}%
  \BibitemOpen
  \bibfield  {author} {\bibinfo {author} {\bibfnamefont {A.~M.}\ \bibnamefont {Dziewonski}}\ and\ \bibinfo {author} {\bibfnamefont {D.~L.}\ \bibnamefont {Anderson}},\ }\bibfield  {title} {\bibinfo {title} {Preliminary reference {E}arth model},\ }\href {https://doi.org/https://doi.org/10.1016/0031-9201(81)90046-7} {\bibfield  {journal} {\bibinfo  {journal} {Physics of the {E}arth and Planetary Interiors}\ }\textbf {\bibinfo {volume} {25}},\ \bibinfo {pages} {297} (\bibinfo {year} {1981})}\BibitemShut {NoStop}%
\bibitem [{\citenamefont {Singh}\ \emph {et~al.}(2023)\citenamefont {Singh}, \citenamefont {Briggs}, \citenamefont {Gorman}, \citenamefont {Benedict}, \citenamefont {Wu}, \citenamefont {Hamel}, \citenamefont {Coleman}, \citenamefont {Coppari}, \citenamefont {Fernandez-Pañella}, \citenamefont {McGuire}, \citenamefont {Sims}, \citenamefont {Wicks}, \citenamefont {Eggert}, \citenamefont {Fratanduono},\ and\ \citenamefont {Smith}}]{singh}%
  \BibitemOpen
  \bibfield  {author} {\bibinfo {author} {\bibfnamefont {S.}~\bibnamefont {Singh}}, \bibinfo {author} {\bibfnamefont {R.}~\bibnamefont {Briggs}}, \bibinfo {author} {\bibfnamefont {M.~G.}\ \bibnamefont {Gorman}}, \bibinfo {author} {\bibfnamefont {L.~X.}\ \bibnamefont {Benedict}}, \bibinfo {author} {\bibfnamefont {C.~J.}\ \bibnamefont {Wu}}, \bibinfo {author} {\bibfnamefont {S.}~\bibnamefont {Hamel}}, \bibinfo {author} {\bibfnamefont {A.~L.}\ \bibnamefont {Coleman}}, \bibinfo {author} {\bibfnamefont {F.}~\bibnamefont {Coppari}}, \bibinfo {author} {\bibfnamefont {A.}~\bibnamefont {Fernandez-Pañella}}, \bibinfo {author} {\bibfnamefont {C.}~\bibnamefont {McGuire}}, \bibinfo {author} {\bibfnamefont {M.}~\bibnamefont {Sims}}, \bibinfo {author} {\bibfnamefont {J.~K.}\ \bibnamefont {Wicks}}, \bibinfo {author} {\bibfnamefont {J.~H.}\ \bibnamefont {Eggert}}, \bibinfo {author} {\bibfnamefont {D.~E.}\ \bibnamefont {Fratanduono}},\ and\ \bibinfo {author} {\bibfnamefont {R.~F.}\ \bibnamefont {Smith}},\ }\bibfield  {title}
  {\bibinfo {title} {{Structural study of hcp and liquid iron under shock compression up to 275 {G}{P}a}},\ }\href {https://doi.org/10.1103/PhysRevB.108.184104} {\bibfield  {journal} {\bibinfo  {journal} {Physical Review B}\ }\textbf {\bibinfo {volume} {108}},\ \bibinfo {pages} {184104} (\bibinfo {year} {2023})}\BibitemShut {NoStop}%
\bibitem [{\citenamefont {Morard}\ \emph {et~al.}(2022)\citenamefont {Morard}, \citenamefont {Antonangeli}, \citenamefont {Bouchet}, \citenamefont {Rivoldini}, \citenamefont {Boccato}, \citenamefont {Miozzi}, \citenamefont {Boulard}, \citenamefont {Bureau}, \citenamefont {Mezouar}, \citenamefont {Prescher}, \citenamefont {Chariton},\ and\ \citenamefont {Greenberg}}]{morard2022}%
  \BibitemOpen
  \bibfield  {author} {\bibinfo {author} {\bibfnamefont {G.}~\bibnamefont {Morard}}, \bibinfo {author} {\bibfnamefont {D.}~\bibnamefont {Antonangeli}}, \bibinfo {author} {\bibfnamefont {J.}~\bibnamefont {Bouchet}}, \bibinfo {author} {\bibfnamefont {A.}~\bibnamefont {Rivoldini}}, \bibinfo {author} {\bibfnamefont {S.}~\bibnamefont {Boccato}}, \bibinfo {author} {\bibfnamefont {F.}~\bibnamefont {Miozzi}}, \bibinfo {author} {\bibfnamefont {E.}~\bibnamefont {Boulard}}, \bibinfo {author} {\bibfnamefont {H.}~\bibnamefont {Bureau}}, \bibinfo {author} {\bibfnamefont {M.}~\bibnamefont {Mezouar}}, \bibinfo {author} {\bibfnamefont {C.}~\bibnamefont {Prescher}}, \bibinfo {author} {\bibfnamefont {S.}~\bibnamefont {Chariton}},\ and\ \bibinfo {author} {\bibfnamefont {E.}~\bibnamefont {Greenberg}},\ }\bibfield  {title} {\bibinfo {title} {Structural and {E}lectronic {T}ransitions in {L}iquid {F}e{O} {U}nder {H}igh {P}ressure},\ }\href {https://doi.org/https://doi.org/10.1029/2022JB025117} {\bibfield  {journal} {\bibinfo
  {journal} {Journal of Geophysical Research: Solid {E}arth}\ }\textbf {\bibinfo {volume} {127}},\ \bibinfo {pages} {e2022JB025117} (\bibinfo {year} {2022})}\BibitemShut {NoStop}%
\bibitem [{\citenamefont {Amouretti}\ \emph {et~al.}(2025)\citenamefont {Amouretti}, \citenamefont {Crepisson}, \citenamefont {Azadi}, \citenamefont {Cabaret}, \citenamefont {Campbell}, \citenamefont {Chin}, \citenamefont {Colin}, \citenamefont {Collins}, \citenamefont {Crandall}, \citenamefont {Fiquet}, \citenamefont {Forte}, \citenamefont {Gawne}, \citenamefont {Guyot}, \citenamefont {Heighway}, \citenamefont {Lee}, \citenamefont {McGonegle}, \citenamefont {Nagler}, \citenamefont {Pintor}, \citenamefont {Polsin}, \citenamefont {Rousse}, \citenamefont {Shi}, \citenamefont {Smith}, \citenamefont {Wark}, \citenamefont {Vinko},\ and\ \citenamefont {Harmand}}]{amourettiPRL}%
  \BibitemOpen
  \bibfield  {author} {\bibinfo {author} {\bibfnamefont {A.}~\bibnamefont {Amouretti}}, \bibinfo {author} {\bibfnamefont {C.}~\bibnamefont {Crepisson}}, \bibinfo {author} {\bibfnamefont {S.}~\bibnamefont {Azadi}}, \bibinfo {author} {\bibfnamefont {D.}~\bibnamefont {Cabaret}}, \bibinfo {author} {\bibfnamefont {T.}~\bibnamefont {Campbell}}, \bibinfo {author} {\bibfnamefont {D.~A.}\ \bibnamefont {Chin}}, \bibinfo {author} {\bibfnamefont {B.}~\bibnamefont {Colin}}, \bibinfo {author} {\bibfnamefont {G.~R.}\ \bibnamefont {Collins}}, \bibinfo {author} {\bibfnamefont {L.}~\bibnamefont {Crandall}}, \bibinfo {author} {\bibfnamefont {G.}~\bibnamefont {Fiquet}}, \bibinfo {author} {\bibfnamefont {A.}~\bibnamefont {Forte}}, \bibinfo {author} {\bibfnamefont {T.}~\bibnamefont {Gawne}}, \bibinfo {author} {\bibfnamefont {F.}~\bibnamefont {Guyot}}, \bibinfo {author} {\bibfnamefont {P.}~\bibnamefont {Heighway}}, \bibinfo {author} {\bibfnamefont {H.}~\bibnamefont {Lee}}, \bibinfo {author} {\bibfnamefont {D.}~\bibnamefont
  {McGonegle}}, \bibinfo {author} {\bibfnamefont {B.}~\bibnamefont {Nagler}}, \bibinfo {author} {\bibfnamefont {J.}~\bibnamefont {Pintor}}, \bibinfo {author} {\bibfnamefont {D.}~\bibnamefont {Polsin}}, \bibinfo {author} {\bibfnamefont {G.}~\bibnamefont {Rousse}}, \bibinfo {author} {\bibfnamefont {Y.}~\bibnamefont {Shi}}, \bibinfo {author} {\bibfnamefont {E.}~\bibnamefont {Smith}}, \bibinfo {author} {\bibfnamefont {J.~S.}\ \bibnamefont {Wark}}, \bibinfo {author} {\bibfnamefont {S.~M.}\ \bibnamefont {Vinko}},\ and\ \bibinfo {author} {\bibfnamefont {M.}~\bibnamefont {Harmand}},\ }\bibfield  {title} {\bibinfo {title} {Phase transitions of \ch{Fe2O3} under laser shock compression},\ }\href {https://doi.org/https://doi.org/10.1103/PhysRevLett.134.176102} {\bibfield  {journal} {\bibinfo  {journal} {Physycal Review Letters}\ }\textbf {\bibinfo {volume} {134}},\ \bibinfo {pages} {176102} (\bibinfo {year} {2025})}\BibitemShut {NoStop}%
\bibitem [{\citenamefont {Renganathan}\ \emph {et~al.}(2023)\citenamefont {Renganathan}, \citenamefont {Sharma}, \citenamefont {Turneaure},\ and\ \citenamefont {Gupta}}]{renganathan}%
  \BibitemOpen
  \bibfield  {author} {\bibinfo {author} {\bibfnamefont {P.}~\bibnamefont {Renganathan}}, \bibinfo {author} {\bibfnamefont {S.~M.}\ \bibnamefont {Sharma}}, \bibinfo {author} {\bibfnamefont {S.~J.}\ \bibnamefont {Turneaure}},\ and\ \bibinfo {author} {\bibfnamefont {Y.~M.}\ \bibnamefont {Gupta}},\ }\bibfield  {title} {\bibinfo {title} {Real-time (nanoseconds) determination of liquid phase growth during shock-induced melting},\ }\href {https://doi.org/https://doi.org/10.1126/sciadv.ade5745} {\bibfield  {journal} {\bibinfo  {journal} {Science Advances}\ }\textbf {\bibinfo {volume} {9}},\ \bibinfo {pages} {eade5745} (\bibinfo {year} {2023})}\BibitemShut {NoStop}%
\bibitem [{\citenamefont {Zastrau}\ \emph {et~al.}(2021)\citenamefont {Zastrau}, \citenamefont {Appel}, \citenamefont {Baehtz}, \citenamefont {Baehr}, \citenamefont {Batchelor}, \citenamefont {Berghauser}, \citenamefont {Banjafar}, \citenamefont {Brambrink}, \citenamefont {Cerantola}, \citenamefont {Cowan}, \citenamefont {Damker}, \citenamefont {Dietrich}, \citenamefont {Cafiso}, \citenamefont {Dreyer}, \citenamefont {Engel}, \citenamefont {Feldmann}, \citenamefont {Findeisen}, \citenamefont {Foese}, \citenamefont {Fulla-{M}arsa}, \citenamefont {Gode}, \citenamefont {Hassan}, \citenamefont {Hauser}, \citenamefont {Herrmannsdorfer}, \citenamefont {Hoppner}, \citenamefont {Kaa}, \citenamefont {Kaever}, \citenamefont {Knofel}, \citenamefont {Konopkova}, \citenamefont {Garcia}, \citenamefont {Liermann}, \citenamefont {Mainberger}, \citenamefont {Makita}, \citenamefont {Martens}, \citenamefont {McBride}, \citenamefont {Moller}, \citenamefont {Nakatsutsumi}, \citenamefont {Pelka}, \citenamefont {Plueckthun},
  \citenamefont {Prescher}, \citenamefont {Preston}, \citenamefont {Roper}, \citenamefont {Schmidt}, \citenamefont {Seidel}, \citenamefont {Schwinkendorf}, \citenamefont {Schoelmerichch{}}, \citenamefont {Schramm}, \citenamefont {Schropp}, \citenamefont {Strohm}, \citenamefont {Sukharnikov}, \citenamefont {Talkovski}, \citenamefont {Thorpe}, \citenamefont {Toncian}, \citenamefont {Toncian}, \citenamefont {Wollenweber}, \citenamefont {Yamamoto},\ and\ \citenamefont {Tschentscher}}]{zastrau}%
  \BibitemOpen
  \bibfield  {author} {\bibinfo {author} {\bibfnamefont {U.}~\bibnamefont {Zastrau}}, \bibinfo {author} {\bibfnamefont {K.}~\bibnamefont {Appel}}, \bibinfo {author} {\bibfnamefont {C.}~\bibnamefont {Baehtz}}, \bibinfo {author} {\bibfnamefont {O.}~\bibnamefont {Baehr}}, \bibinfo {author} {\bibfnamefont {L.}~\bibnamefont {Batchelor}}, \bibinfo {author} {\bibfnamefont {A.}~\bibnamefont {Berghauser}}, \bibinfo {author} {\bibfnamefont {M.}~\bibnamefont {Banjafar}}, \bibinfo {author} {\bibfnamefont {E.}~\bibnamefont {Brambrink}}, \bibinfo {author} {\bibfnamefont {V.}~\bibnamefont {Cerantola}}, \bibinfo {author} {\bibfnamefont {T.}~\bibnamefont {Cowan}}, \bibinfo {author} {\bibfnamefont {H.}~\bibnamefont {Damker}}, \bibinfo {author} {\bibfnamefont {S.}~\bibnamefont {Dietrich}}, \bibinfo {author} {\bibfnamefont {S.~D.~D.}\ \bibnamefont {Cafiso}}, \bibinfo {author} {\bibfnamefont {J.}~\bibnamefont {Dreyer}}, \bibinfo {author} {\bibfnamefont {H.-O.}\ \bibnamefont {Engel}}, \bibinfo {author} {\bibfnamefont
  {T.}~\bibnamefont {Feldmann}}, \bibinfo {author} {\bibfnamefont {S.}~\bibnamefont {Findeisen}}, \bibinfo {author} {\bibfnamefont {M.}~\bibnamefont {Foese}}, \bibinfo {author} {\bibfnamefont {D.}~\bibnamefont {Fulla-{M}arsa}}, \bibinfo {author} {\bibfnamefont {S.}~\bibnamefont {Gode}}, \bibinfo {author} {\bibfnamefont {M.}~\bibnamefont {Hassan}}, \bibinfo {author} {\bibfnamefont {J.}~\bibnamefont {Hauser}}, \bibinfo {author} {\bibfnamefont {T.}~\bibnamefont {Herrmannsdorfer}}, \bibinfo {author} {\bibfnamefont {H.}~\bibnamefont {Hoppner}}, \bibinfo {author} {\bibfnamefont {J.}~\bibnamefont {Kaa}}, \bibinfo {author} {\bibfnamefont {P.}~\bibnamefont {Kaever}}, \bibinfo {author} {\bibfnamefont {K.}~\bibnamefont {Knofel}}, \bibinfo {author} {\bibfnamefont {Z.}~\bibnamefont {Konopkova}}, \bibinfo {author} {\bibfnamefont {A.~L.}\ \bibnamefont {Garcia}}, \bibinfo {author} {\bibfnamefont {H.-P.}\ \bibnamefont {Liermann}}, \bibinfo {author} {\bibfnamefont {J.}~\bibnamefont {Mainberger}}, \bibinfo {author}
  {\bibfnamefont {M.}~\bibnamefont {Makita}}, \bibinfo {author} {\bibfnamefont {E.-C.}\ \bibnamefont {Martens}}, \bibinfo {author} {\bibfnamefont {E.~E.}\ \bibnamefont {McBride}}, \bibinfo {author} {\bibfnamefont {D.}~\bibnamefont {Moller}}, \bibinfo {author} {\bibfnamefont {M.}~\bibnamefont {Nakatsutsumi}}, \bibinfo {author} {\bibfnamefont {A.}~\bibnamefont {Pelka}}, \bibinfo {author} {\bibfnamefont {C.}~\bibnamefont {Plueckthun}}, \bibinfo {author} {\bibfnamefont {C.}~\bibnamefont {Prescher}}, \bibinfo {author} {\bibfnamefont {T.~R.}\ \bibnamefont {Preston}}, \bibinfo {author} {\bibfnamefont {M.}~\bibnamefont {Roper}}, \bibinfo {author} {\bibfnamefont {A.}~\bibnamefont {Schmidt}}, \bibinfo {author} {\bibfnamefont {W.}~\bibnamefont {Seidel}}, \bibinfo {author} {\bibfnamefont {J.-P.}\ \bibnamefont {Schwinkendorf}}, \bibinfo {author} {\bibfnamefont {M.~O.}\ \bibnamefont {Schoelmerichch{}}}, \bibinfo {author} {\bibfnamefont {U.}~\bibnamefont {Schramm}}, \bibinfo {author} {\bibfnamefont {A.}~\bibnamefont
  {Schropp}}, \bibinfo {author} {\bibfnamefont {C.}~\bibnamefont {Strohm}}, \bibinfo {author} {\bibfnamefont {K.}~\bibnamefont {Sukharnikov}}, \bibinfo {author} {\bibfnamefont {P.}~\bibnamefont {Talkovski}}, \bibinfo {author} {\bibfnamefont {I.}~\bibnamefont {Thorpe}}, \bibinfo {author} {\bibfnamefont {M.}~\bibnamefont {Toncian}}, \bibinfo {author} {\bibfnamefont {T.}~\bibnamefont {Toncian}}, \bibinfo {author} {\bibfnamefont {L.}~\bibnamefont {Wollenweber}}, \bibinfo {author} {\bibfnamefont {S.}~\bibnamefont {Yamamoto}},\ and\ \bibinfo {author} {\bibfnamefont {T.}~\bibnamefont {Tschentscher}},\ }\bibfield  {title} {\bibinfo {title} {{The High Energy Density Scientific Instrument at the European {X}{F}{E}{L}}},\ }\href {https://doi.org/10.1107/S1600577521007335} {\bibfield  {journal} {\bibinfo  {journal} {Journal of Synchrotron Radiation}\ }\textbf {\bibinfo {volume} {28}},\ \bibinfo {pages} {1393} (\bibinfo {year} {2021})}\BibitemShut {NoStop}%
\bibitem [{\citenamefont {Bancroft}\ \emph {et~al.}(1956)\citenamefont {Bancroft}, \citenamefont {Peterson},\ and\ \citenamefont {Minshall}}]{bancroft}%
  \BibitemOpen
  \bibfield  {author} {\bibinfo {author} {\bibfnamefont {D.}~\bibnamefont {Bancroft}}, \bibinfo {author} {\bibfnamefont {E.~L.}\ \bibnamefont {Peterson}},\ and\ \bibinfo {author} {\bibfnamefont {S.}~\bibnamefont {Minshall}},\ }\bibfield  {title} {\bibinfo {title} {Polymorphism of iron at high pressure},\ }\href {https://doi.org/10.1063/1.1722359} {\bibfield  {journal} {\bibinfo  {journal} {J. Appl. Phys.}\ }\textbf {\bibinfo {volume} {27}},\ \bibinfo {pages} {291–298} (\bibinfo {year} {1956})}\BibitemShut {NoStop}%
\bibitem [{\citenamefont {Turneaure}\ \emph {et~al.}(2020)\citenamefont {Turneaure}, \citenamefont {Sharma},\ and\ \citenamefont {Gupta}}]{turneaure}%
  \BibitemOpen
  \bibfield  {author} {\bibinfo {author} {\bibfnamefont {S.~J.}\ \bibnamefont {Turneaure}}, \bibinfo {author} {\bibfnamefont {S.~M.}\ \bibnamefont {Sharma}},\ and\ \bibinfo {author} {\bibfnamefont {Y.~M.}\ \bibnamefont {Gupta}},\ }\bibfield  {title} {\bibinfo {title} {Crystal structure and melting of fe shock compressed to 273 {G}{P}a: In situ x-ray diffraction},\ }\href {https://doi.org/10.1103/PhysRevLett.125.215702} {\bibfield  {journal} {\bibinfo  {journal} {Phys. Rev. Lett.}\ }\textbf {\bibinfo {volume} {125}},\ \bibinfo {pages} {215702} (\bibinfo {year} {2020})}\BibitemShut {NoStop}%
\bibitem [{\citenamefont {Fischer}\ \emph {et~al.}(2011)\citenamefont {Fischer}, \citenamefont {Campbell}, \citenamefont {Shofner}, \citenamefont {Lord}, \citenamefont {Dera},\ and\ \citenamefont {Prakapenka}}]{fisher}%
  \BibitemOpen
  \bibfield  {author} {\bibinfo {author} {\bibfnamefont {R.~A.}\ \bibnamefont {Fischer}}, \bibinfo {author} {\bibfnamefont {A.~J.}\ \bibnamefont {Campbell}}, \bibinfo {author} {\bibfnamefont {G.~A.}\ \bibnamefont {Shofner}}, \bibinfo {author} {\bibfnamefont {O.~T.}\ \bibnamefont {Lord}}, \bibinfo {author} {\bibfnamefont {P.}~\bibnamefont {Dera}},\ and\ \bibinfo {author} {\bibfnamefont {V.~B.}\ \bibnamefont {Prakapenka}},\ }\bibfield  {title} {\bibinfo {title} {Equation of state and phase diagram of feo},\ }\href {https://doi.org/10.1016/j.epsl.2011.02.025} {\bibfield  {journal} {\bibinfo  {journal} {{E}arth and Planetary Science Letters}\ }\textbf {\bibinfo {volume} {304}},\ \bibinfo {pages} {496} (\bibinfo {year} {2011})}\BibitemShut {NoStop}%
\bibitem [{\citenamefont {Greenberg}\ \emph {et~al.}(2023)\citenamefont {Greenberg}, \citenamefont {Nazarov}, \citenamefont {Landa}, \citenamefont {Ying}, \citenamefont {Hood}, \citenamefont {Hen}, \citenamefont {Jeanloz}, \citenamefont {Prakapenka}, \citenamefont {Struzhkin}, \citenamefont {Rozenberg},\ and\ \citenamefont {Leonov}}]{greenberg_FeO}%
  \BibitemOpen
  \bibfield  {author} {\bibinfo {author} {\bibfnamefont {E.}~\bibnamefont {Greenberg}}, \bibinfo {author} {\bibfnamefont {R.}~\bibnamefont {Nazarov}}, \bibinfo {author} {\bibfnamefont {A.}~\bibnamefont {Landa}}, \bibinfo {author} {\bibfnamefont {J.}~\bibnamefont {Ying}}, \bibinfo {author} {\bibfnamefont {R.~Q.}\ \bibnamefont {Hood}}, \bibinfo {author} {\bibfnamefont {B.}~\bibnamefont {Hen}}, \bibinfo {author} {\bibfnamefont {R.}~\bibnamefont {Jeanloz}}, \bibinfo {author} {\bibfnamefont {V.~B.}\ \bibnamefont {Prakapenka}}, \bibinfo {author} {\bibfnamefont {V.~V.}\ \bibnamefont {Struzhkin}}, \bibinfo {author} {\bibfnamefont {G.~K.}\ \bibnamefont {Rozenberg}},\ and\ \bibinfo {author} {\bibfnamefont {I.~V.}\ \bibnamefont {Leonov}},\ }\bibfield  {title} {\bibinfo {title} {Phase transitions and spin state of iron in \ch{FeO} under the conditions of {E}arth's deep interior},\ }\href {https://doi.org/10.1103/PhysRevB.107.L241103} {\bibfield  {journal} {\bibinfo  {journal} {Phys. Rev. B}\ }\textbf {\bibinfo {volume}
  {107}},\ \bibinfo {pages} {L241103} (\bibinfo {year} {2023})}\BibitemShut {NoStop}%
\bibitem [{\citenamefont {Jeanloz}\ and\ \citenamefont {Ahrens}(1980)}]{jeanloz}%
  \BibitemOpen
  \bibfield  {author} {\bibinfo {author} {\bibfnamefont {R.}~\bibnamefont {Jeanloz}}\ and\ \bibinfo {author} {\bibfnamefont {T.~J.}\ \bibnamefont {Ahrens}},\ }\bibfield  {title} {\bibinfo {title} {Equations of state of {F}e{O} and {C}a{O}},\ }\href {https://doi.org/https://doi.org/10.1111/j.1365-246X.1980.tb02588.x} {\bibfield  {journal} {\bibinfo  {journal} {Geophysical Journal International}\ }\textbf {\bibinfo {volume} {62}},\ \bibinfo {pages} {505–528} (\bibinfo {year} {1980})}\BibitemShut {NoStop}%
\bibitem [{\citenamefont {Hawreliak}\ \emph {et~al.}(2008)\citenamefont {Hawreliak}, \citenamefont {Kalantar}, \citenamefont {Stölken}, \citenamefont {Remington}, \citenamefont {Lorenzana},\ and\ \citenamefont {Wark}}]{hawreliak2008}%
  \BibitemOpen
  \bibfield  {author} {\bibinfo {author} {\bibfnamefont {J.~A.}\ \bibnamefont {Hawreliak}}, \bibinfo {author} {\bibfnamefont {D.~H.}\ \bibnamefont {Kalantar}}, \bibinfo {author} {\bibfnamefont {J.~S.}\ \bibnamefont {Stölken}}, \bibinfo {author} {\bibfnamefont {B.~A.}\ \bibnamefont {Remington}}, \bibinfo {author} {\bibfnamefont {H.~E.}\ \bibnamefont {Lorenzana}},\ and\ \bibinfo {author} {\bibfnamefont {J.~S.}\ \bibnamefont {Wark}},\ }\bibfield  {title} {\bibinfo {title} {High-pressure nanocrystalline structure of a shock-compressed single crystal of iron},\ }\href {https://doi.org/10.1103/PhysRevB.78.220101} {\bibfield  {journal} {\bibinfo  {journal} {Phys. Rev. B}\ }\textbf {\bibinfo {volume} {78}},\ \bibinfo {pages} {220101(R)} (\bibinfo {year} {2008})}\BibitemShut {NoStop}%
\bibitem [{\citenamefont {Bykova}\ \emph {et~al.}(2016)\citenamefont {Bykova}, \citenamefont {Dubrovinsky}, \citenamefont {Dubrovinskaia}, \citenamefont {Bykov}, \citenamefont {McCammon}, \citenamefont {Ovsyannikov}, \citenamefont {Liermann}, \citenamefont {Kupenko}, \citenamefont {Chumakov}, \citenamefont {Rüffer}, \citenamefont {Hanfland},\ and\ \citenamefont {Prakapenka}}]{bykova}%
  \BibitemOpen
  \bibfield  {author} {\bibinfo {author} {\bibfnamefont {E.}~\bibnamefont {Bykova}}, \bibinfo {author} {\bibfnamefont {L.}~\bibnamefont {Dubrovinsky}}, \bibinfo {author} {\bibfnamefont {N.}~\bibnamefont {Dubrovinskaia}}, \bibinfo {author} {\bibfnamefont {M.}~\bibnamefont {Bykov}}, \bibinfo {author} {\bibfnamefont {C.}~\bibnamefont {McCammon}}, \bibinfo {author} {\bibfnamefont {S.~V.}\ \bibnamefont {Ovsyannikov}}, \bibinfo {author} {\bibfnamefont {H.-P.}\ \bibnamefont {Liermann}}, \bibinfo {author} {\bibfnamefont {I.}~\bibnamefont {Kupenko}}, \bibinfo {author} {\bibfnamefont {A.~I.}\ \bibnamefont {Chumakov}}, \bibinfo {author} {\bibfnamefont {R.}~\bibnamefont {Rüffer}}, \bibinfo {author} {\bibfnamefont {M.}~\bibnamefont {Hanfland}},\ and\ \bibinfo {author} {\bibfnamefont {V.}~\bibnamefont {Prakapenka}},\ }\bibfield  {title} {\bibinfo {title} {Structural complexity of simple {Fe$_2$O$_3$} at high pressures and temperatures},\ }\href {https://doi.org/10.1038/ncomms10661} {\bibfield  {journal} {\bibinfo
  {journal} {Nature Communications}\ }\textbf {\bibinfo {volume} {7}},\ \bibinfo {pages} {10661} (\bibinfo {year} {2016})}\BibitemShut {NoStop}%
\bibitem [{\citenamefont {Barnes}\ and\ \citenamefont {Lyon}(1987)}]{barnes_sesame_1987}%
  \BibitemOpen
  \bibfield  {author} {\bibinfo {author} {\bibfnamefont {J.}~\bibnamefont {Barnes}}\ and\ \bibinfo {author} {\bibfnamefont {S.}~\bibnamefont {Lyon}},\ }\bibfield  {title} {\bibinfo {title} {{SESAME}: the {Los} {Alamos} {National} {Laboratory} equation of state database},\ }\href@noop {} {\bibfield  {journal} {\bibinfo  {journal} {Los Alamos National Laboratory}\ } (\bibinfo {year} {1987})}\BibitemShut {NoStop}%
\bibitem [{\citenamefont {Fu}\ and\ \citenamefont {Hirose}(2024)}]{fu_hirose}%
  \BibitemOpen
  \bibfield  {author} {\bibinfo {author} {\bibfnamefont {S.}~\bibnamefont {Fu}}\ and\ \bibinfo {author} {\bibfnamefont {K.}~\bibnamefont {Hirose}},\ }\bibfield  {title} {\bibinfo {title} {Melting behavior of b1 \ch{FeO} up to 186 {G}{P}a: Existence of feo-rich melts in the lowermost mantle},\ }\href {https://doi.org/10.1029/2023GL106475} {\bibfield  {journal} {\bibinfo  {journal} {Geophysical Research Letters}\ }\textbf {\bibinfo {volume} {51}},\ \bibinfo {pages} {e2023GL106475} (\bibinfo {year} {2024})}\BibitemShut {NoStop}%
\bibitem [{\citenamefont {Dobrosavljevic}\ \emph {et~al.}(2023)\citenamefont {Dobrosavljevic}, \citenamefont {Zhang}, \citenamefont {Sturhahn}, \citenamefont {Chariton}, \citenamefont {Prakapenka}, \citenamefont {Zhao}, \citenamefont {Toellner}, \citenamefont {Pardo},\ and\ \citenamefont {Jackson}}]{dobrosavljevic}%
  \BibitemOpen
  \bibfield  {author} {\bibinfo {author} {\bibfnamefont {V.~V.}\ \bibnamefont {Dobrosavljevic}}, \bibinfo {author} {\bibfnamefont {D.}~\bibnamefont {Zhang}}, \bibinfo {author} {\bibfnamefont {W.}~\bibnamefont {Sturhahn}}, \bibinfo {author} {\bibfnamefont {S.}~\bibnamefont {Chariton}}, \bibinfo {author} {\bibfnamefont {V.~B.}\ \bibnamefont {Prakapenka}}, \bibinfo {author} {\bibfnamefont {J.}~\bibnamefont {Zhao}}, \bibinfo {author} {\bibfnamefont {T.~S.}\ \bibnamefont {Toellner}}, \bibinfo {author} {\bibfnamefont {O.~S.}\ \bibnamefont {Pardo}},\ and\ \bibinfo {author} {\bibfnamefont {J.~M.}\ \bibnamefont {Jackson}},\ }\bibfield  {title} {\bibinfo {title} {Melting and defect transitions in {F}e{O} up to pressures of {E}arth’s core-mantle boundary},\ }\href {https://doi.org/https://doi.org/10.1038/s41467-023-43154-w} {\bibfield  {journal} {\bibinfo  {journal} {Nature Communications}\ }\textbf {\bibinfo {volume} {14}},\ \bibinfo {pages} {7336} (\bibinfo {year} {2023})}\BibitemShut {NoStop}%
\bibitem [{\citenamefont {Eggert}\ \emph {et~al.}(2002)\citenamefont {Eggert}, \citenamefont {Weck}, \citenamefont {Loubeyre},\ and\ \citenamefont {Mezouar}}]{eggert}%
  \BibitemOpen
  \bibfield  {author} {\bibinfo {author} {\bibfnamefont {J.~H.}\ \bibnamefont {Eggert}}, \bibinfo {author} {\bibfnamefont {G.}~\bibnamefont {Weck}}, \bibinfo {author} {\bibfnamefont {P.}~\bibnamefont {Loubeyre}},\ and\ \bibinfo {author} {\bibfnamefont {M.}~\bibnamefont {Mezouar}},\ }\bibfield  {title} {\bibinfo {title} {Quantitative structure factor and density measurements of high-pressure fluids in diamond anvil cells by x-ray diffraction: Argon and water},\ }\href {https://doi.org/https://doi.org/10.1103/PhysRevB.65.174105} {\bibfield  {journal} {\bibinfo  {journal} {Physical Review B}\ }\textbf {\bibinfo {volume} {65}},\ \bibinfo {pages} {174105} (\bibinfo {year} {2002})}\BibitemShut {NoStop}%
\bibitem [{\citenamefont {Shi}\ \emph {et~al.}(2020)\citenamefont {Shi}, \citenamefont {Alderman}, \citenamefont {Tamalonis}, \citenamefont {Weber}, \citenamefont {You},\ and\ \citenamefont {Benmore}}]{shi}%
  \BibitemOpen
  \bibfield  {author} {\bibinfo {author} {\bibfnamefont {C.}~\bibnamefont {Shi}}, \bibinfo {author} {\bibfnamefont {O.~L.~G.}\ \bibnamefont {Alderman}}, \bibinfo {author} {\bibfnamefont {A.}~\bibnamefont {Tamalonis}}, \bibinfo {author} {\bibfnamefont {R.}~\bibnamefont {Weber}}, \bibinfo {author} {\bibfnamefont {J.}~\bibnamefont {You}},\ and\ \bibinfo {author} {\bibfnamefont {C.~J.}\ \bibnamefont {Benmore}},\ }\bibfield  {title} {\bibinfo {title} {Redox-structure dependence of molten iron oxides},\ }\href {https://doi.org/https://doi.org/10.1038/s43246-020-00080-4} {\bibfield  {journal} {\bibinfo  {journal} {Communications Materials}\ }\textbf {\bibinfo {volume} {1}},\ \bibinfo {pages} {80} (\bibinfo {year} {2020})}\BibitemShut {NoStop}%
\bibitem [{\citenamefont {Kaplow}\ \emph {et~al.}(1965)\citenamefont {Kaplow}, \citenamefont {Strong},\ and\ \citenamefont {Averbach}}]{kaplow}%
  \BibitemOpen
  \bibfield  {author} {\bibinfo {author} {\bibfnamefont {R.}~\bibnamefont {Kaplow}}, \bibinfo {author} {\bibfnamefont {S.~L.}\ \bibnamefont {Strong}},\ and\ \bibinfo {author} {\bibfnamefont {B.~L.}\ \bibnamefont {Averbach}},\ }\bibfield  {title} {\bibinfo {title} {Radial density functions for liquid mercury and lead},\ }\bibfield  {journal} {\bibinfo  {journal} {Physical Review}\ }\textbf {\bibinfo {volume} {138}},\ \href {https://doi.org/10.1103/PhysRev.138.A1336} {10.1103/PhysRev.138.A1336} (\bibinfo {year} {1965})\BibitemShut {NoStop}%
\bibitem [{\citenamefont {Sanloup}\ \emph {et~al.}(2000)\citenamefont {Sanloup}, \citenamefont {Guyot}, \citenamefont {Gillet}, \citenamefont {Fiquet}, \citenamefont {Hemley}, \citenamefont {Mezouar},\ and\ \citenamefont {Martinez}}]{sanloup}%
  \BibitemOpen
  \bibfield  {author} {\bibinfo {author} {\bibfnamefont {C.}~\bibnamefont {Sanloup}}, \bibinfo {author} {\bibfnamefont {F.}~\bibnamefont {Guyot}}, \bibinfo {author} {\bibfnamefont {P.}~\bibnamefont {Gillet}}, \bibinfo {author} {\bibfnamefont {G.}~\bibnamefont {Fiquet}}, \bibinfo {author} {\bibfnamefont {R.~J.}\ \bibnamefont {Hemley}}, \bibinfo {author} {\bibfnamefont {M.}~\bibnamefont {Mezouar}},\ and\ \bibinfo {author} {\bibfnamefont {I.}~\bibnamefont {Martinez}},\ }\bibfield  {title} {\bibinfo {title} {Structural changes in liquid {F}e at high pressures and high temperatures from synchrotron x-ray diffraction},\ }\href {https://doi.org/https://doi.org/10.1209/epl/i2000-00417-3} {\bibfield  {journal} {\bibinfo  {journal} {Europhysics Letters}\ }\textbf {\bibinfo {volume} {52}},\ \bibinfo {pages} {151} (\bibinfo {year} {2000})}\BibitemShut {NoStop}%
\bibitem [{\citenamefont {Gonzalez}\ and\ \citenamefont {Gonzalez}(2023)}]{gonzalez}%
  \BibitemOpen
  \bibfield  {author} {\bibinfo {author} {\bibfnamefont {L.}~\bibnamefont {Gonzalez}}\ and\ \bibinfo {author} {\bibfnamefont {D.}~\bibnamefont {Gonzalez}},\ }\bibfield  {title} {\bibinfo {title} {Structure and dynamics in liquid iron at high pressure and temperature. a first principles study},\ }\href {https://doi.org/10.1029/2022JB025119} {\bibfield  {journal} {\bibinfo  {journal} {Journal of Geophysical Research: Solid {E}arth}\ }\textbf {\bibinfo {volume} {128}},\ \bibinfo {pages} {e2022JB025119} (\bibinfo {year} {2023})}\BibitemShut {NoStop}%
\bibitem [{\citenamefont {Wark}\ \emph {et~al.}(2025)\citenamefont {Wark}, \citenamefont {Peake}, \citenamefont {Stevens}, \citenamefont {Heighway}, \citenamefont {Ping}, \citenamefont {Sterne}, \citenamefont {Albertazzi}, \citenamefont {Ali}, \citenamefont {Antonelli}, \citenamefont {Armstrong}, \citenamefont {Baehtz}, \citenamefont {Ball}, \citenamefont {Banerjee}, \citenamefont {Belonoshko}, \citenamefont {Bolme}, \citenamefont {Bouffetier}, \citenamefont {Briggs}, \citenamefont {Buakor}, \citenamefont {Butcher}, \citenamefont {Cafiso}, \citenamefont {Cerantola}, \citenamefont {Chantel}, \citenamefont {Cicco}, \citenamefont {Coleman}, \citenamefont {Collier}, \citenamefont {Collins}, \citenamefont {Comley}, \citenamefont {Coppari}, \citenamefont {Cowan}, \citenamefont {Cristoforetti}, \citenamefont {Cynn}, \citenamefont {Descamps}, \citenamefont {Dorchies}, \citenamefont {Duff}, \citenamefont {Dwivedi}, \citenamefont {Edwards}, \citenamefont {Eggert}, \citenamefont {Errandonea}, \citenamefont {Fiquet},
  \citenamefont {Galtier}, \citenamefont {Garcia}, \citenamefont {Ginestet}, \citenamefont {Gizzi}, \citenamefont {Gleason}, \citenamefont {Goede}, \citenamefont {Gonzalez}, \citenamefont {Gorman}, \citenamefont {Harmand}, \citenamefont {Hartley}, \citenamefont {Hernandez-Gomez}, \citenamefont {Higginbotham}, \citenamefont {Höppner}, \citenamefont {Humphries}, \citenamefont {Husband}, \citenamefont {Hutchinson}, \citenamefont {Hwang}, \citenamefont {Keen}, \citenamefont {Kim}, \citenamefont {Koester}, \citenamefont {Konopkova}, \citenamefont {Kraus}, \citenamefont {Krygier}, \citenamefont {Labate}, \citenamefont {Lazicki}, \citenamefont {Lee}, \citenamefont {Liermann}, \citenamefont {Mason}, \citenamefont {Masruri}, \citenamefont {Massani}, \citenamefont {McBride}, \citenamefont {McGuire}, \citenamefont {McHardy}, \citenamefont {McGonegle}, \citenamefont {McWilliams}, \citenamefont {Merkel}, \citenamefont {Morard}, \citenamefont {Nagler}, \citenamefont {Nakatsutsumi}, \citenamefont {Nguyen-Cong},
  \citenamefont {Norton}, \citenamefont {Oleynik}, \citenamefont {Otzen}, \citenamefont {Ozaki}, \citenamefont {Pandolfi}, \citenamefont {Pelka}, \citenamefont {Pereira}, \citenamefont {Phillips}, \citenamefont {Prescher}, \citenamefont {Preston}, \citenamefont {Randolph}, \citenamefont {Ranjan}, \citenamefont {Ravasio}, \citenamefont {Redmer}, \citenamefont {Rips}, \citenamefont {Santamaria-Perez}, \citenamefont {Savage}, \citenamefont {Schoelmerich}, \citenamefont {Schwinkendorf}, \citenamefont {Singh}, \citenamefont {Smith}, \citenamefont {Smith}, \citenamefont {Sollier}, \citenamefont {Spear}, \citenamefont {Spindloe}, \citenamefont {Stevenson}, \citenamefont {Strohm}, \citenamefont {Suer}, \citenamefont {Tang}, \citenamefont {Toncian}, \citenamefont {Toncian}, \citenamefont {Tracy}, \citenamefont {Trapananti}, \citenamefont {Tschentscher}, \citenamefont {Tyldesley}, \citenamefont {Vennari}, \citenamefont {Vinci}, \citenamefont {Voge}, \citenamefont {Volz}, \citenamefont {Vorberger}, \citenamefont
  {Willman}, \citenamefont {Wollenweber}, \citenamefont {Zastrau}, \citenamefont {Brambrink}, \citenamefont {Appel},\ and\ \citenamefont {McMahon}}]{wark}%
  \BibitemOpen
  \bibfield  {author} {\bibinfo {author} {\bibfnamefont {J.~S.}\ \bibnamefont {Wark}}, \bibinfo {author} {\bibfnamefont {D.~J.}\ \bibnamefont {Peake}}, \bibinfo {author} {\bibfnamefont {T.}~\bibnamefont {Stevens}}, \bibinfo {author} {\bibfnamefont {P.~G.}\ \bibnamefont {Heighway}}, \bibinfo {author} {\bibfnamefont {Y.}~\bibnamefont {Ping}}, \bibinfo {author} {\bibfnamefont {P.}~\bibnamefont {Sterne}}, \bibinfo {author} {\bibfnamefont {B.}~\bibnamefont {Albertazzi}}, \bibinfo {author} {\bibfnamefont {S.~J.}\ \bibnamefont {Ali}}, \bibinfo {author} {\bibfnamefont {L.}~\bibnamefont {Antonelli}}, \bibinfo {author} {\bibfnamefont {M.~R.}\ \bibnamefont {Armstrong}}, \bibinfo {author} {\bibfnamefont {C.}~\bibnamefont {Baehtz}}, \bibinfo {author} {\bibfnamefont {O.~B.}\ \bibnamefont {Ball}}, \bibinfo {author} {\bibfnamefont {S.}~\bibnamefont {Banerjee}}, \bibinfo {author} {\bibfnamefont {A.~B.}\ \bibnamefont {Belonoshko}}, \bibinfo {author} {\bibfnamefont {C.~A.}\ \bibnamefont {Bolme}}, \bibinfo {author}
  {\bibfnamefont {V.}~\bibnamefont {Bouffetier}}, \bibinfo {author} {\bibfnamefont {R.}~\bibnamefont {Briggs}}, \bibinfo {author} {\bibfnamefont {K.}~\bibnamefont {Buakor}}, \bibinfo {author} {\bibfnamefont {T.}~\bibnamefont {Butcher}}, \bibinfo {author} {\bibfnamefont {S.~D.~D.}\ \bibnamefont {Cafiso}}, \bibinfo {author} {\bibfnamefont {V.}~\bibnamefont {Cerantola}}, \bibinfo {author} {\bibfnamefont {J.}~\bibnamefont {Chantel}}, \bibinfo {author} {\bibfnamefont {A.~D.}\ \bibnamefont {Cicco}}, \bibinfo {author} {\bibfnamefont {A.~L.}\ \bibnamefont {Coleman}}, \bibinfo {author} {\bibfnamefont {J.}~\bibnamefont {Collier}}, \bibinfo {author} {\bibfnamefont {G.}~\bibnamefont {Collins}}, \bibinfo {author} {\bibfnamefont {A.~J.}\ \bibnamefont {Comley}}, \bibinfo {author} {\bibfnamefont {F.}~\bibnamefont {Coppari}}, \bibinfo {author} {\bibfnamefont {T.~E.}\ \bibnamefont {Cowan}}, \bibinfo {author} {\bibfnamefont {G.}~\bibnamefont {Cristoforetti}}, \bibinfo {author} {\bibfnamefont {H.}~\bibnamefont {Cynn}}, \bibinfo
  {author} {\bibfnamefont {A.}~\bibnamefont {Descamps}}, \bibinfo {author} {\bibfnamefont {F.}~\bibnamefont {Dorchies}}, \bibinfo {author} {\bibfnamefont {M.~J.}\ \bibnamefont {Duff}}, \bibinfo {author} {\bibfnamefont {A.}~\bibnamefont {Dwivedi}}, \bibinfo {author} {\bibfnamefont {C.}~\bibnamefont {Edwards}}, \bibinfo {author} {\bibfnamefont {J.~H.}\ \bibnamefont {Eggert}}, \bibinfo {author} {\bibfnamefont {D.}~\bibnamefont {Errandonea}}, \bibinfo {author} {\bibfnamefont {G.}~\bibnamefont {Fiquet}}, \bibinfo {author} {\bibfnamefont {E.}~\bibnamefont {Galtier}}, \bibinfo {author} {\bibfnamefont {A.~L.}\ \bibnamefont {Garcia}}, \bibinfo {author} {\bibfnamefont {H.}~\bibnamefont {Ginestet}}, \bibinfo {author} {\bibfnamefont {L.}~\bibnamefont {Gizzi}}, \bibinfo {author} {\bibfnamefont {A.}~\bibnamefont {Gleason}}, \bibinfo {author} {\bibfnamefont {S.}~\bibnamefont {Goede}}, \bibinfo {author} {\bibfnamefont {J.~M.}\ \bibnamefont {Gonzalez}}, \bibinfo {author} {\bibfnamefont {M.~G.}\ \bibnamefont {Gorman}},
  \bibinfo {author} {\bibfnamefont {M.}~\bibnamefont {Harmand}}, \bibinfo {author} {\bibfnamefont {N.}~\bibnamefont {Hartley}}, \bibinfo {author} {\bibfnamefont {C.}~\bibnamefont {Hernandez-Gomez}}, \bibinfo {author} {\bibfnamefont {A.}~\bibnamefont {Higginbotham}}, \bibinfo {author} {\bibfnamefont {H.}~\bibnamefont {Höppner}}, \bibinfo {author} {\bibfnamefont {O.~S.}\ \bibnamefont {Humphries}}, \bibinfo {author} {\bibfnamefont {R.~J.}\ \bibnamefont {Husband}}, \bibinfo {author} {\bibfnamefont {T.~M.}\ \bibnamefont {Hutchinson}}, \bibinfo {author} {\bibfnamefont {H.}~\bibnamefont {Hwang}}, \bibinfo {author} {\bibfnamefont {D.}~\bibnamefont {Keen}}, \bibinfo {author} {\bibfnamefont {J.}~\bibnamefont {Kim}}, \bibinfo {author} {\bibfnamefont {P.}~\bibnamefont {Koester}}, \bibinfo {author} {\bibfnamefont {Z.}~\bibnamefont {Konopkova}}, \bibinfo {author} {\bibfnamefont {D.}~\bibnamefont {Kraus}}, \bibinfo {author} {\bibfnamefont {A.}~\bibnamefont {Krygier}}, \bibinfo {author} {\bibfnamefont {L.}~\bibnamefont
  {Labate}}, \bibinfo {author} {\bibfnamefont {A.~E.}\ \bibnamefont {Lazicki}}, \bibinfo {author} {\bibfnamefont {Y.}~\bibnamefont {Lee}}, \bibinfo {author} {\bibfnamefont {H.-P.}\ \bibnamefont {Liermann}}, \bibinfo {author} {\bibfnamefont {P.}~\bibnamefont {Mason}}, \bibinfo {author} {\bibfnamefont {M.}~\bibnamefont {Masruri}}, \bibinfo {author} {\bibfnamefont {B.}~\bibnamefont {Massani}}, \bibinfo {author} {\bibfnamefont {E.~E.}\ \bibnamefont {McBride}}, \bibinfo {author} {\bibfnamefont {C.}~\bibnamefont {McGuire}}, \bibinfo {author} {\bibfnamefont {J.~D.}\ \bibnamefont {McHardy}}, \bibinfo {author} {\bibfnamefont {D.}~\bibnamefont {McGonegle}}, \bibinfo {author} {\bibfnamefont {R.~S.}\ \bibnamefont {McWilliams}}, \bibinfo {author} {\bibfnamefont {S.}~\bibnamefont {Merkel}}, \bibinfo {author} {\bibfnamefont {G.}~\bibnamefont {Morard}}, \bibinfo {author} {\bibfnamefont {B.}~\bibnamefont {Nagler}}, \bibinfo {author} {\bibfnamefont {M.}~\bibnamefont {Nakatsutsumi}}, \bibinfo {author} {\bibfnamefont
  {K.}~\bibnamefont {Nguyen-Cong}}, \bibinfo {author} {\bibfnamefont {A.-M.}\ \bibnamefont {Norton}}, \bibinfo {author} {\bibfnamefont {I.~I.}\ \bibnamefont {Oleynik}}, \bibinfo {author} {\bibfnamefont {C.}~\bibnamefont {Otzen}}, \bibinfo {author} {\bibfnamefont {N.}~\bibnamefont {Ozaki}}, \bibinfo {author} {\bibfnamefont {S.}~\bibnamefont {Pandolfi}}, \bibinfo {author} {\bibfnamefont {A.}~\bibnamefont {Pelka}}, \bibinfo {author} {\bibfnamefont {K.~A.}\ \bibnamefont {Pereira}}, \bibinfo {author} {\bibfnamefont {J.~P.}\ \bibnamefont {Phillips}}, \bibinfo {author} {\bibfnamefont {C.}~\bibnamefont {Prescher}}, \bibinfo {author} {\bibfnamefont {T.~R.}\ \bibnamefont {Preston}}, \bibinfo {author} {\bibfnamefont {L.}~\bibnamefont {Randolph}}, \bibinfo {author} {\bibfnamefont {D.}~\bibnamefont {Ranjan}}, \bibinfo {author} {\bibfnamefont {A.}~\bibnamefont {Ravasio}}, \bibinfo {author} {\bibfnamefont {R.}~\bibnamefont {Redmer}}, \bibinfo {author} {\bibfnamefont {J.}~\bibnamefont {Rips}}, \bibinfo {author}
  {\bibfnamefont {D.}~\bibnamefont {Santamaria-Perez}}, \bibinfo {author} {\bibfnamefont {D.~J.}\ \bibnamefont {Savage}}, \bibinfo {author} {\bibfnamefont {M.}~\bibnamefont {Schoelmerich}}, \bibinfo {author} {\bibfnamefont {J.-P.}\ \bibnamefont {Schwinkendorf}}, \bibinfo {author} {\bibfnamefont {S.}~\bibnamefont {Singh}}, \bibinfo {author} {\bibfnamefont {J.}~\bibnamefont {Smith}}, \bibinfo {author} {\bibfnamefont {R.~F.}\ \bibnamefont {Smith}}, \bibinfo {author} {\bibfnamefont {A.}~\bibnamefont {Sollier}}, \bibinfo {author} {\bibfnamefont {J.}~\bibnamefont {Spear}}, \bibinfo {author} {\bibfnamefont {C.}~\bibnamefont {Spindloe}}, \bibinfo {author} {\bibfnamefont {M.}~\bibnamefont {Stevenson}}, \bibinfo {author} {\bibfnamefont {C.}~\bibnamefont {Strohm}}, \bibinfo {author} {\bibfnamefont {T.-A.}\ \bibnamefont {Suer}}, \bibinfo {author} {\bibfnamefont {M.}~\bibnamefont {Tang}}, \bibinfo {author} {\bibfnamefont {M.}~\bibnamefont {Toncian}}, \bibinfo {author} {\bibfnamefont {T.}~\bibnamefont {Toncian}}, \bibinfo
  {author} {\bibfnamefont {S.~J.}\ \bibnamefont {Tracy}}, \bibinfo {author} {\bibfnamefont {A.}~\bibnamefont {Trapananti}}, \bibinfo {author} {\bibfnamefont {T.}~\bibnamefont {Tschentscher}}, \bibinfo {author} {\bibfnamefont {M.}~\bibnamefont {Tyldesley}}, \bibinfo {author} {\bibfnamefont {C.~E.}\ \bibnamefont {Vennari}}, \bibinfo {author} {\bibfnamefont {T.}~\bibnamefont {Vinci}}, \bibinfo {author} {\bibfnamefont {S.~C.}\ \bibnamefont {Voge}}, \bibinfo {author} {\bibfnamefont {T.~J.}\ \bibnamefont {Volz}}, \bibinfo {author} {\bibfnamefont {J.}~\bibnamefont {Vorberger}}, \bibinfo {author} {\bibfnamefont {J.~T.}\ \bibnamefont {Willman}}, \bibinfo {author} {\bibfnamefont {L.}~\bibnamefont {Wollenweber}}, \bibinfo {author} {\bibfnamefont {U.}~\bibnamefont {Zastrau}}, \bibinfo {author} {\bibfnamefont {E.}~\bibnamefont {Brambrink}}, \bibinfo {author} {\bibfnamefont {K.}~\bibnamefont {Appel}},\ and\ \bibinfo {author} {\bibfnamefont {M.~I.}\ \bibnamefont {McMahon}},\ }\bibfield  {title} {\bibinfo {title}
  {{Femtosecond temperature measurements of laser-shocked copper deduced from the intensity of the x-ray thermal diffuse scattering}},\ }\bibfield  {journal} {\bibinfo  {journal} {Journal of Applied Physics}\ }\textbf {\bibinfo {volume} {137}},\ \href {https://doi.org/10.1063/5.0256844} {10.1063/5.0256844} (\bibinfo {year} {2025})\BibitemShut {NoStop}%
\bibitem [{\citenamefont {Leydier}(2010)}]{leydier2010}%
  \BibitemOpen
  \bibfield  {author} {\bibinfo {author} {\bibfnamefont {M.}~\bibnamefont {Leydier}},\ }\bibfield  {title} {\bibinfo {title} {Méthodes complémentaires pour l’étude de verres et liquides fondus sur grands instruments : structure et dynamique.},\ }\href {https://doi.org/https://hal.science/tel-00623032/} {\bibfield  {journal} {\bibinfo  {journal} {Thesis}\ } (\bibinfo {year} {2010})}\BibitemShut {NoStop}%
\bibitem [{\citenamefont {Harmand}\ \emph {et~al.}(2015)\citenamefont {Harmand}, \citenamefont {Ravasio}, \citenamefont {Mazevet}, \citenamefont {Bouchet}, \citenamefont {Denoeud}, \citenamefont {Dorchies}, \citenamefont {Feng}, \citenamefont {Fourment}, \citenamefont {Galtier}, \citenamefont {Gaudin}, \citenamefont {Guyot}, \citenamefont {Kodama}, \citenamefont {Koenig}, \citenamefont {Lee}, \citenamefont {Miyanishi}, \citenamefont {Morard}, \citenamefont {Musella}, \citenamefont {Nagler}, \citenamefont {Nakatsutsumi}, \citenamefont {Ozaki}, \citenamefont {Recoules}, \citenamefont {Toleikis}, \citenamefont {Vinci}, \citenamefont {Zastrau}, \citenamefont {Zhu},\ and\ \citenamefont {Benuzzi-Mounaix}}]{harmand}%
  \BibitemOpen
  \bibfield  {author} {\bibinfo {author} {\bibfnamefont {M.}~\bibnamefont {Harmand}}, \bibinfo {author} {\bibfnamefont {A.}~\bibnamefont {Ravasio}}, \bibinfo {author} {\bibfnamefont {S.}~\bibnamefont {Mazevet}}, \bibinfo {author} {\bibfnamefont {J.}~\bibnamefont {Bouchet}}, \bibinfo {author} {\bibfnamefont {A.}~\bibnamefont {Denoeud}}, \bibinfo {author} {\bibfnamefont {F.}~\bibnamefont {Dorchies}}, \bibinfo {author} {\bibfnamefont {Y.}~\bibnamefont {Feng}}, \bibinfo {author} {\bibfnamefont {C.}~\bibnamefont {Fourment}}, \bibinfo {author} {\bibfnamefont {E.}~\bibnamefont {Galtier}}, \bibinfo {author} {\bibfnamefont {J.}~\bibnamefont {Gaudin}}, \bibinfo {author} {\bibfnamefont {F.}~\bibnamefont {Guyot}}, \bibinfo {author} {\bibfnamefont {R.}~\bibnamefont {Kodama}}, \bibinfo {author} {\bibfnamefont {M.}~\bibnamefont {Koenig}}, \bibinfo {author} {\bibfnamefont {H.~J.}\ \bibnamefont {Lee}}, \bibinfo {author} {\bibfnamefont {K.}~\bibnamefont {Miyanishi}}, \bibinfo {author} {\bibfnamefont {G.}~\bibnamefont
  {Morard}}, \bibinfo {author} {\bibfnamefont {R.}~\bibnamefont {Musella}}, \bibinfo {author} {\bibfnamefont {B.}~\bibnamefont {Nagler}}, \bibinfo {author} {\bibfnamefont {M.}~\bibnamefont {Nakatsutsumi}}, \bibinfo {author} {\bibfnamefont {N.}~\bibnamefont {Ozaki}}, \bibinfo {author} {\bibfnamefont {V.}~\bibnamefont {Recoules}}, \bibinfo {author} {\bibfnamefont {S.}~\bibnamefont {Toleikis}}, \bibinfo {author} {\bibfnamefont {T.}~\bibnamefont {Vinci}}, \bibinfo {author} {\bibfnamefont {U.}~\bibnamefont {Zastrau}}, \bibinfo {author} {\bibfnamefont {D.}~\bibnamefont {Zhu}},\ and\ \bibinfo {author} {\bibfnamefont {A.}~\bibnamefont {Benuzzi-Mounaix}},\ }\bibfield  {title} {\bibinfo {title} {X-ray absorption spectroscopy of iron at multimegabar pressures in laser shock experiments},\ }\href {https://doi.org/10.1103/PhysRevB.92.024108} {\bibfield  {journal} {\bibinfo  {journal} {Physycal Review B}\ }\textbf {\bibinfo {volume} {92}},\ \bibinfo {pages} {024108} (\bibinfo {year} {2015})}\BibitemShut {NoStop}%
\bibitem [{\citenamefont {Brown}\ \emph {et~al.}(2000)\citenamefont {Brown}, \citenamefont {Fritz},\ and\ \citenamefont {Hixson}}]{brown}%
  \BibitemOpen
  \bibfield  {author} {\bibinfo {author} {\bibfnamefont {J.~M.}\ \bibnamefont {Brown}}, \bibinfo {author} {\bibfnamefont {J.~N.}\ \bibnamefont {Fritz}},\ and\ \bibinfo {author} {\bibfnamefont {R.~S.}\ \bibnamefont {Hixson}},\ }\bibfield  {title} {\bibinfo {title} {Hugoniot data for iron},\ }\href {https://doi.org/110.1063/1.1319320} {\bibfield  {journal} {\bibinfo  {journal} {J. Appl. Phys.}\ }\textbf {\bibinfo {volume} {88}},\ \bibinfo {pages} {5496–5498} (\bibinfo {year} {2000})}\BibitemShut {NoStop}%
\bibitem [{\citenamefont {Benedict}\ \emph {et~al.}(2022)\citenamefont {Benedict}, \citenamefont {Kraus}, \citenamefont {Hamel},\ and\ \citenamefont {Belof}}]{benedict}%
  \BibitemOpen
  \bibfield  {author} {\bibinfo {author} {\bibfnamefont {L.~X.}\ \bibnamefont {Benedict}}, \bibinfo {author} {\bibfnamefont {R.~G.}\ \bibnamefont {Kraus}}, \bibinfo {author} {\bibfnamefont {S.}~\bibnamefont {Hamel}},\ and\ \bibinfo {author} {\bibfnamefont {J.~L.}\ \bibnamefont {Belof}},\ }\bibfield  {title} {\bibinfo {title} {A limited-ranged two-phase iron equation of state model},\ }\bibfield  {journal} {\bibinfo  {journal} {LLNL Technical Report}\ }\href {https://doi.org/10.2172/1843135} {10.2172/1843135} (\bibinfo {year} {2022})\BibitemShut {NoStop}%
\bibitem [{\citenamefont {Young}\ \emph {et~al.}(2021)\citenamefont {Young}, \citenamefont {Fan}, \citenamefont {Zhao}, \citenamefont {Chen}, \citenamefont {Liu},\ and\ \citenamefont {Huang}}]{young}%
  \BibitemOpen
  \bibfield  {author} {\bibinfo {author} {\bibfnamefont {G.}~\bibnamefont {Young}}, \bibinfo {author} {\bibfnamefont {L.}~\bibnamefont {Fan}}, \bibinfo {author} {\bibfnamefont {B.}~\bibnamefont {Zhao}}, \bibinfo {author} {\bibfnamefont {X.}~\bibnamefont {Chen}}, \bibinfo {author} {\bibfnamefont {X.}~\bibnamefont {Liu}},\ and\ \bibinfo {author} {\bibfnamefont {H.}~\bibnamefont {Huang}},\ }\bibfield  {title} {\bibinfo {title} {Equation of state for {F}e-9.0 wt\% {O} up to 246 {G}{P}a: Implications for oxygen in the {E}arth's outer core},\ }\href {https://doi.org/10.1029/2020JB021056} {\bibfield  {journal} {\bibinfo  {journal} {Journal of Geophysical Research: Solid {E}arth}\ }\textbf {\bibinfo {volume} {126}},\ \bibinfo {pages} {e2020JB021056} (\bibinfo {year} {2021})}\BibitemShut {NoStop}%
\bibitem [{\citenamefont {Amouretti}\ \emph {et~al.}(2024)\citenamefont {Amouretti}, \citenamefont {Harmand}, \citenamefont {Albertazzi}, \citenamefont {Boury}, \citenamefont {Benuzzi-Mounaix}, \citenamefont {Chin}, \citenamefont {Guyot}, \citenamefont {Koenig}, \citenamefont {Vinci},\ and\ \citenamefont {Fiquet}}]{LULI}%
  \BibitemOpen
  \bibfield  {author} {\bibinfo {author} {\bibfnamefont {A.}~\bibnamefont {Amouretti}}, \bibinfo {author} {\bibfnamefont {M.}~\bibnamefont {Harmand}}, \bibinfo {author} {\bibfnamefont {B.}~\bibnamefont {Albertazzi}}, \bibinfo {author} {\bibfnamefont {A.}~\bibnamefont {Boury}}, \bibinfo {author} {\bibfnamefont {A.}~\bibnamefont {Benuzzi-Mounaix}}, \bibinfo {author} {\bibfnamefont {D.~A.}\ \bibnamefont {Chin}}, \bibinfo {author} {\bibfnamefont {F.}~\bibnamefont {Guyot}}, \bibinfo {author} {\bibfnamefont {M.}~\bibnamefont {Koenig}}, \bibinfo {author} {\bibfnamefont {T.}~\bibnamefont {Vinci}},\ and\ \bibinfo {author} {\bibfnamefont {G.}~\bibnamefont {Fiquet}},\ }\bibfield  {title} {\bibinfo {title} {Laser-driven shock compression and equation of state of \ch{Fe2O3} up to 700 {G}{P}a},\ }\href {https://doi.org/10.1103/PhysRevB.110.144110} {\bibfield  {journal} {\bibinfo  {journal} {Phys. Rev. B}\ }\textbf {\bibinfo {volume} {110}},\ \bibinfo {pages} {144110} (\bibinfo {year} {2024})}\BibitemShut {NoStop}%
\bibitem [{\citenamefont {Liu}\ \emph {et~al.}(2023)\citenamefont {Liu}, \citenamefont {Sun}, \citenamefont {Lv}, \citenamefont {Zhang}, \citenamefont {Fu}, \citenamefont {Prakapenka}, \citenamefont {Wang}, \citenamefont {Ho}, \citenamefont {Lin}, ,\ and\ \citenamefont {Wentzcovitch}}]{liu}%
  \BibitemOpen
  \bibfield  {author} {\bibinfo {author} {\bibfnamefont {J.}~\bibnamefont {Liu}}, \bibinfo {author} {\bibfnamefont {Y.}~\bibnamefont {Sun}}, \bibinfo {author} {\bibfnamefont {C.}~\bibnamefont {Lv}}, \bibinfo {author} {\bibfnamefont {F.}~\bibnamefont {Zhang}}, \bibinfo {author} {\bibfnamefont {S.}~\bibnamefont {Fu}}, \bibinfo {author} {\bibfnamefont {V.~B.}\ \bibnamefont {Prakapenka}}, \bibinfo {author} {\bibfnamefont {C.}~\bibnamefont {Wang}}, \bibinfo {author} {\bibfnamefont {K.}~\bibnamefont {Ho}}, \bibinfo {author} {\bibfnamefont {J.}~\bibnamefont {Lin}}, ,\ and\ \bibinfo {author} {\bibfnamefont {R.~M.}\ \bibnamefont {Wentzcovitch}},\ }\bibfield  {title} {\bibinfo {title} {Iron-rich {F}e–o compounds at {E}arth’s core pressures},\ }\bibfield  {journal} {\bibinfo  {journal} {The Innovation}\ }\textbf {\bibinfo {volume} {4}},\ \href {https://doi.org/10.1016/j.xinn.2022.100354} {10.1016/j.xinn.2022.100354} (\bibinfo {year} {2023})\BibitemShut {NoStop}%
\bibitem [{\citenamefont {Gautam}\ and\ \citenamefont {Carter}(2018)}]{gautam2018}%
  \BibitemOpen
  \bibfield  {author} {\bibinfo {author} {\bibfnamefont {G.~S.}\ \bibnamefont {Gautam}}\ and\ \bibinfo {author} {\bibfnamefont {E.~A.}\ \bibnamefont {Carter}},\ }\bibfield  {title} {\bibinfo {title} {Evaluating transition metal oxides within {D}{F}{T}-{S}{C}{A}{N} and {S}{C}{A}{N}+{U} frameworks for solar thermochemical applications},\ }\href {https://doi.org/10.1103/PhysRevMaterials.2.095401} {\bibfield  {journal} {\bibinfo  {journal} {Physical Review Materials}\ }\textbf {\bibinfo {volume} {2}},\ \bibinfo {pages} {095401} (\bibinfo {year} {2018})}\BibitemShut {NoStop}%
\bibitem [{\citenamefont {Umemoto}\ and\ \citenamefont {Wentzcovitch}(2011)}]{umemoto}%
  \BibitemOpen
  \bibfield  {author} {\bibinfo {author} {\bibfnamefont {K.}~\bibnamefont {Umemoto}}\ and\ \bibinfo {author} {\bibfnamefont {R.~M.}\ \bibnamefont {Wentzcovitch}},\ }\bibfield  {title} {\bibinfo {title} {Two-stage dissociation in \ch{MgSiO3} post-perovskite},\ }\href {https://doi.org/10.1016/j.epsl.2011.09.032} {\bibfield  {journal} {\bibinfo  {journal} {{E}arth and Planetary Science Letters}\ }\textbf {\bibinfo {volume} {311}},\ \bibinfo {pages} {225} (\bibinfo {year} {2011})}\BibitemShut {NoStop}%
\bibitem [{\citenamefont {Frost}\ \emph {et~al.}(2004)\citenamefont {Frost}, \citenamefont {Liebske}, \citenamefont {Langenhorst}, \citenamefont {McCammon}, \citenamefont {Trønnes},\ and\ \citenamefont {Rubie}}]{frostnature}%
  \BibitemOpen
  \bibfield  {author} {\bibinfo {author} {\bibfnamefont {D.~J.}\ \bibnamefont {Frost}}, \bibinfo {author} {\bibfnamefont {C.}~\bibnamefont {Liebske}}, \bibinfo {author} {\bibfnamefont {F.}~\bibnamefont {Langenhorst}}, \bibinfo {author} {\bibfnamefont {C.~A.}\ \bibnamefont {McCammon}}, \bibinfo {author} {\bibfnamefont {R.~G.}\ \bibnamefont {Trønnes}},\ and\ \bibinfo {author} {\bibfnamefont {D.~C.}\ \bibnamefont {Rubie}},\ }\bibfield  {title} {\bibinfo {title} {Experimental evidence for the existence of iron-rich metal in the {E}arth's lower mantle},\ }\href {https://doi.org/https://doi.org/10.1038/nature02413} {\bibfield  {journal} {\bibinfo  {journal} {Nature}\ }\textbf {\bibinfo {volume} {428}},\ \bibinfo {pages} {409–412} (\bibinfo {year} {2004})}\BibitemShut {NoStop}%
\bibitem [{\citenamefont {Xu}\ \emph {et~al.}(2015)\citenamefont {Xu}, \citenamefont {Shim},\ and\ \citenamefont {Morgan}}]{xu2015}%
  \BibitemOpen
  \bibfield  {author} {\bibinfo {author} {\bibfnamefont {S.}~\bibnamefont {Xu}}, \bibinfo {author} {\bibfnamefont {S.-H.}\ \bibnamefont {Shim}},\ and\ \bibinfo {author} {\bibfnamefont {D.}~\bibnamefont {Morgan}},\ }\bibfield  {title} {\bibinfo {title} {Origin of {F}e$^{3+} $ in {F}e-containing, {A}l-free mantle silicate perovskite},\ }\href {https://doi.org/https://doi.org/10.1016/j.epsl.2014.11.006} {\bibfield  {journal} {\bibinfo  {journal} {{E}arth and Planetary Science Letters}\ }\textbf {\bibinfo {volume} {409}},\ \bibinfo {pages} {319} (\bibinfo {year} {2015})}\BibitemShut {NoStop}%
\bibitem [{\citenamefont {Wade}\ and\ \citenamefont {Wood}(2005)}]{wade2005}%
  \BibitemOpen
  \bibfield  {author} {\bibinfo {author} {\bibfnamefont {J.}~\bibnamefont {Wade}}\ and\ \bibinfo {author} {\bibfnamefont {B.}~\bibnamefont {Wood}},\ }\bibfield  {title} {\bibinfo {title} {Core formation and the oxidation state of the {E}arth},\ }\href {https://doi.org/10.1016/j.epsl.2005.05.017} {\bibfield  {journal} {\bibinfo  {journal} {{E}arth and Planetary Science Letters}\ }\textbf {\bibinfo {volume} {236}},\ \bibinfo {pages} {78} (\bibinfo {year} {2005})}\BibitemShut {NoStop}%
\bibitem [{\citenamefont {Brazhkin}\ \emph {et~al.}(2010)\citenamefont {Brazhkin}, \citenamefont {Farnan}, \citenamefont {ichi Funakoshi}, \citenamefont {Kanzaki}, \citenamefont {Katayama}, \citenamefont {Lyapin},\ and\ \citenamefont {Saito}}]{brazhkin}%
  \BibitemOpen
  \bibfield  {author} {\bibinfo {author} {\bibfnamefont {V.}~\bibnamefont {Brazhkin}}, \bibinfo {author} {\bibfnamefont {I.}~\bibnamefont {Farnan}}, \bibinfo {author} {\bibfnamefont {K.}~\bibnamefont {ichi Funakoshi}}, \bibinfo {author} {\bibfnamefont {M.}~\bibnamefont {Kanzaki}}, \bibinfo {author} {\bibfnamefont {Y.}~\bibnamefont {Katayama}}, \bibinfo {author} {\bibfnamefont {A.~G.}\ \bibnamefont {Lyapin}},\ and\ \bibinfo {author} {\bibfnamefont {H.}~\bibnamefont {Saito}},\ }\bibfield  {title} {\bibinfo {title} {Structural transformations and anomalous viscosity in the \ch{B2O3} melt under high pressure},\ }\href {https://doi.org/10.1103/PhysRevX.9.041018} {\bibfield  {journal} {\bibinfo  {journal} {Physical Review Letters}\ }\textbf {\bibinfo {volume} {105}},\ \bibinfo {pages} {115701} (\bibinfo {year} {2010})}\BibitemShut {NoStop}%
\bibitem [{\citenamefont {Sapnik}\ \emph {et~al.}(2025)\citenamefont {Sapnik}, \citenamefont {Chater}, \citenamefont {Keeble}, \citenamefont {Evans}, \citenamefont {Bertolotti}, \citenamefont {Guagliardi}, \citenamefont {Støckler}, \citenamefont {Harbourne}, \citenamefont {Borup}, \citenamefont {Silberg}, \citenamefont {Descamps}, \citenamefont {Prescher}, \citenamefont {Klee}, \citenamefont {Phelipeau}, \citenamefont {Ullah}, \citenamefont {Medina}, \citenamefont {Bird}, \citenamefont {Kaznelson}, \citenamefont {Lynn}, \citenamefont {Goodwin}, \citenamefont {Iversen}, \citenamefont {Crepisson}, \citenamefont {Bozin}, \citenamefont {Jensen}, \citenamefont {McBride}, \citenamefont {Neder}, \citenamefont {Robinson}, \citenamefont {Wark}, \citenamefont {Andrzejewski}, \citenamefont {Boesenberg}, \citenamefont {Brambrink}, \citenamefont {Camarda}, \citenamefont {Cerantola}, \citenamefont {Goede}, \citenamefont {Höppner}, \citenamefont {Humphries}, \citenamefont {Konopkova}, \citenamefont {Kujala}, \citenamefont
  {Michelat}, \citenamefont {Nakatsutsumi}, \citenamefont {Preston}, \citenamefont {Randolph}, \citenamefont {Schmidt}, \citenamefont {Strohm}, \citenamefont {Tang}, \citenamefont {Zastrau}, \citenamefont {Appel},\ and\ \citenamefont {Keen}}]{sapnik}%
  \BibitemOpen
  \bibfield  {author} {\bibinfo {author} {\bibfnamefont {A.~F.}\ \bibnamefont {Sapnik}}, \bibinfo {author} {\bibfnamefont {P.~A.}\ \bibnamefont {Chater}}, \bibinfo {author} {\bibfnamefont {D.~S.}\ \bibnamefont {Keeble}}, \bibinfo {author} {\bibfnamefont {J.~S.~O.}\ \bibnamefont {Evans}}, \bibinfo {author} {\bibfnamefont {F.}~\bibnamefont {Bertolotti}}, \bibinfo {author} {\bibfnamefont {A.}~\bibnamefont {Guagliardi}}, \bibinfo {author} {\bibfnamefont {L.~J.}\ \bibnamefont {Støckler}}, \bibinfo {author} {\bibfnamefont {E.~A.}\ \bibnamefont {Harbourne}}, \bibinfo {author} {\bibfnamefont {A.~B.}\ \bibnamefont {Borup}}, \bibinfo {author} {\bibfnamefont {R.~S.}\ \bibnamefont {Silberg}}, \bibinfo {author} {\bibfnamefont {A.}~\bibnamefont {Descamps}}, \bibinfo {author} {\bibfnamefont {C.}~\bibnamefont {Prescher}}, \bibinfo {author} {\bibfnamefont {B.~D.}\ \bibnamefont {Klee}}, \bibinfo {author} {\bibfnamefont {A.}~\bibnamefont {Phelipeau}}, \bibinfo {author} {\bibfnamefont {I.}~\bibnamefont {Ullah}}, \bibinfo
  {author} {\bibfnamefont {K.~G.}\ \bibnamefont {Medina}}, \bibinfo {author} {\bibfnamefont {T.~A.}\ \bibnamefont {Bird}}, \bibinfo {author} {\bibfnamefont {V.}~\bibnamefont {Kaznelson}}, \bibinfo {author} {\bibfnamefont {W.}~\bibnamefont {Lynn}}, \bibinfo {author} {\bibfnamefont {A.~L.}\ \bibnamefont {Goodwin}}, \bibinfo {author} {\bibfnamefont {B.~B.}\ \bibnamefont {Iversen}}, \bibinfo {author} {\bibfnamefont {C.}~\bibnamefont {Crepisson}}, \bibinfo {author} {\bibfnamefont {E.~S.}\ \bibnamefont {Bozin}}, \bibinfo {author} {\bibfnamefont {K.~M.}\ \bibnamefont {Jensen}}, \bibinfo {author} {\bibfnamefont {E.~E.}\ \bibnamefont {McBride}}, \bibinfo {author} {\bibfnamefont {R.~B.}\ \bibnamefont {Neder}}, \bibinfo {author} {\bibfnamefont {I.}~\bibnamefont {Robinson}}, \bibinfo {author} {\bibfnamefont {J.}~\bibnamefont {Wark}}, \bibinfo {author} {\bibfnamefont {M.}~\bibnamefont {Andrzejewski}}, \bibinfo {author} {\bibfnamefont {U.}~\bibnamefont {Boesenberg}}, \bibinfo {author} {\bibfnamefont {E.}~\bibnamefont
  {Brambrink}}, \bibinfo {author} {\bibfnamefont {C.}~\bibnamefont {Camarda}}, \bibinfo {author} {\bibfnamefont {V.}~\bibnamefont {Cerantola}}, \bibinfo {author} {\bibfnamefont {S.}~\bibnamefont {Goede}}, \bibinfo {author} {\bibfnamefont {H.}~\bibnamefont {Höppner}}, \bibinfo {author} {\bibfnamefont {O.~S.}\ \bibnamefont {Humphries}}, \bibinfo {author} {\bibfnamefont {Z.}~\bibnamefont {Konopkova}}, \bibinfo {author} {\bibfnamefont {N.}~\bibnamefont {Kujala}}, \bibinfo {author} {\bibfnamefont {T.}~\bibnamefont {Michelat}}, \bibinfo {author} {\bibfnamefont {M.}~\bibnamefont {Nakatsutsumi}}, \bibinfo {author} {\bibfnamefont {T.~R.}\ \bibnamefont {Preston}}, \bibinfo {author} {\bibfnamefont {L.}~\bibnamefont {Randolph}}, \bibinfo {author} {\bibfnamefont {A.}~\bibnamefont {Schmidt}}, \bibinfo {author} {\bibfnamefont {C.}~\bibnamefont {Strohm}}, \bibinfo {author} {\bibfnamefont {M.}~\bibnamefont {Tang}}, \bibinfo {author} {\bibfnamefont {U.}~\bibnamefont {Zastrau}}, \bibinfo {author} {\bibfnamefont
  {K.}~\bibnamefont {Appel}},\ and\ \bibinfo {author} {\bibfnamefont {D.}~\bibnamefont {Keen}},\ }\bibfield  {title} {\bibinfo {title} {High-quality ultra-fast total scattering and pair distribution function data using an x-ray free electron laser},\ }\href {https://doi.org/10.1107/S205225252500538X} {\bibfield  {journal} {\bibinfo  {journal} {IUCrJ}\ }\textbf {\bibinfo {volume} {12}},\ \bibinfo {pages} {531} (\bibinfo {year} {2025})}\BibitemShut {NoStop}%
\bibitem [{\citenamefont {Mason}\ \emph {et~al.}(2018)\citenamefont {Mason}, \citenamefont {Banerjee}, \citenamefont {Smith}, \citenamefont {Butcher}, \citenamefont {Phillips}, \citenamefont {Höppner}, \citenamefont {Möller}, \citenamefont {Ertel}, \citenamefont {Vido}, \citenamefont {Hollingham}, \citenamefont {Norton}, \citenamefont {Tomlinson}, \citenamefont {Zata}, \citenamefont {Merchan}, \citenamefont {Hooker}, \citenamefont {Tyldesley}, \citenamefont {Toncian}, \citenamefont {Hernandez-Gomez}, \citenamefont {Edwards},\ and\ \citenamefont {Collier}}]{mason}%
  \BibitemOpen
  \bibfield  {author} {\bibinfo {author} {\bibfnamefont {P.}~\bibnamefont {Mason}}, \bibinfo {author} {\bibfnamefont {S.}~\bibnamefont {Banerjee}}, \bibinfo {author} {\bibfnamefont {J.}~\bibnamefont {Smith}}, \bibinfo {author} {\bibfnamefont {T.}~\bibnamefont {Butcher}}, \bibinfo {author} {\bibfnamefont {J.}~\bibnamefont {Phillips}}, \bibinfo {author} {\bibfnamefont {H.}~\bibnamefont {Höppner}}, \bibinfo {author} {\bibfnamefont {D.}~\bibnamefont {Möller}}, \bibinfo {author} {\bibfnamefont {K.}~\bibnamefont {Ertel}}, \bibinfo {author} {\bibfnamefont {M.~D.}\ \bibnamefont {Vido}}, \bibinfo {author} {\bibfnamefont {I.}~\bibnamefont {Hollingham}}, \bibinfo {author} {\bibfnamefont {A.}~\bibnamefont {Norton}}, \bibinfo {author} {\bibfnamefont {S.}~\bibnamefont {Tomlinson}}, \bibinfo {author} {\bibfnamefont {T.}~\bibnamefont {Zata}}, \bibinfo {author} {\bibfnamefont {J.~S.}\ \bibnamefont {Merchan}}, \bibinfo {author} {\bibfnamefont {C.}~\bibnamefont {Hooker}}, \bibinfo {author} {\bibfnamefont {M.}~\bibnamefont
  {Tyldesley}}, \bibinfo {author} {\bibfnamefont {T.}~\bibnamefont {Toncian}}, \bibinfo {author} {\bibfnamefont {C.}~\bibnamefont {Hernandez-Gomez}}, \bibinfo {author} {\bibfnamefont {C.}~\bibnamefont {Edwards}},\ and\ \bibinfo {author} {\bibfnamefont {J.}~\bibnamefont {Collier}},\ }\bibfield  {title} {\bibinfo {title} {{Development of a 100 J, 10 Hz laser for compression experiments at the High Energy Density instrument at the European {X}{F}{E}{L}}},\ }\bibfield  {journal} {\bibinfo  {journal} {High Power Laser Science and Engineering}\ }\textbf {\bibinfo {volume} {6}},\ \href {https://doi.org/10.1017/hpl.2018.56} {10.1017/hpl.2018.56} (\bibinfo {year} {2018})\BibitemShut {NoStop}%
\bibitem [{\citenamefont {Gorman}\ \emph {et~al.}(2024)\citenamefont {Gorman}, \citenamefont {McGonegle}, \citenamefont {Smith}, \citenamefont {Singh}, \citenamefont {Jenkins}, \citenamefont {McWilliams}, \citenamefont {Albertazzi}, \citenamefont {Ali}, \citenamefont {Antonelli}, \citenamefont {Armstrong}, \citenamefont {Baehtz}, \citenamefont {Ball}, \citenamefont {Banerjee}, \citenamefont {Belonoshko}, \citenamefont {Benuzzi-Mounaix}, \citenamefont {Bolme}, \citenamefont {Bouffetier}, \citenamefont {Briggs}, \citenamefont {Buakor}, \citenamefont {Butcher}, \citenamefont {Cafiso}, \citenamefont {Cerantola}, \citenamefont {Chantel}, \citenamefont {Cicco}, \citenamefont {Clarke}, \citenamefont {Coleman}, \citenamefont {Collier}, \citenamefont {Collins}, \citenamefont {Comley}, \citenamefont {Coppari}, \citenamefont {Cowan}, \citenamefont {Cristoforetti}, \citenamefont {Cynn}, \citenamefont {Descamps}, \citenamefont {Dorchies}, \citenamefont {Duff}, \citenamefont {Dwivedi}, \citenamefont {Edwards},
  \citenamefont {Eggert}, \citenamefont {Errandonea}, \citenamefont {Fiquet}, \citenamefont {Galtier}, \citenamefont {Garcia}, \citenamefont {Ginestet}, \citenamefont {Gizzi}, \citenamefont {Gleason}, \citenamefont {Goede}, \citenamefont {Gonzalez}, \citenamefont {Harmand}, \citenamefont {Hartley}, \citenamefont {Heighway}, \citenamefont {Hernandez-Gomez}, \citenamefont {Higginbotham}, \citenamefont {Höppner}, \citenamefont {Husband}, \citenamefont {Hutchinson}, \citenamefont {Hwang}, \citenamefont {Lazicki}, \citenamefont {Keen}, \citenamefont {Kim}, \citenamefont {Koester}, \citenamefont {Konopkova}, \citenamefont {Kraus}, \citenamefont {Krygier}, \citenamefont {Labate}, \citenamefont {Lee}, \citenamefont {Liermann}, \citenamefont {Mason}, \citenamefont {Masruri}, \citenamefont {Massani}, \citenamefont {McBride}, \citenamefont {McGuire}, \citenamefont {McHardy}, \citenamefont {Merkel}, \citenamefont {Morard}, \citenamefont {Nagler}, \citenamefont {Nakatsutsumi}, \citenamefont {Nguyen-Cong}, \citenamefont
  {Norton}, \citenamefont {Oleynik}, \citenamefont {Otzen}, \citenamefont {Ozaki}, \citenamefont {Pandolfi}, \citenamefont {Peake}, \citenamefont {Pelka}, \citenamefont {Pereira}, \citenamefont {Phillips}, \citenamefont {Prescher}, \citenamefont {Preston}, \citenamefont {Randolph}, \citenamefont {Ranjan}, \citenamefont {Ravasio}, \citenamefont {Redmer}, \citenamefont {Rips}, \citenamefont {Santamaria-Perez}, \citenamefont {Savage}, \citenamefont {Schoelmerich}, \citenamefont {Schwinkendorf}, \citenamefont {Smith}, \citenamefont {Sollier}, \citenamefont {Spear}, \citenamefont {Spindloe}, \citenamefont {Stevenson}, \citenamefont {Strohm}, \citenamefont {Suer}, \citenamefont {Tang}, \citenamefont {Toncian}, \citenamefont {Toncian}, \citenamefont {Tracy}, \citenamefont {Trapananti}, \citenamefont {Tschentscher}, \citenamefont {Tyldesley}, \citenamefont {Vennari}, \citenamefont {Vinci}, \citenamefont {Vogel}, \citenamefont {Volz}, \citenamefont {Vorberger}, \citenamefont {Walsh}, \citenamefont {Wark},
  \citenamefont {Willman}, \citenamefont {Wollenweber}, \citenamefont {Zastrau}, \citenamefont {Brambrink}, \citenamefont {Appel},\ and\ \citenamefont {McMahon}}]{gormanSn}%
  \BibitemOpen
  \bibfield  {author} {\bibinfo {author} {\bibfnamefont {M.~G.}\ \bibnamefont {Gorman}}, \bibinfo {author} {\bibfnamefont {D.}~\bibnamefont {McGonegle}}, \bibinfo {author} {\bibfnamefont {R.~F.}\ \bibnamefont {Smith}}, \bibinfo {author} {\bibfnamefont {S.}~\bibnamefont {Singh}}, \bibinfo {author} {\bibfnamefont {T.}~\bibnamefont {Jenkins}}, \bibinfo {author} {\bibfnamefont {R.~S.}\ \bibnamefont {McWilliams}}, \bibinfo {author} {\bibfnamefont {B.}~\bibnamefont {Albertazzi}}, \bibinfo {author} {\bibfnamefont {S.~J.}\ \bibnamefont {Ali}}, \bibinfo {author} {\bibfnamefont {L.}~\bibnamefont {Antonelli}}, \bibinfo {author} {\bibfnamefont {M.~R.}\ \bibnamefont {Armstrong}}, \bibinfo {author} {\bibfnamefont {C.}~\bibnamefont {Baehtz}}, \bibinfo {author} {\bibfnamefont {O.~B.}\ \bibnamefont {Ball}}, \bibinfo {author} {\bibfnamefont {S.}~\bibnamefont {Banerjee}}, \bibinfo {author} {\bibfnamefont {A.~B.}\ \bibnamefont {Belonoshko}}, \bibinfo {author} {\bibfnamefont {A.}~\bibnamefont {Benuzzi-Mounaix}}, \bibinfo {author}
  {\bibfnamefont {C.~A.}\ \bibnamefont {Bolme}}, \bibinfo {author} {\bibfnamefont {V.}~\bibnamefont {Bouffetier}}, \bibinfo {author} {\bibfnamefont {R.}~\bibnamefont {Briggs}}, \bibinfo {author} {\bibfnamefont {K.}~\bibnamefont {Buakor}}, \bibinfo {author} {\bibfnamefont {T.}~\bibnamefont {Butcher}}, \bibinfo {author} {\bibfnamefont {S.~D.~D.}\ \bibnamefont {Cafiso}}, \bibinfo {author} {\bibfnamefont {V.}~\bibnamefont {Cerantola}}, \bibinfo {author} {\bibfnamefont {J.}~\bibnamefont {Chantel}}, \bibinfo {author} {\bibfnamefont {A.~D.}\ \bibnamefont {Cicco}}, \bibinfo {author} {\bibfnamefont {S.}~\bibnamefont {Clarke}}, \bibinfo {author} {\bibfnamefont {A.~L.}\ \bibnamefont {Coleman}}, \bibinfo {author} {\bibfnamefont {J.}~\bibnamefont {Collier}}, \bibinfo {author} {\bibfnamefont {G.~W.}\ \bibnamefont {Collins}}, \bibinfo {author} {\bibfnamefont {A.~J.}\ \bibnamefont {Comley}}, \bibinfo {author} {\bibfnamefont {F.}~\bibnamefont {Coppari}}, \bibinfo {author} {\bibfnamefont {T.~E.}\ \bibnamefont {Cowan}},
  \bibinfo {author} {\bibfnamefont {G.}~\bibnamefont {Cristoforetti}}, \bibinfo {author} {\bibfnamefont {H.}~\bibnamefont {Cynn}}, \bibinfo {author} {\bibfnamefont {A.}~\bibnamefont {Descamps}}, \bibinfo {author} {\bibfnamefont {F.}~\bibnamefont {Dorchies}}, \bibinfo {author} {\bibfnamefont {M.~J.}\ \bibnamefont {Duff}}, \bibinfo {author} {\bibfnamefont {A.}~\bibnamefont {Dwivedi}}, \bibinfo {author} {\bibfnamefont {C.}~\bibnamefont {Edwards}}, \bibinfo {author} {\bibfnamefont {J.~H.}\ \bibnamefont {Eggert}}, \bibinfo {author} {\bibfnamefont {D.}~\bibnamefont {Errandonea}}, \bibinfo {author} {\bibfnamefont {G.}~\bibnamefont {Fiquet}}, \bibinfo {author} {\bibfnamefont {E.}~\bibnamefont {Galtier}}, \bibinfo {author} {\bibfnamefont {A.~L.}\ \bibnamefont {Garcia}}, \bibinfo {author} {\bibfnamefont {H.}~\bibnamefont {Ginestet}}, \bibinfo {author} {\bibfnamefont {L.}~\bibnamefont {Gizzi}}, \bibinfo {author} {\bibfnamefont {A.}~\bibnamefont {Gleason}}, \bibinfo {author} {\bibfnamefont {S.}~\bibnamefont {Goede}},
  \bibinfo {author} {\bibfnamefont {J.~M.}\ \bibnamefont {Gonzalez}}, \bibinfo {author} {\bibfnamefont {M.}~\bibnamefont {Harmand}}, \bibinfo {author} {\bibfnamefont {N.~J.}\ \bibnamefont {Hartley}}, \bibinfo {author} {\bibfnamefont {P.~G.}\ \bibnamefont {Heighway}}, \bibinfo {author} {\bibfnamefont {C.}~\bibnamefont {Hernandez-Gomez}}, \bibinfo {author} {\bibfnamefont {A.}~\bibnamefont {Higginbotham}}, \bibinfo {author} {\bibfnamefont {H.}~\bibnamefont {Höppner}}, \bibinfo {author} {\bibfnamefont {R.~J.}\ \bibnamefont {Husband}}, \bibinfo {author} {\bibfnamefont {T.~M.}\ \bibnamefont {Hutchinson}}, \bibinfo {author} {\bibfnamefont {H.}~\bibnamefont {Hwang}}, \bibinfo {author} {\bibfnamefont {A.~E.}\ \bibnamefont {Lazicki}}, \bibinfo {author} {\bibfnamefont {D.}~\bibnamefont {Keen}}, \bibinfo {author} {\bibfnamefont {J.}~\bibnamefont {Kim}}, \bibinfo {author} {\bibfnamefont {P.}~\bibnamefont {Koester}}, \bibinfo {author} {\bibfnamefont {Z.}~\bibnamefont {Konopkova}}, \bibinfo {author} {\bibfnamefont
  {D.}~\bibnamefont {Kraus}}, \bibinfo {author} {\bibfnamefont {A.}~\bibnamefont {Krygier}}, \bibinfo {author} {\bibfnamefont {L.}~\bibnamefont {Labate}}, \bibinfo {author} {\bibfnamefont {Y.}~\bibnamefont {Lee}}, \bibinfo {author} {\bibfnamefont {H.-P.}\ \bibnamefont {Liermann}}, \bibinfo {author} {\bibfnamefont {P.}~\bibnamefont {Mason}}, \bibinfo {author} {\bibfnamefont {M.}~\bibnamefont {Masruri}}, \bibinfo {author} {\bibfnamefont {B.}~\bibnamefont {Massani}}, \bibinfo {author} {\bibfnamefont {E.~E.}\ \bibnamefont {McBride}}, \bibinfo {author} {\bibfnamefont {C.}~\bibnamefont {McGuire}}, \bibinfo {author} {\bibfnamefont {J.~D.}\ \bibnamefont {McHardy}}, \bibinfo {author} {\bibfnamefont {S.}~\bibnamefont {Merkel}}, \bibinfo {author} {\bibfnamefont {G.}~\bibnamefont {Morard}}, \bibinfo {author} {\bibfnamefont {B.}~\bibnamefont {Nagler}}, \bibinfo {author} {\bibfnamefont {M.}~\bibnamefont {Nakatsutsumi}}, \bibinfo {author} {\bibfnamefont {K.}~\bibnamefont {Nguyen-Cong}}, \bibinfo {author} {\bibfnamefont
  {A.-M.}\ \bibnamefont {Norton}}, \bibinfo {author} {\bibfnamefont {I.~I.}\ \bibnamefont {Oleynik}}, \bibinfo {author} {\bibfnamefont {C.}~\bibnamefont {Otzen}}, \bibinfo {author} {\bibfnamefont {N.}~\bibnamefont {Ozaki}}, \bibinfo {author} {\bibfnamefont {S.}~\bibnamefont {Pandolfi}}, \bibinfo {author} {\bibfnamefont {D.~J.}\ \bibnamefont {Peake}}, \bibinfo {author} {\bibfnamefont {A.}~\bibnamefont {Pelka}}, \bibinfo {author} {\bibfnamefont {K.~A.}\ \bibnamefont {Pereira}}, \bibinfo {author} {\bibfnamefont {J.~P.}\ \bibnamefont {Phillips}}, \bibinfo {author} {\bibfnamefont {C.}~\bibnamefont {Prescher}}, \bibinfo {author} {\bibfnamefont {T.~R.}\ \bibnamefont {Preston}}, \bibinfo {author} {\bibfnamefont {L.}~\bibnamefont {Randolph}}, \bibinfo {author} {\bibfnamefont {D.}~\bibnamefont {Ranjan}}, \bibinfo {author} {\bibfnamefont {A.}~\bibnamefont {Ravasio}}, \bibinfo {author} {\bibfnamefont {R.}~\bibnamefont {Redmer}}, \bibinfo {author} {\bibfnamefont {J.}~\bibnamefont {Rips}}, \bibinfo {author} {\bibfnamefont
  {D.}~\bibnamefont {Santamaria-Perez}}, \bibinfo {author} {\bibfnamefont {D.~J.}\ \bibnamefont {Savage}}, \bibinfo {author} {\bibfnamefont {M.}~\bibnamefont {Schoelmerich}}, \bibinfo {author} {\bibfnamefont {J.-P.}\ \bibnamefont {Schwinkendorf}}, \bibinfo {author} {\bibfnamefont {J.}~\bibnamefont {Smith}}, \bibinfo {author} {\bibfnamefont {A.}~\bibnamefont {Sollier}}, \bibinfo {author} {\bibfnamefont {J.}~\bibnamefont {Spear}}, \bibinfo {author} {\bibfnamefont {C.}~\bibnamefont {Spindloe}}, \bibinfo {author} {\bibfnamefont {M.}~\bibnamefont {Stevenson}}, \bibinfo {author} {\bibfnamefont {C.}~\bibnamefont {Strohm}}, \bibinfo {author} {\bibfnamefont {T.-A.}\ \bibnamefont {Suer}}, \bibinfo {author} {\bibfnamefont {M.}~\bibnamefont {Tang}}, \bibinfo {author} {\bibfnamefont {M.}~\bibnamefont {Toncian}}, \bibinfo {author} {\bibfnamefont {T.}~\bibnamefont {Toncian}}, \bibinfo {author} {\bibfnamefont {S.~J.}\ \bibnamefont {Tracy}}, \bibinfo {author} {\bibfnamefont {A.}~\bibnamefont {Trapananti}}, \bibinfo {author}
  {\bibfnamefont {T.}~\bibnamefont {Tschentscher}}, \bibinfo {author} {\bibfnamefont {M.}~\bibnamefont {Tyldesley}}, \bibinfo {author} {\bibfnamefont {C.~E.}\ \bibnamefont {Vennari}}, \bibinfo {author} {\bibfnamefont {T.}~\bibnamefont {Vinci}}, \bibinfo {author} {\bibfnamefont {S.~C.}\ \bibnamefont {Vogel}}, \bibinfo {author} {\bibfnamefont {T.~J.}\ \bibnamefont {Volz}}, \bibinfo {author} {\bibfnamefont {J.}~\bibnamefont {Vorberger}}, \bibinfo {author} {\bibfnamefont {J.~P.~S.}\ \bibnamefont {Walsh}}, \bibinfo {author} {\bibfnamefont {J.~S.}\ \bibnamefont {Wark}}, \bibinfo {author} {\bibfnamefont {J.~T.}\ \bibnamefont {Willman}}, \bibinfo {author} {\bibfnamefont {L.}~\bibnamefont {Wollenweber}}, \bibinfo {author} {\bibfnamefont {U.}~\bibnamefont {Zastrau}}, \bibinfo {author} {\bibfnamefont {E.}~\bibnamefont {Brambrink}}, \bibinfo {author} {\bibfnamefont {K.}~\bibnamefont {Appel}},\ and\ \bibinfo {author} {\bibfnamefont {M.~I.}\ \bibnamefont {McMahon}},\ }\bibfield  {title} {\bibinfo {title} {{Coordination
  changes in liquid tin under shock compression determined using in situ femtosecond x-ray diffraction}},\ }\href {https://doi.org/10.1063/5.0201702} {\bibfield  {journal} {\bibinfo  {journal} {Journal of Applied Physics}\ }\textbf {\bibinfo {volume} {135}},\ \bibinfo {pages} {165902} (\bibinfo {year} {2024})}\BibitemShut {NoStop}%
\bibitem [{\citenamefont {Kieffer}\ and\ \citenamefont {Karkoulis}(2013)}]{kieffer}%
  \BibitemOpen
  \bibfield  {author} {\bibinfo {author} {\bibfnamefont {J.}~\bibnamefont {Kieffer}}\ and\ \bibinfo {author} {\bibfnamefont {D.}~\bibnamefont {Karkoulis}},\ }\bibfield  {title} {\bibinfo {title} {{P}y{F}{A}{I}, a versatile library for azimuthal regrouping},\ }\href {https://doi.org/10.1088/1742-6596/425/20/202012} {\bibfield  {journal} {\bibinfo  {journal} {Journal of Physics: Conference Series}\ }\textbf {\bibinfo {volume} {425}},\ \bibinfo {pages} {202012} (\bibinfo {year} {2013})}\BibitemShut {NoStop}%
\bibitem [{\citenamefont {Heinen}\ and\ \citenamefont {Drewitt}(2022)}]{liquiddiffract}%
  \BibitemOpen
  \bibfield  {author} {\bibinfo {author} {\bibfnamefont {B.~J.}\ \bibnamefont {Heinen}}\ and\ \bibinfo {author} {\bibfnamefont {J.~W.~E.}\ \bibnamefont {Drewitt}},\ }\bibfield  {title} {\bibinfo {title} {Liquiddiffract: Software for liquid total scattering analysis},\ }\href {https://doi.org/https://doi.org/10.1007/s00269-022-01186-6} {\bibfield  {journal} {\bibinfo  {journal} {Physics and Chemistry of Minerals}\ }\textbf {\bibinfo {volume} {49}},\ \bibinfo {pages} {1432} (\bibinfo {year} {2022})}\BibitemShut {NoStop}%
\bibitem [{\citenamefont {Barker}\ and\ \citenamefont {Hollenbach}(1972)}]{barker}%
  \BibitemOpen
  \bibfield  {author} {\bibinfo {author} {\bibfnamefont {L.~M.}\ \bibnamefont {Barker}}\ and\ \bibinfo {author} {\bibfnamefont {R.~E.}\ \bibnamefont {Hollenbach}},\ }\bibfield  {title} {\bibinfo {title} {Laser interferometer for measuring high velocities of any reflecting surface},\ }\href {https://doi.org/https://doi.org/10.1063/1.1660986} {\bibfield  {journal} {\bibinfo  {journal} {Journal of Applied Physics}\ }\textbf {\bibinfo {volume} {43}},\ \bibinfo {pages} {4669} (\bibinfo {year} {1972})}\BibitemShut {NoStop}%
\bibitem [{\citenamefont {Descamps}\ \emph {et~al.}(2025)\citenamefont {Descamps}, \citenamefont {Hutchinson}, \citenamefont {Briggs}, \citenamefont {McBride}, \citenamefont {Millot}, \citenamefont {Michelat}, \citenamefont {Eggert}, \citenamefont {Albertazzi}, \citenamefont {Antonelli}, \citenamefont {Armstrong}, \citenamefont {Baehtz}, \citenamefont {Ball}, \citenamefont {Banerjee}, \citenamefont {Belonoshko}, \citenamefont {Benuzzi-Mounaix}, \citenamefont {Bolme}, \citenamefont {Bouffetier}, \citenamefont {Buakor}, \citenamefont {Butcher}, \citenamefont {Cerantola}, \citenamefont {Chantel}, \citenamefont {Coleman}, \citenamefont {Collier}, \citenamefont {Collins}, \citenamefont {Comley}, \citenamefont {Coppari}, \citenamefont {Cowan}, \citenamefont {Crépisson}, \citenamefont {Cristoforetti}, \citenamefont {Cynn}, \citenamefont {Cafiso}, \citenamefont {Dorchies}, \citenamefont {Duff}, \citenamefont {Dwivedi}, \citenamefont {Errandonea}, \citenamefont {Galtier}, \citenamefont {Ginestet}, \citenamefont
  {Gizzi}, \citenamefont {Gleason}, \citenamefont {Goede}, \citenamefont {Gonzalez}, \citenamefont {Gorman}, \citenamefont {Harmand}, \citenamefont {Hartley}, \citenamefont {Heighway}, \citenamefont {Hernandez-Gomez}, \citenamefont {Higginbotham}, \citenamefont {Höppner}, \citenamefont {Husband}, \citenamefont {Hwang}, \citenamefont {Kim}, \citenamefont {Koester}, \citenamefont {Konopkova}, \citenamefont {Kraus}, \citenamefont {Krygier}, \citenamefont {Labate}, \citenamefont {Garcia}, \citenamefont {Lazicki}, \citenamefont {Lee}, \citenamefont {Mason}, \citenamefont {Masruri}, \citenamefont {Massani}, \citenamefont {McGonegle}, \citenamefont {McGuire}, \citenamefont {McHardy}, \citenamefont {McWilliams}, \citenamefont {Merkel}, \citenamefont {Morard}, \citenamefont {Nagler}, \citenamefont {Nakatsutsumi}, \citenamefont {Nguyen-Cong}, \citenamefont {Norton}, \citenamefont {Oleynik}, \citenamefont {Otzen}, \citenamefont {Ozaki}, \citenamefont {Pandolfi}, \citenamefont {Peake}, \citenamefont {Pelka},
  \citenamefont {Pereira}, \citenamefont {Phillips}, \citenamefont {Prescher}, \citenamefont {Preston}, \citenamefont {Randolph}, \citenamefont {Ranjan}, \citenamefont {Ravasio}, \citenamefont {Redmer}, \citenamefont {Rips}, \citenamefont {Santamaria-Perez}, \citenamefont {Savage}, \citenamefont {Schoelmerich}, \citenamefont {Schwinkendorf}, \citenamefont {Singh}, \citenamefont {Smith}, \citenamefont {Smith}, \citenamefont {Sollier}, \citenamefont {Spear}, \citenamefont {Spindloe}, \citenamefont {Stevenson}, \citenamefont {Strohm}, \citenamefont {Suer}, \citenamefont {Tang}, \citenamefont {Tschentscher}, \citenamefont {Toncian}, \citenamefont {Toncian}, \citenamefont {Tracy}, \citenamefont {Tyldesley}, \citenamefont {Vennari}, \citenamefont {Vinci}, \citenamefont {Volz}, \citenamefont {Vorberger}, \citenamefont {Walsh}, \citenamefont {Wark}, \citenamefont {Willman}, \citenamefont {Wollenweber}, \citenamefont {Zastrau}, \citenamefont {Brambrink}, \citenamefont {Appel},\ and\ \citenamefont
  {McMahon}}]{descamps_VISAR}%
  \BibitemOpen
  \bibfield  {author} {\bibinfo {author} {\bibfnamefont {A.}~\bibnamefont {Descamps}}, \bibinfo {author} {\bibfnamefont {T.~M.}\ \bibnamefont {Hutchinson}}, \bibinfo {author} {\bibfnamefont {R.}~\bibnamefont {Briggs}}, \bibinfo {author} {\bibfnamefont {E.~E.}\ \bibnamefont {McBride}}, \bibinfo {author} {\bibfnamefont {M.}~\bibnamefont {Millot}}, \bibinfo {author} {\bibfnamefont {T.}~\bibnamefont {Michelat}}, \bibinfo {author} {\bibfnamefont {J.~H.}\ \bibnamefont {Eggert}}, \bibinfo {author} {\bibfnamefont {B.}~\bibnamefont {Albertazzi}}, \bibinfo {author} {\bibfnamefont {L.}~\bibnamefont {Antonelli}}, \bibinfo {author} {\bibfnamefont {M.~R.}\ \bibnamefont {Armstrong}}, \bibinfo {author} {\bibfnamefont {C.}~\bibnamefont {Baehtz}}, \bibinfo {author} {\bibfnamefont {O.~B.}\ \bibnamefont {Ball}}, \bibinfo {author} {\bibfnamefont {S.}~\bibnamefont {Banerjee}}, \bibinfo {author} {\bibfnamefont {A.~B.}\ \bibnamefont {Belonoshko}}, \bibinfo {author} {\bibfnamefont {A.}~\bibnamefont {Benuzzi-Mounaix}}, \bibinfo
  {author} {\bibfnamefont {C.~A.}\ \bibnamefont {Bolme}}, \bibinfo {author} {\bibfnamefont {V.}~\bibnamefont {Bouffetier}}, \bibinfo {author} {\bibfnamefont {K.}~\bibnamefont {Buakor}}, \bibinfo {author} {\bibfnamefont {T.}~\bibnamefont {Butcher}}, \bibinfo {author} {\bibfnamefont {V.}~\bibnamefont {Cerantola}}, \bibinfo {author} {\bibfnamefont {J.}~\bibnamefont {Chantel}}, \bibinfo {author} {\bibfnamefont {A.~L.}\ \bibnamefont {Coleman}}, \bibinfo {author} {\bibfnamefont {J.}~\bibnamefont {Collier}}, \bibinfo {author} {\bibfnamefont {G.}~\bibnamefont {Collins}}, \bibinfo {author} {\bibfnamefont {A.~J.}\ \bibnamefont {Comley}}, \bibinfo {author} {\bibfnamefont {F.}~\bibnamefont {Coppari}}, \bibinfo {author} {\bibfnamefont {T.~E.}\ \bibnamefont {Cowan}}, \bibinfo {author} {\bibfnamefont {C.}~\bibnamefont {Crépisson}}, \bibinfo {author} {\bibfnamefont {G.}~\bibnamefont {Cristoforetti}}, \bibinfo {author} {\bibfnamefont {H.}~\bibnamefont {Cynn}}, \bibinfo {author} {\bibfnamefont {S.~D.~D.}\ \bibnamefont
  {Cafiso}}, \bibinfo {author} {\bibfnamefont {F.}~\bibnamefont {Dorchies}}, \bibinfo {author} {\bibfnamefont {M.~J.}\ \bibnamefont {Duff}}, \bibinfo {author} {\bibfnamefont {A.}~\bibnamefont {Dwivedi}}, \bibinfo {author} {\bibfnamefont {D.}~\bibnamefont {Errandonea}}, \bibinfo {author} {\bibfnamefont {E.}~\bibnamefont {Galtier}}, \bibinfo {author} {\bibfnamefont {H.}~\bibnamefont {Ginestet}}, \bibinfo {author} {\bibfnamefont {L.}~\bibnamefont {Gizzi}}, \bibinfo {author} {\bibfnamefont {A.}~\bibnamefont {Gleason}}, \bibinfo {author} {\bibfnamefont {S.}~\bibnamefont {Goede}}, \bibinfo {author} {\bibfnamefont {J.~M.}\ \bibnamefont {Gonzalez}}, \bibinfo {author} {\bibfnamefont {M.~G.}\ \bibnamefont {Gorman}}, \bibinfo {author} {\bibfnamefont {M.}~\bibnamefont {Harmand}}, \bibinfo {author} {\bibfnamefont {N.~J.}\ \bibnamefont {Hartley}}, \bibinfo {author} {\bibfnamefont {P.~G.}\ \bibnamefont {Heighway}}, \bibinfo {author} {\bibfnamefont {C.}~\bibnamefont {Hernandez-Gomez}}, \bibinfo {author} {\bibfnamefont
  {A.}~\bibnamefont {Higginbotham}}, \bibinfo {author} {\bibfnamefont {H.}~\bibnamefont {Höppner}}, \bibinfo {author} {\bibfnamefont {R.~J.}\ \bibnamefont {Husband}}, \bibinfo {author} {\bibfnamefont {H.}~\bibnamefont {Hwang}}, \bibinfo {author} {\bibfnamefont {J.}~\bibnamefont {Kim}}, \bibinfo {author} {\bibfnamefont {P.}~\bibnamefont {Koester}}, \bibinfo {author} {\bibfnamefont {Z.}~\bibnamefont {Konopkova}}, \bibinfo {author} {\bibfnamefont {D.}~\bibnamefont {Kraus}}, \bibinfo {author} {\bibfnamefont {A.}~\bibnamefont {Krygier}}, \bibinfo {author} {\bibfnamefont {L.}~\bibnamefont {Labate}}, \bibinfo {author} {\bibfnamefont {A.~L.}\ \bibnamefont {Garcia}}, \bibinfo {author} {\bibfnamefont {A.~E.}\ \bibnamefont {Lazicki}}, \bibinfo {author} {\bibfnamefont {Y.}~\bibnamefont {Lee}}, \bibinfo {author} {\bibfnamefont {P.}~\bibnamefont {Mason}}, \bibinfo {author} {\bibfnamefont {M.}~\bibnamefont {Masruri}}, \bibinfo {author} {\bibfnamefont {B.}~\bibnamefont {Massani}}, \bibinfo {author} {\bibfnamefont
  {D.}~\bibnamefont {McGonegle}}, \bibinfo {author} {\bibfnamefont {C.}~\bibnamefont {McGuire}}, \bibinfo {author} {\bibfnamefont {J.~D.}\ \bibnamefont {McHardy}}, \bibinfo {author} {\bibfnamefont {R.~S.}\ \bibnamefont {McWilliams}}, \bibinfo {author} {\bibfnamefont {S.}~\bibnamefont {Merkel}}, \bibinfo {author} {\bibfnamefont {G.}~\bibnamefont {Morard}}, \bibinfo {author} {\bibfnamefont {B.}~\bibnamefont {Nagler}}, \bibinfo {author} {\bibfnamefont {M.}~\bibnamefont {Nakatsutsumi}}, \bibinfo {author} {\bibfnamefont {K.}~\bibnamefont {Nguyen-Cong}}, \bibinfo {author} {\bibfnamefont {A.-M.}\ \bibnamefont {Norton}}, \bibinfo {author} {\bibfnamefont {I.~I.}\ \bibnamefont {Oleynik}}, \bibinfo {author} {\bibfnamefont {C.}~\bibnamefont {Otzen}}, \bibinfo {author} {\bibfnamefont {N.}~\bibnamefont {Ozaki}}, \bibinfo {author} {\bibfnamefont {S.}~\bibnamefont {Pandolfi}}, \bibinfo {author} {\bibfnamefont {D.~J.}\ \bibnamefont {Peake}}, \bibinfo {author} {\bibfnamefont {A.}~\bibnamefont {Pelka}}, \bibinfo {author}
  {\bibfnamefont {K.~A.}\ \bibnamefont {Pereira}}, \bibinfo {author} {\bibfnamefont {J.~P.}\ \bibnamefont {Phillips}}, \bibinfo {author} {\bibfnamefont {C.}~\bibnamefont {Prescher}}, \bibinfo {author} {\bibfnamefont {T.~R.}\ \bibnamefont {Preston}}, \bibinfo {author} {\bibfnamefont {L.}~\bibnamefont {Randolph}}, \bibinfo {author} {\bibfnamefont {D.}~\bibnamefont {Ranjan}}, \bibinfo {author} {\bibfnamefont {A.}~\bibnamefont {Ravasio}}, \bibinfo {author} {\bibfnamefont {R.}~\bibnamefont {Redmer}}, \bibinfo {author} {\bibfnamefont {J.}~\bibnamefont {Rips}}, \bibinfo {author} {\bibfnamefont {D.}~\bibnamefont {Santamaria-Perez}}, \bibinfo {author} {\bibfnamefont {D.~J.}\ \bibnamefont {Savage}}, \bibinfo {author} {\bibfnamefont {M.}~\bibnamefont {Schoelmerich}}, \bibinfo {author} {\bibfnamefont {J.-P.}\ \bibnamefont {Schwinkendorf}}, \bibinfo {author} {\bibfnamefont {S.}~\bibnamefont {Singh}}, \bibinfo {author} {\bibfnamefont {J.}~\bibnamefont {Smith}}, \bibinfo {author} {\bibfnamefont {R.~F.}\ \bibnamefont
  {Smith}}, \bibinfo {author} {\bibfnamefont {A.}~\bibnamefont {Sollier}}, \bibinfo {author} {\bibfnamefont {J.}~\bibnamefont {Spear}}, \bibinfo {author} {\bibfnamefont {C.}~\bibnamefont {Spindloe}}, \bibinfo {author} {\bibfnamefont {M.}~\bibnamefont {Stevenson}}, \bibinfo {author} {\bibfnamefont {C.}~\bibnamefont {Strohm}}, \bibinfo {author} {\bibfnamefont {T.-A.}\ \bibnamefont {Suer}}, \bibinfo {author} {\bibfnamefont {M.}~\bibnamefont {Tang}}, \bibinfo {author} {\bibfnamefont {T.}~\bibnamefont {Tschentscher}}, \bibinfo {author} {\bibfnamefont {M.}~\bibnamefont {Toncian}}, \bibinfo {author} {\bibfnamefont {T.}~\bibnamefont {Toncian}}, \bibinfo {author} {\bibfnamefont {S.~J.}\ \bibnamefont {Tracy}}, \bibinfo {author} {\bibfnamefont {M.}~\bibnamefont {Tyldesley}}, \bibinfo {author} {\bibfnamefont {C.~E.}\ \bibnamefont {Vennari}}, \bibinfo {author} {\bibfnamefont {T.}~\bibnamefont {Vinci}}, \bibinfo {author} {\bibfnamefont {T.~J.}\ \bibnamefont {Volz}}, \bibinfo {author} {\bibfnamefont {J.}~\bibnamefont
  {Vorberger}}, \bibinfo {author} {\bibfnamefont {J.~P.~S.}\ \bibnamefont {Walsh}}, \bibinfo {author} {\bibfnamefont {J.~S.}\ \bibnamefont {Wark}}, \bibinfo {author} {\bibfnamefont {J.~T.}\ \bibnamefont {Willman}}, \bibinfo {author} {\bibfnamefont {L.}~\bibnamefont {Wollenweber}}, \bibinfo {author} {\bibfnamefont {U.}~\bibnamefont {Zastrau}}, \bibinfo {author} {\bibfnamefont {E.}~\bibnamefont {Brambrink}}, \bibinfo {author} {\bibfnamefont {K.}~\bibnamefont {Appel}},\ and\ \bibinfo {author} {\bibfnamefont {M.~I.}\ \bibnamefont {McMahon}},\ }\bibfield  {title} {\bibinfo {title} {Calibration and characterization of the line-visar diagnostic at the {H}{E}{D}-{H}{I}{B}{E}{F} instrument at the european {X}{F}{E}{L}},\ }\href {https://doi.org/10.1063/5.0271027} {\bibfield  {journal} {\bibinfo  {journal} {Rev. Sci. Instrum.}\ }\textbf {\bibinfo {volume} {96}},\ \bibinfo {pages} {075206} (\bibinfo {year} {2025})}\BibitemShut {NoStop}%
\bibitem [{\citenamefont {Katagiri}\ \emph {et~al.}(2022)\citenamefont {Katagiri}, \citenamefont {Ozaki}, \citenamefont {Murayama}, \citenamefont {Nonaka}, \citenamefont {Hironaka}, \citenamefont {Inubushi}, \citenamefont {Miyanishi}, \citenamefont {Nakamura}, \citenamefont {Okuchi}, \citenamefont {Sano}, \citenamefont {Seto}, \citenamefont {Shigemori}, \citenamefont {Sueda}, \citenamefont {Togashi}, \citenamefont {Umeda}, \citenamefont {Yabashi}, \citenamefont {Yabuuchi},\ and\ \citenamefont {Kodama}}]{katagiri}%
  \BibitemOpen
  \bibfield  {author} {\bibinfo {author} {\bibfnamefont {K.}~\bibnamefont {Katagiri}}, \bibinfo {author} {\bibfnamefont {N.}~\bibnamefont {Ozaki}}, \bibinfo {author} {\bibfnamefont {D.}~\bibnamefont {Murayama}}, \bibinfo {author} {\bibfnamefont {K.}~\bibnamefont {Nonaka}}, \bibinfo {author} {\bibfnamefont {Y.}~\bibnamefont {Hironaka}}, \bibinfo {author} {\bibfnamefont {Y.}~\bibnamefont {Inubushi}}, \bibinfo {author} {\bibfnamefont {K.}~\bibnamefont {Miyanishi}}, \bibinfo {author} {\bibfnamefont {H.}~\bibnamefont {Nakamura}}, \bibinfo {author} {\bibfnamefont {T.}~\bibnamefont {Okuchi}}, \bibinfo {author} {\bibfnamefont {T.}~\bibnamefont {Sano}}, \bibinfo {author} {\bibfnamefont {Y.}~\bibnamefont {Seto}}, \bibinfo {author} {\bibfnamefont {K.}~\bibnamefont {Shigemori}}, \bibinfo {author} {\bibfnamefont {K.}~\bibnamefont {Sueda}}, \bibinfo {author} {\bibfnamefont {T.}~\bibnamefont {Togashi}}, \bibinfo {author} {\bibfnamefont {Y.}~\bibnamefont {Umeda}}, \bibinfo {author} {\bibfnamefont {M.}~\bibnamefont
  {Yabashi}}, \bibinfo {author} {\bibfnamefont {T.}~\bibnamefont {Yabuuchi}},\ and\ \bibinfo {author} {\bibfnamefont {R.}~\bibnamefont {Kodama}},\ }\bibfield  {title} {\bibinfo {title} {Hugoniot equation-of-state and structure of laser-shocked polyimide \ch{C22H10N2O5}},\ }\href {https://doi.org/10.1103/PhysRevB.105.054103} {\bibfield  {journal} {\bibinfo  {journal} {Physical Review B}\ }\textbf {\bibinfo {volume} {105}},\ \bibinfo {pages} {054103} (\bibinfo {year} {2022})}\BibitemShut {NoStop}%
\bibitem [{\citenamefont {Zeldovitch}\ and\ \citenamefont {Raizer}(1967)}]{zeldovitch}%
  \BibitemOpen
  \bibfield  {author} {\bibinfo {author} {\bibfnamefont {Y.~B.}\ \bibnamefont {Zeldovitch}}\ and\ \bibinfo {author} {\bibfnamefont {Y.~P.}\ \bibnamefont {Raizer}},\ }\bibfield  {title} {\bibinfo {title} {Shock waves in solids, in {P}hysics of shock waves and high-temperature hydrodynamic phenomena},\ }\href@noop {} {\bibfield  {journal} {\bibinfo  {journal} {Elsevier}\ ,\ \bibinfo {pages} {685}} (\bibinfo {year} {1967})}\BibitemShut {NoStop}%
\bibitem [{\citenamefont {McQueen}\ and\ \citenamefont {{M}arsh}(1966)}]{Mcqueen}%
  \BibitemOpen
  \bibfield  {author} {\bibinfo {author} {\bibfnamefont {R.~G.}\ \bibnamefont {McQueen}}\ and\ \bibinfo {author} {\bibfnamefont {S.~P.}\ \bibnamefont {{M}arsh}},\ }\bibfield  {title} {\bibinfo {title} {Handbook of {Physical} {Constants} (unpublished data)},\ }\href@noop {} {\bibfield  {journal} {\bibinfo  {journal} {Geological Society of America Memoir}\ }\textbf {\bibinfo {volume} {97}},\ \bibinfo {pages} {153} (\bibinfo {year} {1966})}\BibitemShut {NoStop}%
\bibitem [{\citenamefont {Ramis}\ \emph {et~al.}(2009)\citenamefont {Ramis}, \citenamefont {ter Vehn},\ and\ \citenamefont {Ramírez}}]{ramis}%
  \BibitemOpen
  \bibfield  {author} {\bibinfo {author} {\bibfnamefont {R.}~\bibnamefont {Ramis}}, \bibinfo {author} {\bibfnamefont {J.~M.}\ \bibnamefont {ter Vehn}},\ and\ \bibinfo {author} {\bibfnamefont {J.}~\bibnamefont {Ramírez}},\ }\bibfield  {title} {\bibinfo {title} {Multi2d – a computer code for two-dimensional radiation hydrodynamics},\ }\href {https://doi.org/https://doi.org/10.1016/j.cpc.2008.12.033} {\bibfield  {journal} {\bibinfo  {journal} {Computer Physics Communications}\ }\textbf {\bibinfo {volume} {180}},\ \bibinfo {pages} {977} (\bibinfo {year} {2009})}\BibitemShut {NoStop}%
\end{thebibliography}%

\end{document}